\pdfoutput=1
\documentclass[11pt,twoside,a4paper,cmspaper,final,collab]{cms-tdr}
\def\svnVersion{1d83b82-D}\def\svnDate{2021/05/11}\def\cmsCernNoTag{CERN-EP-2020-219}\def\cmsCernDate{\today}\def\cmsMessage{"Published in the Journal of Instrumentation as \href{http://dx.doi.org/10.1088/1748-0221/16/05/P05014}{\doi{10.1088/1748-0221/16/05/P05014}.}"}
\begin{document}\cmsNoteHeader{EGM-17-001}

\newlength\cmsFigWidth
\setlength\cmsFigWidth{0.48\textwidth}
\providecommand{\cmsLeft}{left\xspace}
\providecommand{\cmsRight}{right\xspace}
\providecommand{\NA}{\ensuremath{\text{---}}\xspace}
\providecommand{\cmsTable}[1]{\resizebox{\textwidth}{!}{#1}}
\newlength\cmsTabSkip\setlength{\cmsTabSkip}{1ex}

\cmsNoteHeader{EGM-17-001} 

\title{Electron and photon reconstruction and identification with the CMS experiment at the CERN LHC}

\date{\today}

\abstract{
The performance is presented of the reconstruction and identification algorithms for electrons and photons with the CMS experiment at the LHC.
The reported results are based on proton-proton collision data collected at a center-of-mass energy of 13\TeV and recorded in 2016--2018, corresponding to an integrated luminosity of 136\fbinv. Results obtained from lead-lead collision data collected at $\sqrtsNN=5.02\TeV$ are also presented. Innovative techniques are used to reconstruct the electron and photon signals in the detector and to optimize the energy resolution. Events with electrons and photons in the final state are used to measure the energy resolution and energy scale uncertainty in the recorded events.
The measured energy resolution for electrons produced in $\PZ$ boson decays in proton-proton collision data ranges from 2 to 5\%, depending on electron pseudorapidity and energy loss through bremsstrahlung in the detector material.  
The energy scale in the same range of energies is measured with an uncertainty smaller than 0.1 (0.3)\% in the barrel (endcap) region in proton-proton collisions and better than 1 (3)\% in the barrel (endcap) region in heavy ion collisions.
The timing resolution for electrons from $\PZ$ boson decays with the full 2016--2018 proton-proton collision data set is measured to be 200\unit{ps}. 
}

\hypersetup{
pdfauthor={CMS Collaboration},
pdftitle={Electron and photon reconstruction and identification with the CMS experiment at the CERN LHC},
pdfsubject={CMS},
pdfkeywords={CMS, electron, photon, software},
}

\maketitle 

\tableofcontents
\clearpage
\section{Introduction}
Electrons and photons are reconstructed with high purity and efficiency in the CMS experiment, one of the two general-purpose detectors operating at the CERN LHC~\cite{Evans:2008zzb}.
These electromagnetically interacting particles  leave a distinctive signal in the electromagnetic calorimeter (ECAL) as an isolated energy deposit that is also associated with a trace in the silicon tracker in the case of electrons.
These properties, together with the excellent energy resolution of the ECAL,
make electrons and photons ideal to use both in precision measurements
and in searches for physics beyond the standard model with the CMS detector.

After a very successful Run 1, at 7 and 8\TeV during the years 2009--2012, which culminated in the discovery of the Higgs boson in July 2012~\cite{Chatrchyan:2013lba,Aad:2012tfa}
and a two-year maintenance period, the LHC resumed its operations in 2015 with LHC Run 2, providing proton-proton ($\Pp\Pp$) collisions at an increased center-of-mass-energy of 13\TeV.
In this paper, the performance of the reconstruction and identification of electrons and photons with the CMS detector in Run 2 is presented. The Run 1 results are reported in Refs.~\cite{Khachatryan:2015hwa, Khachatryan:2015iwa}.
The new results are based on $\Pp\Pp$ collision data collected during 2016--2018, and correspond to a total integrated luminosity of 136\fbinv~\cite{CMS-PAS-LUM-17-001,CMS-PAS-LUM-17-004,CMS-PAS-LUM-18-002}. The $\Pp\Pp$ collisions were delivered with a 25\unit{ns} bunch spacing, and an average number of interactions per beam crossing (pileup or PU) increasing through the years from 22 to 32.
In addition, the reconstruction of electrons and photons in lead-lead (PbPb) ion collisions is presented, which requires specific updates because of the significantly higher particle multiplicity compared with $\Pp\Pp$ collisions. The PbPb collisions were recorded in 2018 at a nucleon-nucleon (NN) center-of-mass energy of $\sqrtsNN = 5.02$\TeV, corresponding to an integrated luminosity of $1.7$\nbinv.

Table \ref{tb:objectives} lists the main objectives described in the paper concerning electrons and photons, the summary of the methods used to achieve them, as well as the reference to the sections in the paper where they are described.

\begin{table}[hbp]
\centering
\topcaption{List of the main objectives described in the paper concerning electrons and photons, a summary of the methods used to achieve them, and a reference to the section where they are detailed. }
\begin{tabular}{p{0.35\textwidth}p{0.45\textwidth}c}
\hline
Objective & Method & Section \\ \hline
Offline reconstruction& Clustering and tracking algorithms integrated in the ``particle-flow'' framework&\ref{sec:sec4} \\
Online reconstruction&Clustering and tracking algorithms with minimal differences with respect to the offline reconstruction, but not integrated in the ``particle-flow'' framework& \ref{sec:sec5}\\
Energy regression & Multivariate technique  &\ref{sec:EnergyReg}\\
Energy scale and spreading  & ``Fit method'' and ``spreading method'' & \ref{sec:SS}\\ 
Identification& Cut-based and multivariate selections& \ref{sec:sec8}\\ 
Performance comparison among the years& Energy reconstruction and object identification& \ref{sec:perfUL2017}\\ 
Timing& Comparison of arrival time of electrons from $\PZ$ decay& \ref{sec:sec8timing}\\
Performance in PbPb collisions& Clustering and tracking algorithms integrated in the modified ``particle-flow'' framework&\ref{sec:sec10} \\
\hline
\end{tabular}
\label{tb:objectives}
\end{table}

\section{The CMS detector}

This section describes in detail the parts and features of the CMS detector relevant for this paper. A more detailed description of the CMS detector, together with a definition of  the relevant kinematic variables, can be found in Ref.~\cite{Chatrchyan:2008zzk}.
The right-handed coordinate system adopted by CMS is centered in the nominal collision point inside the experiment, the $y$ axis pointing vertically upward, and the $x$ axis pointing radially inward towards the LHC center. The azimuthal angle $\phi$ is measured in radians relative to the $x$-axis in the $x$-$y$ plane. The polar angle $\theta$ is measured relative to the $z$ axis. Pseudorapidity $\eta$ is defined as $\eta \equiv -\ln \left (\tan{\theta/2}\right )$.

The central feature of the CMS apparatus is a superconducting solenoid with an internal diameter of 6\unit{m}, providing a magnetic field of 3.8\unit{T}. Within the solenoid volume are a silicon pixel and strip tracker, a lead tungstate (PbWO$_{4}$) crystal electromagnetic calorimeter, and a brass and scintillator hadron calorimeter (HCAL), each composed of a barrel and two endcap sections. Forward calorimeters (HF) extend the $\eta$ coverage provided by the barrel and endcap detectors. Muons are detected in gas-ionization chambers embedded in the steel flux-return yoke outside of the solenoid.

The silicon tracker measures charged particles within $\abs{\eta} < 2.5$ and is composed by silicon pixels and strips.
The CMS Phase~1 pixel detector~\cite{Dominguez:1481838}, installed during the 2016--17 winter shutdown, is designed to cope with an instantaneous luminosity of $2\times10^{34}\percms$  at 25\unit{ns} bunch spacing
and to maintain an excellent reconstruction efficiency.
The original (Phase~0) pixel detector had three layers in the barrel and two disks in each of the endcaps, whereas the Phase~1 pixel detector has one more layer and disk each in the barrel and endcaps, respectively, with a total of 124 million pixels.

The amount of material located upstream, i.e., in front of the ECAL, mainly consisting of the tracker, the mechanical support structure, and the cooling system, is expressed in units of radiation lengths $X_0$ and ranges from ${\simeq}0.39 X_0$ at $\abs{\eta} = 0$ to ${\simeq}1.94 X_0$ at $\abs{\eta} = 1.4$, decreasing to ${\simeq}1.53 X_0$ at $\abs{\eta} = 2$~\cite{Khachatryan:2015hwa}. The quoted numbers correspond to the Phase~1 upgraded detector that achieves an overall reduction in the tracker material budget of 0.1--0.3 $X_0$  (or 4--20\%) in the pixel region corresponding to $1.4 <\abs{\eta}< 2.0$.

For charged particles of transverse momentum $\pt$ in the range $1 < \pt < 10\GeV$ and $\abs{\eta} < 1.4$, the track resolutions are typically 1.5\% in \pt~\cite{Chatrchyan:2014fea}.

The ECAL consists of 75\,848 PbWO$_{4}$ crystals, which cover the range $\abs{\eta} < 1.48 $ in the barrel region (EB) and $1.48 < \abs{\eta} < 3.00$ in the two endcap regions (EE). The crystals are $25.8 X_0$ deep in the barrel and $24.7 X_0$ deep in the endcaps.
Preshower detectors consisting of two planes of silicon sensors interleaved with a total of $3 X_0$ of lead are located in front of each EE detector.
The energy deposited in the ECAL crystals is detected in the form of scintillation light by avalanche photodiodes (APDs) in the EB and by vacuum phototriodes (VPTs) in the EE.
The electrical signal from the photodetectors is amplified and shaped using a multigain preamplifier (MGPA), which provides three simultaneous analogue outputs that are shaped to have a rise time of approximately 50\unit{ns} and fall to 10\% of the peak value in 400\unit{ns}~\cite{CERN-LHCC-97-033}. The shaped signals are sampled at the LHC bunch crossing frequency of 40\unit{MHz} and digitized by a system of three channels of floating-point Analog-to-Digital Converters (ADCs) .
To maximize the dynamic range (40\MeV to ${\sim}$1.5--3\TeV),  three different preamplifiers with different gain settings are used for each of the ECAL crystals, each with its own ADC~\cite{Chatrchyan:2008zzk}. The largest unsaturated digitization from the 3 ADCs is used to reconstruct electromagnetic objects.

The CMS particle-flow (PF) event reconstruction~\cite{Sirunyan:2017ulk}, used to reconstruct and identify each physics-object/particle in an event,
combines optimally the information from all subdetectors.
In this process, the identification of the particle type (photon, electron, muon, charged or neutral hadron) plays an important role in the determination of the particle direction and energy. Photons (e.g., direct or coming from $\pi^{0}$ decays or from electron bremsstrahlung) are identified as ECAL energy deposits (clusters) not linked to any extrapolated track.

Electrons (e.g., direct or coming from photon conversions in the tracker material or from semileptonic decays of hadrons) are identified as primary charged-particle tracks and potentially as ECAL energy clusters. These clusters correspond to the electron tracks extrapolated to the ECAL surface and to possible bremsstrahlung photons emitted by the electron when traversing the tracker material. Muons are identified as tracks in the central tracker consistent with either tracks or several hits in the muon system, and potentially  associated with calorimeter deposits compatible with the hypothesis of being a muon.
Charged and neutral hadrons may initiate a hadronic shower in the ECAL, and are subsequently fully absorbed in the HCAL. The corresponding
clusters are used to estimate their energies and directions.

The reconstructed vertex with the largest value of summed physics-objects $\pt^2$ is the primary $\Pp\Pp$ interaction vertex. The physics objects are the jets, clustered using the anti-$\kt$ algorithm with a distance parameter of $R=0.4$~\cite{Cacciari:2008gp, Cacciari:2011ma} with the tracks assigned to the primary vertex as inputs, and the associated missing transverse momentum, which is the negative vector \pt sum of those jets and leptons.

Events of interest are selected using a two-tiered trigger system~\cite{Khachatryan:2016bia}. The first level (L1), composed of custom hardware processors, uses information from the calorimeters and muon detectors
to select events with a latency of 4\mus of the collision and with a total average rate of about 100\unit{kHz}~\cite{Perrotta:2015jyu}.
The second level, known as the high-level trigger (HLT), consists of a farm of processors running a version of the full event reconstruction software optimized for fast processing, and reduces the event rate to around 1\unit{kHz} before data storage.
Dedicated techniques~\cite{Sirunyan:2020foa} are used in all detector subsystems to reject signals from electronic noise, from pileup,  or from particles that do not originate from $\Pp\Pp$ collisions in the bunch crossing of interest, such as particles arriving from $\Pp\Pp$ collisions that occur in adjacent bunch crossings (so-called out-of-time pileup).

\section{Data and simulated event samples}\label{sec:sec3}
The data used in this paper were collected from $\Pp\Pp$ collisions at 13\TeV, satisfying a trigger requirement of an isolated single electron with $\ET$ thresholds at 27, 32, and 32\GeV in 2016, 2017 and 2018, corresponding to integrated luminosities of 35.9, 41.5, and 58.7\fbinv, respectively.

The best detector alignment, energy calibrations and corrections are being performed for the full Run 2 data for each year separately; 
they are obtained using the procedures described in Refs.~\cite{Chatrchyan:2013dga,Chatrchyan:2014wfa}. 
For this paper, only the 2017 data use these most updated conditions and the best calibrations since they were already available at the time of writing. This paper documents the performance and results that are used in more than 90\% of CMS physics analyses based on Run 2 data. 
In the later sections, the recalibrated data set of 2017 is referred to as the ``Legacy'' data set, whereas the 2016 and 2018 data samples are referred to as ``EOY'' (end of year).
The improvements brought by the recently recalibrated 2017 data are discussed in Section ~\ref{sec:EOYvsUL}.

Samples of Monte Carlo (MC) simulated events are used to compare the measured and expected performance. Drell--Yan (DY) $\PZ/\gamma^*$ + jets and $\PZ\to \mu \mu \gamma$ events are simulated at next-to-leading order (NLO) with the \MGvATNLO~(v2.2.2, 2.6.1 and 2.4.2 for 2016, 2017, and 2018 conditions, respectively)~\cite{Alwall:2014hca} event generator,  interfaced with \PYTHIA v8.212~\cite{Sjostrand:2014zea} for parton showers and hadronization. The CUETP8M1 underlying event tune~\cite{Khachatryan:2015pea} is used
for 2016 MC samples and the CP5~\cite{Sirunyan2020} tune is used for 2017 and 2018 MC samples.
The matrix elements are computed at NLO for the three processes $\Pp\Pp \to \PZ + \text{N}_{\text{jets}}$, where $\text{N}_{\text{jets}} = 0$, 1, 2, and merged with the parton showers using the FxFx~\cite{Frederix:2012ps} scheme with a merging scale of 30\GeV. The NNPDF 3.0 (2016) and 3.1 (2017--2018) with leading order (LO), in 2016, and next-to-next-leading order (NNLO), in 2017--2018, parton distribution functions (PDFs)~\cite{Ball:2014uwa} are used.
Simulated event samples for $\gamma$ + jet final states from direct photon production are generated at LO with \PYTHIA. The NNPDF2.3 LO PDFs~\cite{Ball:2012cx} are used for these samples.

A detailed detector simulation based on the \GEANTfour~(v9.4.3)~\cite{Agostinelli:2002hh} package is applied to all generated events. The presence of multiple $\Pp\Pp$ interactions in the same and nearby bunch crossings is incorporated by simulating additional interactions (including also those out-of-time coming from neighbouring bunch crossings) with a multiplicity that matches that observed in data.

\section{Offline electron and photon reconstruction}\label{sec:sec4}
\subsection{Overview of strategy and methods}
\label{sec:strategyReco}
Electrons and photons deposit almost all of their energy in the
ECAL, whereas hadrons are expected to deposit most of their energy in the HCAL. In addition, electrons
produce hits in the tracker layers. The signals in the ECAL crystals are reconstructed by fitting the signal pulse with multiple
template functions to subtract the contribution from out-of-time pileup. This procedure~\cite{Sirunyan:2020pmc} has
been used for the whole LHC Run 2 data-taking period, for both the HLT and offline event
reconstruction. As during Run 1, the signal amplitudes are corrected by time-dependent crystal response corrections and per-channel intercalibrations.

As an electron or photon propagates through the material in front of the ECAL, it may interact with the material with the electron emitting bremsstrahlung photons and the photon converting into an electron-positron pair.
Thus, by the time the electron or photon reaches the ECAL, it may no longer be a single particle, but it could consist of a shower of multiple electrons and photons.

A~dedicated algorithm is used to combine the clusters from the individual particles into a single object to recover the energy of the primary electron or photon.
Additionally, the trajectory of an electron losing momentum by emitting brems\-strah\-lung photons changes the curvature in the tracker. A~dedicated tracking algorithm, based on the Gaussian sum filter (GSF), is used for electrons to estimate the track parameters~\cite{Adam_2005}.

Electron and photon reconstruction in CMS is fully integrated into the PF framework, and is based on the same basic building blocks as other particles.
This is a major change with respect to the Run 1 reconstruction, where different reconstruction algorithms for electrons and photons were used~\cite{Khachatryan:2015hwa}.
A brief outline of the reconstruction steps is presented below and a detailed description is given in the following sections.

\begin{enumerate}

\item The energy reconstruction algorithm starts with the formation of clusters~\cite{Sirunyan:2017ulk} by grouping together crystals with energies exceeding a predefined threshold (typically ${\sim}80\MeV$ in EB and ${\sim}300\MeV$ in EE), which is generally 2 or 3 times bigger than the electronic noise expected for these crystals. 
A seed cluster is then defined as the one containing most of the energy deposited in any
  specific region, with a minimum transverse energy ($\ET^{\text{seed}}$) above 1\GeV.

We define $\ET$ as $\ET=\sqrt{\smash[b]{m^2+\pt^2}}$ for an object of mass $m$ and transverse momentum \pt.

\item ECAL clusters within a certain geometric area (``window'') around the seed cluster

 are combined into superclusters (SC) to include photon conversions and bremsstrahlung losses. This procedure is referred to as ``superclustering''.

\item Trajectory seeds in the pixel detector that are compatible with the SC position and the trajectory of an electron are used to seed the GSF tracking step.

\item In parallel to the above steps, all tracks reconstructed in the event are tested for compatibility with an
  electron trajectory hypothesis; if successful they are also used to seed the GSF tracking step. The ``generic tracks'' are a collection of tracks (not specific to electrons) selected with $ \pt > 2\GeV $, reconstructed from hits in the tracker through an iterative algorithm known as the Kalman filter (KF)~\cite{Sirunyan:2017ulk}.

\item A dedicated algorithm~\cite{Khachatryan:2015iwa} is used to find the generic tracks that are likely to originate from photons converting into $\Pe^+\Pe^-$ pairs.

\item ECAL clusters, SCs, GSF tracks and generic tracks associated with electrons, as well as conversion tracks and associated clusters, are all imported into the PF algorithm that links the elements together into blocks of particles.

\item These blocks are resolved into electron and photon (\Pe and \PGg) objects, starting from either a GSF track or a SC, respectively. At this point, there is no differentiation between electron and photon candidates. The final list of linked ECAL clusters for each candidate is promoted to a refined supercluster.

\item Electron or photon objects are built from the refined SCs based on loose selection requirements. All objects passing the selection with an associated GSF track are labeled as electrons; without a GSF track they are labeled as photons. This collection is known as the unbiased \Pe/\PGg collection and is used as a starting point by the vast majority of analyses involving electrons and photons.

\item To separate electrons and photons from hadrons in the PF framework, a tighter selection is applied to these \Pe/\PGg objects to decide if they are accepted as an electron
or an isolated photon. If the \Pe/\PGg object passes both the electron and the photon selection criteria, its object type is determined by whether it has a GSF track with a hit in the first layer of the pixel detector. If it fails the electron and photon selection criteria, its basic elements (ECAL clusters and generic tracks) are further considered to form neutral hadrons, charged hadrons or nonisolated photons in the PF framework. This is discussed further in Sec.~\ref{subsec:ged}.

\end{enumerate}

\subsection{Superclustering in the ECAL}\label{sec:ecal-clustering}

Energy deposits in several ECAL channels are clustered under the assumption that each local maximum above a certain energy threshold (1\GeV) corresponds to a single particle incident on the detector. An ECAL energy deposit may be shared between overlapping clusters, and a Gaussian shower profile is used to determine the fraction of the energy deposit to be assigned to each of the clusters.
Because electrons and photons have a significant probability of showering when traversing the CMS tracker, by the time the particle reaches the ECAL, the original object may consist of several electrons and/or photons produced from bremsstrahlung and/or pair production. The multiple ECAL clusters need to be combined into a single SC that captures the energy of the original electron/photon. This step is known as superclustering and the combining process uses two algorithms.

\begin{figure}[h!]
  \centering
    \includegraphics[width=0.5\textwidth]{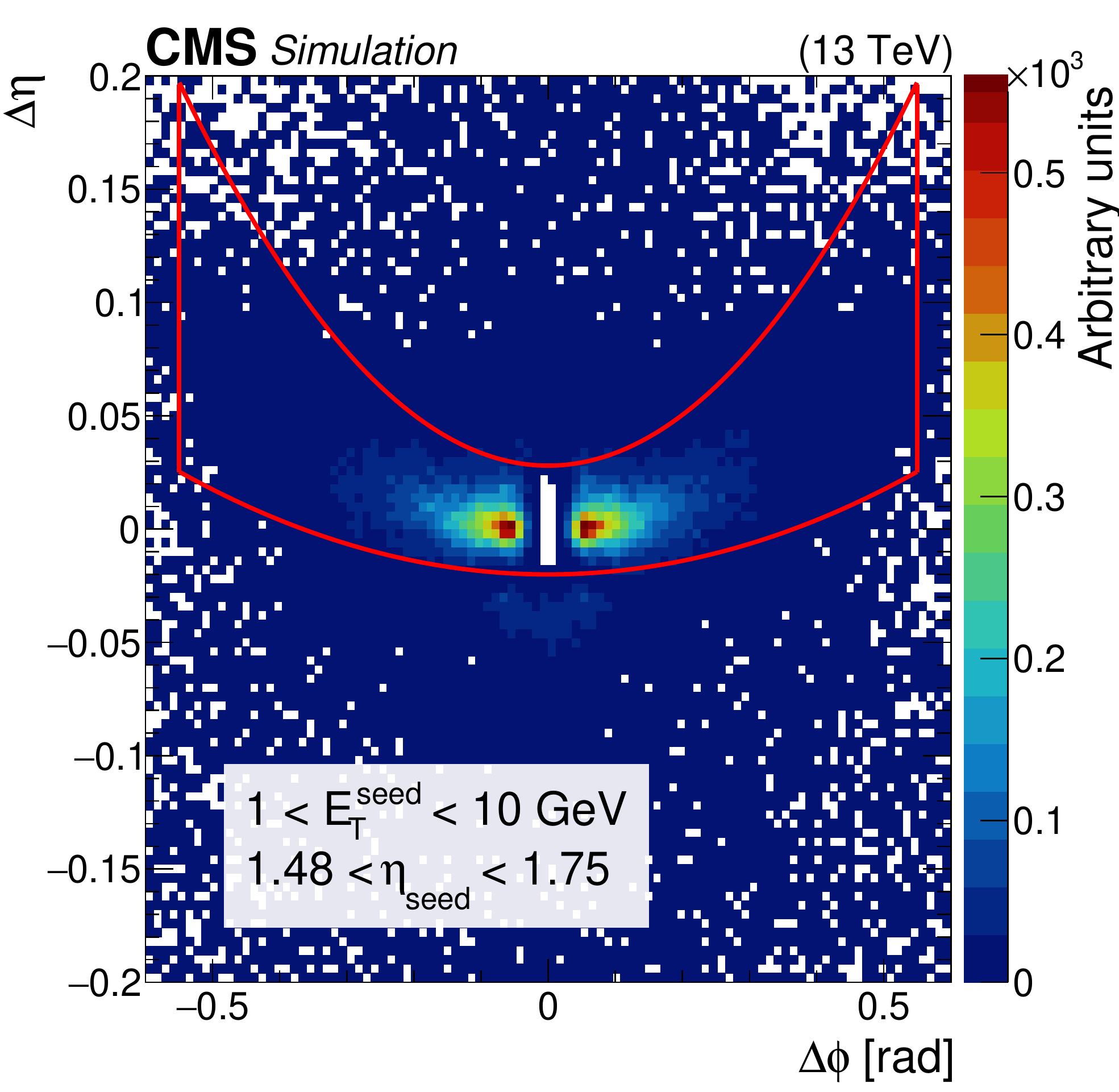}
      \caption{Distribution of $\Delta\eta=\eta_{\text{seed-cluster}}-\eta_{\text{cluster}}$ versus $\Delta\phi=\phi_{\text{seed-cluster}}-\phi_{\text{cluster}}$ for simulated electrons with $1 < \ET^{\text{seed}} <10\GeV$ and $1.48 <  \eta_{\text{seed}} <1.75$. The $z$ axis represents the occupancy of the number of PF clusters matched with the simulation (requiring to share at least 1\% of the simulated electron energy) around the seed. The red line contains approximately the set of clusters selected by the mustache algorithm. The white region at the centre of the plot represents the $\eta$-$\phi$ footprint of the seed cluster.}
    \label{fig:mustache}
\end{figure}

The first is the ``mustache'' algorithm, which is particularly useful to properly measure low-energy deposits. It uses information only from the ECAL and the preshower detector.
The algorithm starts from a cluster above a given threshold, called seed cluster. Additional clusters are added if falling into a zone, whose shape is similar to a mustache in the transverse plane.
The name mustache is used because the distribution of $\Delta\eta=\eta_{\text{seed-cluster}}-\eta_{\text{cluster}}$ versus $\Delta\phi=\phi_{\text{seed-cluster}}-\phi_{\text{cluster}}$  has a slight bend because of the solenoidal structure of the CMS magnetic
field, which tends to spread this radiated energy along $\phi$, rather than along $\eta$.
An example of the mustache SC distribution can be seen in Fig.~\ref{fig:mustache}, for simulated electrons with $1< \ET^{\text{seed}} < 10\GeV$. A similar shape is observed in the case of a photon.

The size of the mustache region depends on $\ET$, since the tracks of particles with larger transverse momenta get less bent by the magnetic field.
The mustache SCs are used to seed electrons, photons, and conversion-finding algorithms.

The second superclustering algorithm is known as the ``refined'' algorithm, and is described in more detail in Section~\ref{sec:ECALSuperclusterRefinement}.
It utilizes tracking information to extrapolate bremsstrahlung tangents and conversion tracks to decide whether a cluster should belong to a SC. It uses mustache SCs as a starting point, but is also capable of creating its own SCs. The refined SCs are used for the determination of all ECAL-based quantities of electron and photon objects.

\subsection{Electron track reconstruction and association}
\label{sec:electron-track-reconstruction}

Electrons use the GSF tracking algorithm to include radiative losses from bremsstrahlung. There have been no significant changes to the tracking algorithm from Run 1~\cite{Khachatryan:2015hwa} and any differences arise primarily from a different ECAL superclustering algorithm. Therefore, the algorithms involved in electron tracking are only briefly summarized here, with additional details available in Ref.~\cite{Khachatryan:2015hwa}.

\subsubsection{Electron seeding}
The GSF track fitting algorithm is CPU intensive and cannot be run on all reconstructed hits in the tracker.
The reconstruction of electron tracks therefore begins with the identification of a hit pattern that might lie on
an electron trajectory (``seeding'').
The electron trajectory seed can be either ``ECAL-driven'' or ``tracker-driven''. The tracker-driven seeding
has an efficiency of ${\sim}50$\% for electrons from $\PZ$ decay with $\pt\sim 3\GeV$ and drops to less than
5\% for $\pt>10\GeV$~\cite{Sirunyan:2017ulk}.

The ECAL-driven seeding first selects mustache SCs with transverse energy $E_{\text{SC,T}} > 4\GeV $ and $H/E_{\text{SC}} < 0.15$, where $E_{\text{SC}}$ and $H$ are the SC energy and the sum of the energy deposits in the HCAL towers within a cone of $\Delta R = \sqrt{\smash[b]{(\Delta\eta)^2+(\Delta\phi)^2}} = 0.15$ centered on the SC position.
Each mustache SC is then compared in $\phi$ and $z$ (or in transverse distance $r$ in the forward regions
where hits occur only in the disks) with a collection of track seeds that are formed by combining multiple hits in the inner tracker detector: triplets or doublets.
The hits of these track seeds must be located in the barrel pixel detector layers, the forward pixel layers, or the endcap tracker.
For a given SC, the trajectory of its corresponding electron is assumed to be helical and is calculated from the SC position, its $\ET^{\text{seed}}$, and the magnetic field strength. This extrapolation towards the collision vertex neglects the effect of any photon emission.
If the first two hits of a tracker seed are matched (within a certain charge-dependent $ \Delta z {\times} \Delta\phi $ window for the barrel pixel detectors, and a $ \Delta r {\times} \Delta\phi $ window for the forward pixel disks and endcap tracker) to the predicted trajectory for a SC
under any charge hypothesis, it is selected for seeding a GSF track~\cite{Khachatryan:2015hwa}.

The tracker-driven approach iterates over all generic tracks.
If any of these KF tracks is compatible with an ECAL cluster, its track seed is used to seed a GSF
track~\cite{Khachatryan:2015hwa}. The compatibility criterion is the logical OR of a cut-based selection and a multivariate selection
based on a boosted decision tree (BDT)~\cite{Hocker:2007ht,QUINLAN1987221}, using track quality and track-cluster matching variables as inputs.
Since it is computationally expensive to reconstruct all tracks in an event, tracker-driven seeding is performed only
in the offline reconstruction and not in HLT.

The ECAL-driven approach performs better for high-$\ET$ isolated electrons with a larger than 95$\%$ seeding efficiency for $\ET>10\GeV$ for electrons from $\PZ$ boson decay. The tracker-driven approach is designed to recover efficiency for low-\pt or nonisolated electrons with a seeding efficiency higher than
${\sim}50$\% for electrons with $\pt>3\GeV$~\cite{Khachatryan:2015hwa}.
It also helps to recover efficiency in the ECAL regions with less precise energy measurements, such as in the barrel-endcap transition region and/or in the gaps between supermodules.

The GSF tracking algorithm is run on all ECAL- and tracker-driven seeds.
If an ECAL-driven seed shares all but one of its hits with a tracker-driven seed, the resulting track candidate is
considered as both ECAL and tracker-seeded.
This is also the case for ECAL-driven seeds, which share all hits with a
tracker-driven seed, but in this case the tracker-driven seed is discarded before the track-finding step.
The majority of electrons fall into one of these two cases.

\subsubsection{Tracking}
The final collection of selected electron seeds (obtained by combining the ECAL-driven and tracker-driven seeds) is used to initiate the reconstruction of electron tracks. For a given seed, the track parameters evaluated at each successive tracker layer are used by the KF algorithm to iteratively build the electron trajectory, with the electron energy loss modeled using a Bethe--Heitler distribution~\cite{PhysRev.93.768}.
If the algorithm finds multiple hits compatible with the predicted position in the next layer, it creates multiple candidate trajectories by doing a $ \chi^{2} $ fit, up to a maximum of five for each tracker layer and for a given initial trajectory.
The candidate trajectories are restricted to those with at most one missing hit, and a penalty is applied to the trajectories with one missing hit by increasing the track $ \chi^{2} $. This penalty helps to minimize the inclusion of hits from converted bremsstrahlung photons in the primary-electron trajectory.
Any ambiguities that arise when a given tracker hit is assigned to multiple track candidates are resolved by dropping the track with fewer hits, or the track with the larger $\chi^2$ value if the number of hits is the same~\cite{Chatrchyan:2014fea}.

Once the track candidates are reconstructed by the KF algorithm, their parameters are estimated at each layer with a GSF fit in which the energy loss is approximated by an admixture of Gaussian distributions~\cite{Khachatryan:2015hwa}.
The GSF tracks obtained from this procedure are extrapolated toward the ECAL under the assumption of a homogeneous magnetic field to perform track-cluster associations.

\subsubsection{Track-cluster association}
\label{sec:ele_trk-clus_assoc}

The electron candidates are constructed by associating the GSF tracks with the SCs, where the position of the
SC is defined as the energy-weighted average of the constituent ECAL cluster positions.
A BDT is used to decide whether to associate a GSF track to an ECAL cluster.
The BDT combines track information, supercluster observables, and track-cluster matching variables.
The track information covers both kinematical and quality-related features.
The SC information includes the spread in $\eta$ and $\phi$ of the full SC, as well as transverse shape variables inferred from a $5{\times}5$ crystal matrix around the cluster seed.

For tracker-driven electrons, only the BDT is used to decide whether to associate a GSF track to an ECAL cluster.
Electron candidates reconstructed from ECAL-driven seeds are required to pass either the
same BDT requirements as for tracker-driven electrons or the following track-cluster matching criteria:
\begin{itemize}
        \item $ \abs{\Delta\eta} = \abs{\eta_{\text{SC}} - \eta^{\text{extrap }}_{\text{trk-in}}} < 0.02 $, with $\eta_{\text{SC}}$ being the SC $ \eta $, and $ \eta^{\text{extrap }}_{\text{trk-in}} $ the track $\eta $ at the position of closest approach to the SC (obtained by extrapolating the innermost track position and direction),
        \item $ \abs{\Delta\phi} = \abs{\phi_{\text{SC}} - \phi^{\text{extrap }}_{\text{trk-in}}} < 0.15 $, with analogous
        definitions for $ \phi $. The wider window in $ \phi $ accounts for the material effect and bending of electrons in the magnetic field.
\end{itemize}

\subsection{Supercluster refinement in the ECAL}
\label{sec:ECALSuperclusterRefinement}

The mustache SCs can be refined using the information from detector subsystems beyond the ECAL crystal and
preshower detectors. Additional conversion and bremsstrahlung clusters are recovered using information from the tracker,
with minimal risk of inclusion of spurious clusters.
A conversion-finding algorithm~\cite{Khachatryan:2015iwa} is employed to identify pairs of tracks consistent with a photon conversion.
A BDT is employed to identify tracks from photon conversions where only one leg has been reconstructed.
The input variables to this BDT include the number of missing hits on the track (for prompt electrons no missing hits are expected), the radius of the first track hit, the signed impact parameter or the distance of closest approach ($d_0$).
The identified conversion tracks can then be linked to the compatible ECAL clusters. Additionally at each tracker layer, the trajectory of the GSF track is extrapolated to form a ``bremsstrahlung tangent'', which can be linked to a compatible ECAL cluster.

Mustache SCs, ECAL clusters, primary generic tracks, GSF tracks, and conversion-flagged tracks are all inputs to the PF algorithm, which builds the \Pe/\PGg objects, as described in Ref.~\cite{Sirunyan:2017ulk}. An \Pe/\PGg object must start from either a mustache SC or a GSF track. To
reduce the CPU time, a mustache SC must either be associated with a GSF track or satisfy $E_{\text{SC,T}} > 10\GeV$ and
($H/E_\text{SC}<0.5$ or $E_{\text{SC,T}} > 100\GeV$).
The ECAL clusters must not be linked to any track from the primary vertex unless that track is associated with the object's GSF track. ECAL clusters already added by the mustache algorithm are exempted from this requirement.
ECAL clusters linked to secondary conversion tracks and bremsstrahlung tangents are then provisionally added to the so
called refined supercluster.
However, in the final step, they can be withdrawn from the refined
SC if this makes the total energy more compatible with the GSF track momentum at the inner layer.
ECAL clusters that are within $\abs{\eta}<0.05$ of the GSF track outermost position extrapolated to the ECAL or within
$\abs{\eta}<0.015$ of a bremsstrahlung tangent are exempted from this removal. Finally, a given ECAL cluster can belong
to only one refined SC.

\subsection{Integration in the global event description}
\label{subsec:ged}

Electrons and photons present a unique challenge in the PF framework because they can be composite objects consisting of several clusters and tracks. This can lead to incorrect results when an object that is not an \Pe/\PGg object is reconstructed under the \Pe/\PGg hypothesis. For example, the photons, charged hadrons, and neutral hadrons in a jet can be reconstructed as \Pe/\PGg objects instead of being reconstructed individually, and can potentially cause a large mismeasurement of the reconstructed jet energy.
Therefore, a minimal selection, as reported in Ref.~\cite{Sirunyan:2017ulk}, is applied to correctly identify hadrons and \Pe/\PGg objects and to improve the measurement of jets and missing transverse momenta.

Because of computing constraints, it is not currently feasible to rerun the PF algorithm using multiple \Pe/\PGg identification
requirements, and hence a common ``loose'' identification selection is used for electrons and photons.
A loose requirement on the BDT classifier is applied for electrons, with a different BDT used for isolated and nonisolated electrons.
Both the BDTs use various shower-shape, detector-based isolation, and tracker-related variables as input.
The BDT selection for nonisolated electrons is the one used for the selection of electron candidates, as explained in Section~\ref{sec:ele_trk-clus_assoc}.
Additionally, selection requirements on $E/p$ (the ratio between the electron energy and its momentum), $H/E$, and on quantities based on the associated generic tracks are applied to reject candidates that are problematic for jet algorithms. Occasionally, an electron can be selected by the PF algorithm, but with
its additional tracks released for charged hadron reconstruction in PF. Photon candidates are required to be isolated, and their shower-shape variables must be compatible with genuine photons.

\subsection{Bremsstrahlung and photon conversion recovery}

\begin{figure}[hbtp]
  \centering
    \includegraphics[ width=\cmsFigWidth]{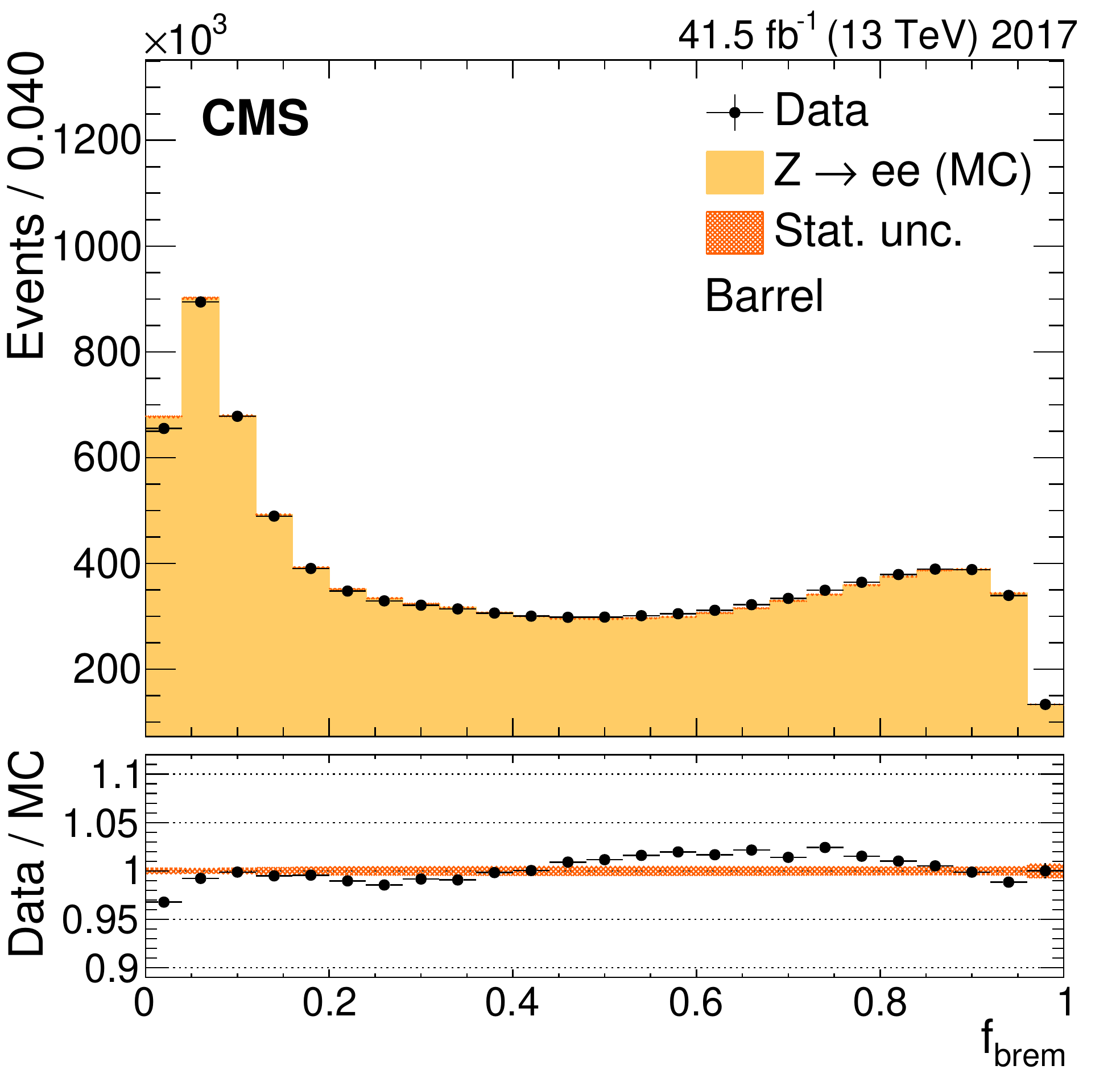}
    \includegraphics[ width=\cmsFigWidth]{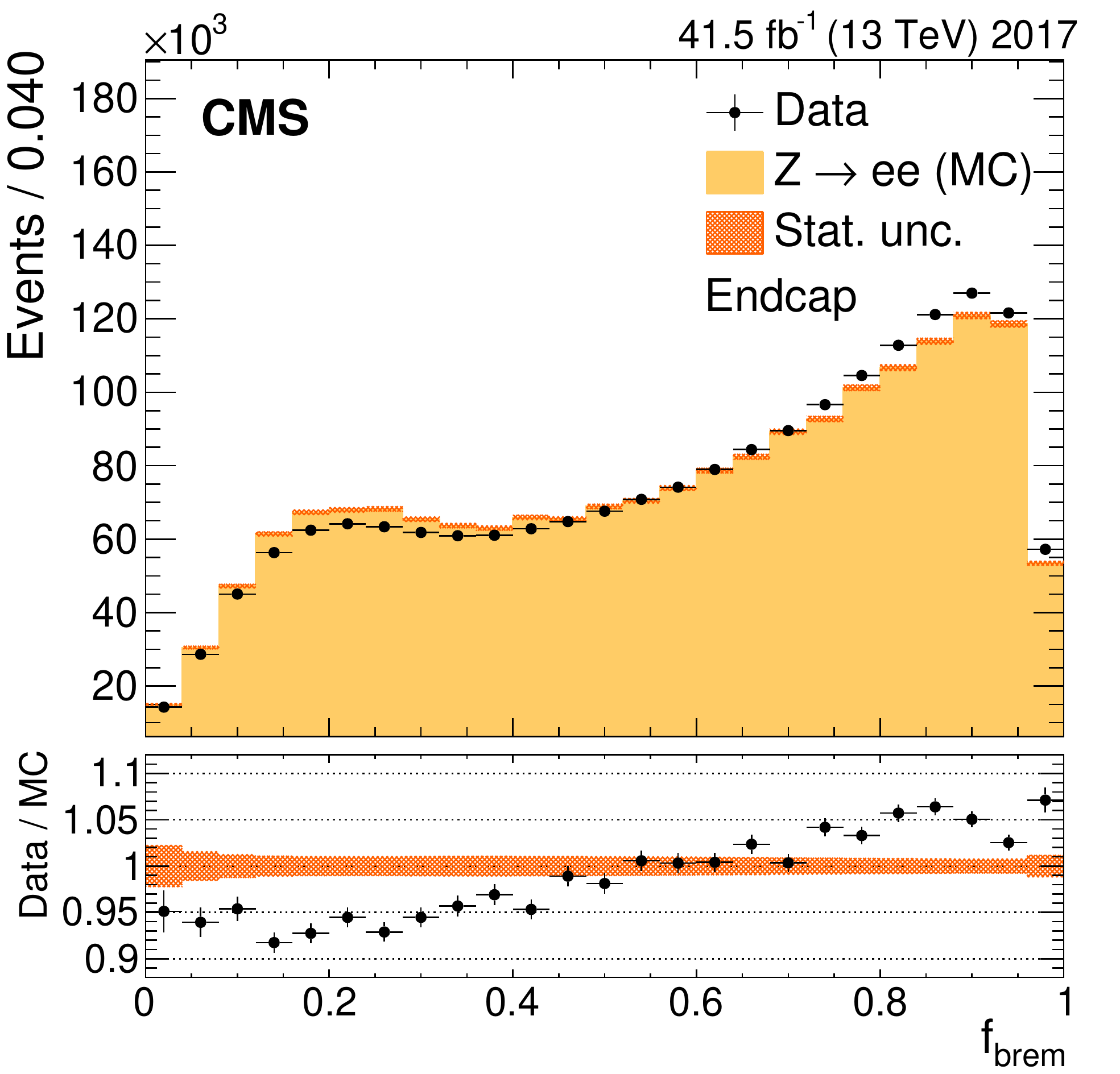}
    \caption{Fraction of the momentum lost by bremsstrahlung between the inner and outer parts of the tracker for electrons from $\PZ$ boson decays in the barrel (\cmsLeft) and in the endcaps (\cmsRight). The upper panels show the comparison between data and simulation. The simulation is shown with the filled histograms and data are represented by the markers. The vertical bars on the markers represent the statistical uncertainties in data. The
hatched regions show the statistical uncertainty in the simulation. The lower panels show the data-to-simulation ratio. }
    \label{fig:fbrem}
\end{figure}

To collect the energy of photons emitted by bremsstrahlung, tangents to the GSF tracks
are extrapolated to the ECAL surface from the track positions. A cluster that is linked to the track is considered as a potential bremsstrahlung photon if the
extrapolated tangent position is within the boundaries of the cluster, as defined above, provided that the distance
between the cluster and the GSF track extrapolation in $\eta$ is smaller than 0.05.
The fraction of the momentum lost by bremsstrahlung, as measured by the tracker is defined as:

\begin{equation}
    f_{\text{brem}} = 1 - \frac{\abs{p_\text{trk-out}}}{\abs{p_\text{trk-in}}}.
\end{equation}
where $p_\text{trk-in}$ is the momentum at the point of closest
approach to the primary vertex, and $p_\text{trk-out}$ is the momentum extrapolated to the surface of the ECAL
from the outermost tracker  layer.
Its distribution is shown in Fig.~\ref{fig:fbrem} for the barrel and the endcaps. Bremsstrahlung photons, as well as prompt photons, have a significant probability to further convert into an $\Pep\Pem$ pair in the tracker material.
Because of higher tracker material budget in the endcaps, $f_{\text{brem}}$ has a higher peak at large values, close to 1, compared
to the distribution in the barrel.
The disagreement observed between data and simulation in the endcap region is attributed to an imperfect modelling of the material in simulation.
According to simulation, the fraction of photon conversions occurring before the last tracker layer is as high as
60\% in the $\eta$ regions with the largest amount of tracker material in front of the
ECAL. A conversion-finder was therefore developed to create links between any two tracks compatible with a photon conversion~\cite{Khachatryan:2015iwa}. To recover converted bremsstrahlung photons, the vector sum of any possible bremsstrahlung pair conversion candidate track momenta is checked for compatibility with the aforementioned electron track tangents.

The photon conversion-finding algorithm is validated by reconstructing the $\PGm\PGm\PGg$ invariant mass from events in which a conversion track pair is matched to the photon, as discussed in Section~\ref{sec:PerformanceAndValidationWithData}.

\subsection{Reconstruction performance}
\label{sec:recoperf}\label{sec:tnp}
Photons are reconstructed as SCs in the ECAL
after applying a very loose selection requirement on $H/E_\text{SC}<0.5$, for which
100\% SC reconstruction efficiency is assumed.
Since electrons are additionally required to have a track matching
with the SC, the reconstruction efficiency for a SC having a matching track is computed, as described below.

Electron reconstruction efficiency is defined as the ratio between the
number of reconstructed SCs matched to reconstructed electrons and the number of all reconstructed SCs.
The electron reconstruction efficiency is computed with a tag-and-probe method using $\PZ\to \Pe\Pe$  events~\cite{Khachatryan:2010xn} as a function of the electron  $\eta$ and $\ET$, and covers all reconstruction effects. This reconstruction efficiency  is higher than 95\% for $\ET > 20\GeV$, and is compatible between data and simulation within 2\%.

The tag-and-probe technique is a generic tool to measure efficiency that exploits dileptons from the decays of resonances, such as a $\PZ$ boson or \PJgy meson. In this technique, one electron of the resonance decay, the tag, is required to pass a tight identification criterion (whose requirements are listed in detail in Sec.~\ref{sec:eleID}) and the other electron, the probe, is used to probe the efficiency under study.
The estimated efficiencies are almost insensitive to variations in the definition of the tag.
For the results in this paper, tag electrons are required to satisfy $\ET > 30$ (35)\GeV for the 2016 (2017--2018) data-taking years, respectively.
The probe is then required to pass the selection criteria (either reconstruction or identification) whose efficiency is under test.
A requirement for having oppositely charged leptons is also applied.
When there are two or more probe candidates corresponding to a given tag within the invariant mass range considered, only the probe with the highest $\ET$ is kept.
In data, the events used in the tag-and-probe procedure are required to satisfy HLT paths that do not bias the efficiency under study.

Backgrounds are estimated by fitting. The invariant mass distributions of the (tag, passing probe) and (tag, failing probe) pairs are fitted separately with a signal plus background model around the $\PZ$ boson mass in the range $[60,120]\GeV$. This range extends sufficiently far from the peak region to enable the background component to be extracted from the fit.
The efficiency under study is computed from the ratio of the signal yields extracted from the two fits. This
procedure is usually performed in bins of $\ET$ and $\eta$ of the probe electron, to measure efficiencies as a function of those variables.

\begin{figure}[hbtp]
  \centering
    \includegraphics[width=0.8\textwidth]{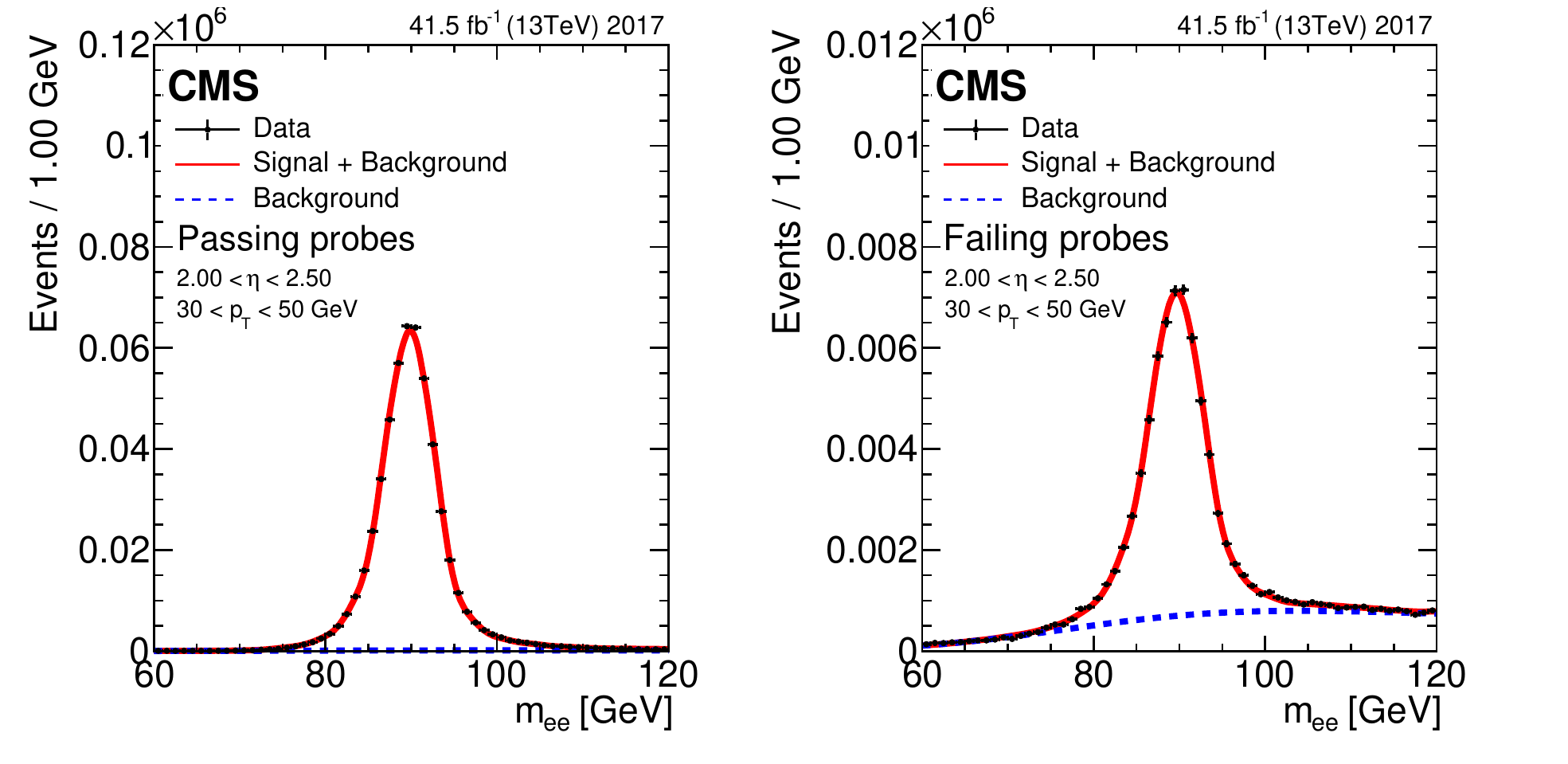}
     \caption{Example $\PZ \to \Pe\Pe$ invariant mass fits for passing (left) and failing (right) probes. Black markers show data while red solid lines show the signal + background fitting model and the blue dotted lines represent the background only component. The vertical bars on the markers represent the statistical uncertainties of the data. 
}
    \label{fig:electronID_fit_example}
\end{figure}

Different models can be used in the fit to disentangle the signal and background components.
In the absence of any kinematic selection on the
tag-and-probe candidates, the background component in the mass spectrum is well described by
a falling exponential. However, the kinematic restrictions on the $\PZ$ candidates in each $\ET$ and $\eta$
range of the probe candidate distort the mass spectrum in a way that is well described by an error
function. Consequently, the background component of the mass spectrum is described by a falling
exponential multiplied by an error function as $\mathrm{f}(m_{\Pe\Pe})= \erf[(a-m_{\Pe\Pe})b]\exp[- (m_{\Pe\Pe}-c)d]$, where $a$ and $c$ are in \GeV and $b$ and $d$ are in \GeV$^{-1}$. All parameters of the exponential and of the error function are free parameters of the fit. The model for the signal component can use analytic expressions, or
be based on templates from simulation. When using analytic functions, a Breit--Wigner (BW) function~\cite{Bohm:2004zi} with the world-average $\PZ$ boson mass and intrinsic width~\cite{Zyla2020} is convolved with a one-sided Crystal Ball (OSCB)
function~\cite{Oreglia:1980cs} that acts as the resolution function. If a template from simulation is
used, the signal part of the distribution is modeled through a sample of simulated electrons from
$\PZ$ boson decays, convolved with a resolution function to account for any remaining differences in
resolution between data and simulation.
An example fit is shown in Fig.~\ref{fig:electronID_fit_example}.
The tag-and-probe technique is applied to data and simulated events to compare efficiencies, and
evaluate data-to-simulation ratios (``scale factors''). In many analyses, these scale factors are applied as
corrections to the simulation, or are used to assess systematic uncertainties. The efficiency in
simulation is estimated from a $\PZ\to \Pe\Pe$ sample that contains no background, since a
spatial match with the generator-level electrons is required.
Several sources of systematic uncertainties are considered. The main uncertainty is
related to the model used in the fit, and is estimated by comparing alternative distributions for signal
and background, in addition to comparing analytic functions with templates from simulation. Only
a small dependence is found on the number of bins used in the fits and on the definition of the tag.

\begin{figure*}[hbtp]
\centering
\includegraphics[width=0.5\textwidth]{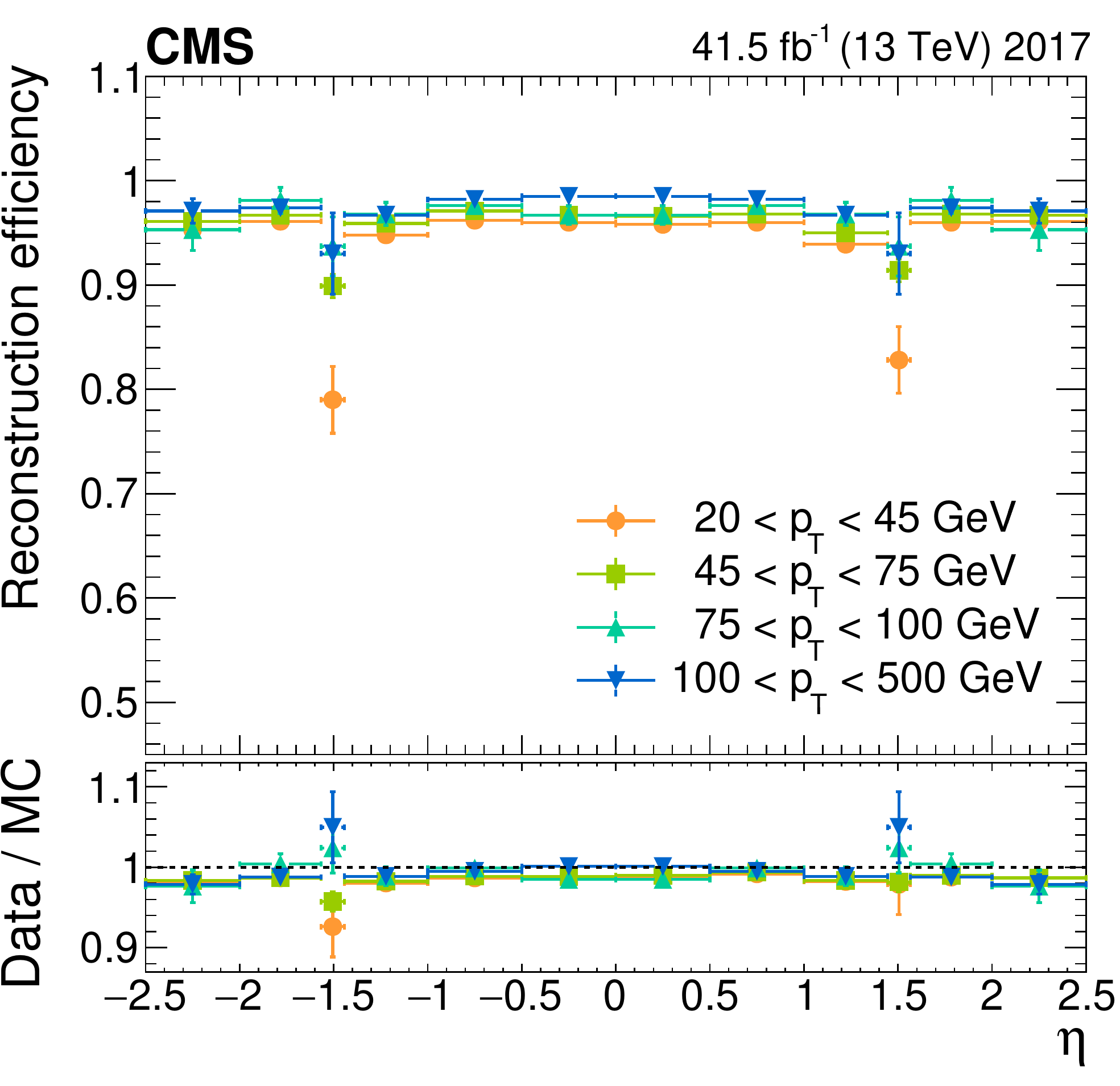}
\caption{Electron reconstruction efficiency versus $\eta$ in data (upper panel) and data-to-simulation efficiency ratios (lower panel) for the 2017 data taking period. The vertical bars on the markers represent the combined statistical and systematic uncertainties. The region $1.44  <\abs{\eta } <  1.57$ corresponds to the transition between the barrel and endcap regions of ECAL and is not considered in physics analyses. }

\label{fig:eleRecoEff2017}
\end{figure*}

The electron reconstruction efficiencies measured in 2017 data
and in simulated DY samples are shown in Fig.~\ref{fig:eleRecoEff2017}, together with the scale factors for different $\pt$ bins as a function of $\eta$.
 They are compatible in data and simulation, giving scale factors close to unity in almost the entire range.
 The region $1.44 < \abs{\eta} < 1.57$ corresponds to the transition between the barrel and endcap regions of ECAL and is not considered in a large number of physics analyses.
 The uncertainties shown in the plots correspond to the quadratic sum of the statistical and systematic contributions, dominated by the latter. The main uncertainty is related to the modeling of the signal.

Other objects, such as hadronic jets, may also produce electron-like signals, leading to such objects being misidentified as electron candidates.

\begin{figure}[h!]
  \centering
    \includegraphics[ width=\cmsFigWidth]{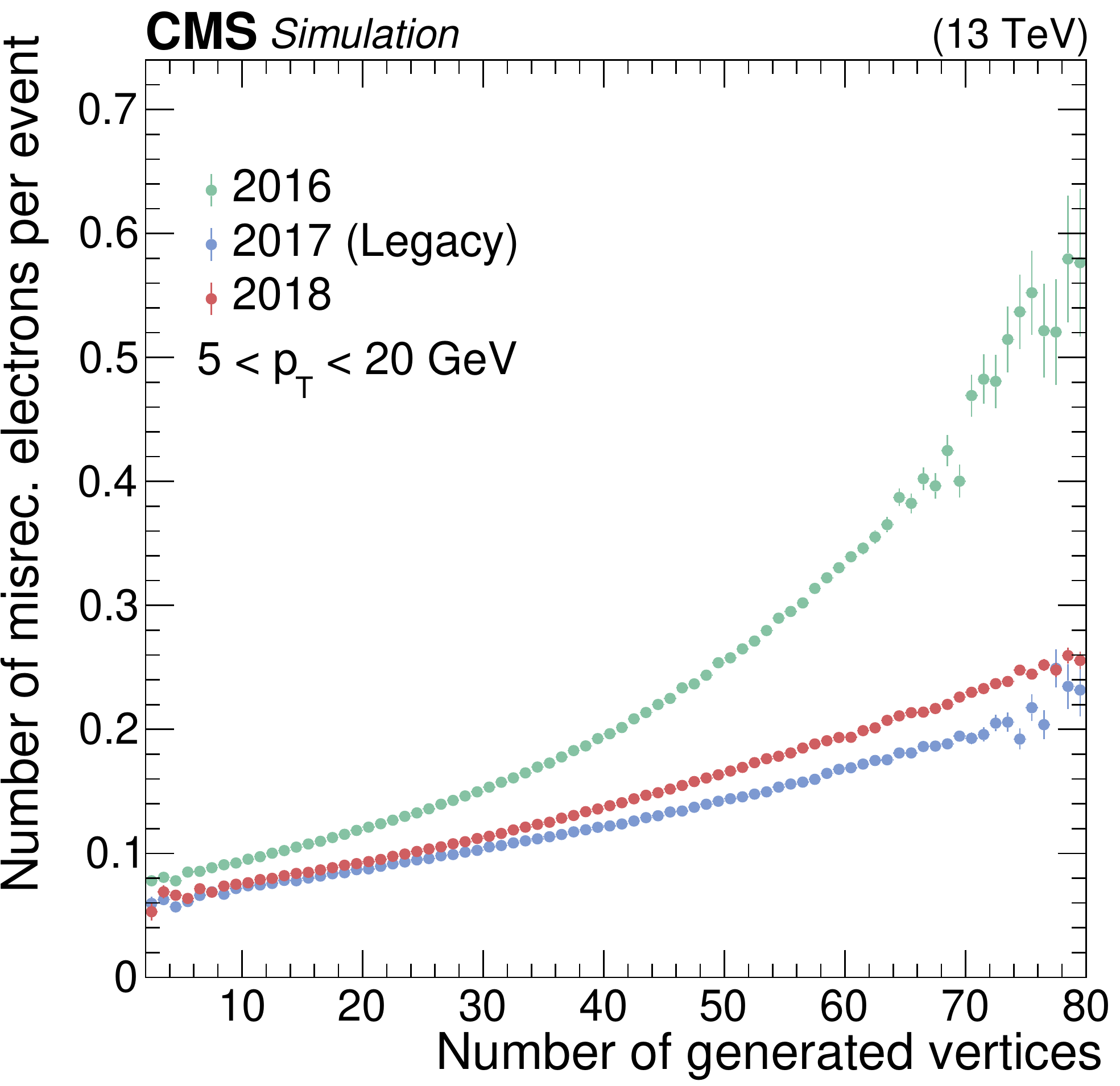}
    \includegraphics[ width=\cmsFigWidth]{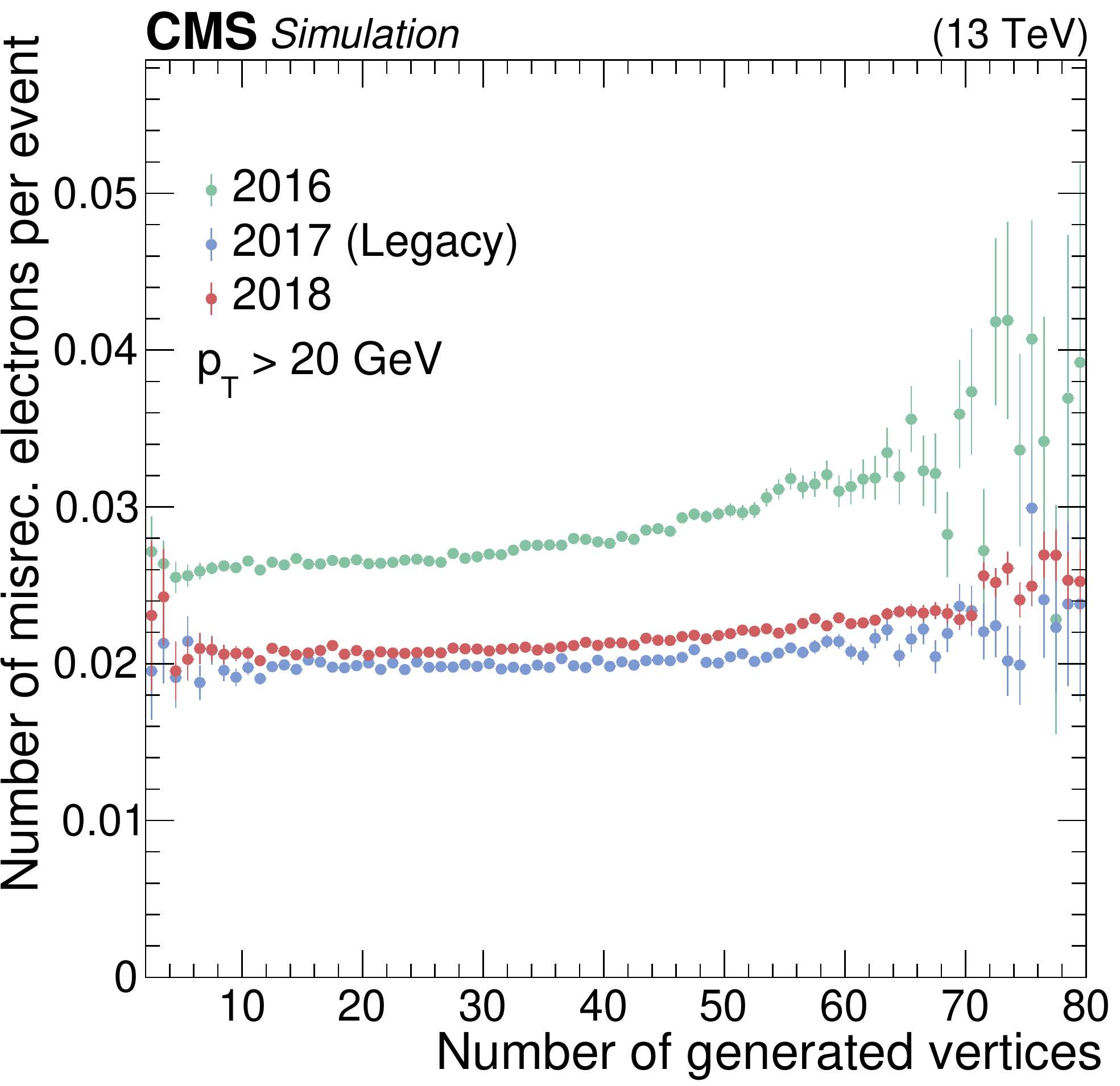}
    \caption{Number of misidentified electron candidates per event as a function of the number of generated vertices in DY + jets MC events simulated with the different detector conditions of the Run 2 data taking period. Results are shown for electrons
      with \pt in the range 5--20\GeV (left) and electrons with $\pt > 20\GeV$ (right) without further selection. The vertical bars on the markers represent the statistical uncertainties of the MC sample.}
    \label{fig:fakeVsPU}
\end{figure}

The better the reconstruction algorithm, the lower the misidentification rate per event. The larger the number of multiple interactions in an event, the larger the misidentification rate.
Figure~\ref{fig:fakeVsPU} shows the number of misidentified electron candidates per event in different \pt ranges (for
DY + jets MC events simulated with the different detector conditions corresponding to the three years of the Run 2 data taking period), as a function of the
number of pileup vertices. The significant suppression of the misidentification rate in 2017 and 2018 is due to the new pixel detector. The slightly better results in 2017 with respect to 2018 are due to the better conditions and calibrations used in the Legacy data set.

\subsection{Electron charge sign measurement}
The measurement of the electron charge sign is affected by potential bremsstrahlung followed by photon conversions. In particular, when the bremsstrahlung photons convert upstream in the detector,
the initiated showers lead to complex hit patterns, and the contributions from conversion electrons can be wrongly
included in the electron track fit.
A direct charge sign estimate is the sign of the GSF track curvature, which can be altered by the presence of conversions, especially for $\abs{\eta} > 2$, where the misidentification probability can reach 10\% for reconstructed electrons from $\PZ$ boson decays without any further selection. This is improved by combining this measurement with the estimates from two other methods. A second method is
based on the associated KF track that is matched to a GSF track when there is at least one shared hit in the innermost region. A third method evaluates the charge sign using the sign of the $\phi$ angle difference between the vector joining the nominal interaction point to the SC position and the vector connecting the nominal interaction point to the innermost hit of the electron GSF track. A~detailed description of the three methods can be found in Ref.~\cite{Khachatryan:2015hwa}.

\begin{figure*}[hbtp]
\centering
        \includegraphics[ width=\cmsFigWidth]{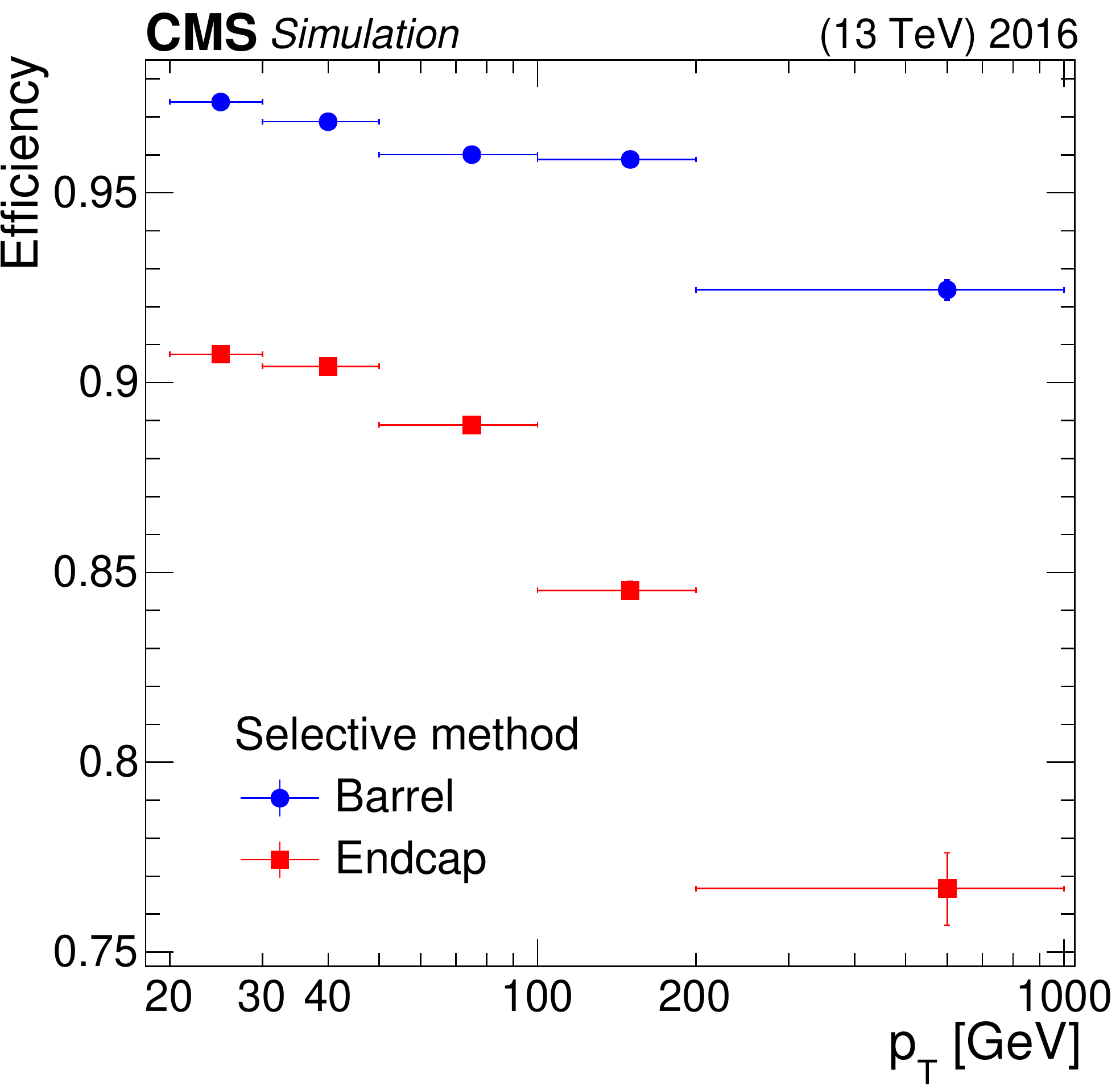}
        \caption{Efficiency of the selective method for the electron charge sign measurement as a function of \pt for electrons in the barrel and endcap regions, as measured using simulated $\PZ\to \Pe\Pe$ events. Electrons are required to satisfy the loose identification requirements described in Section~\ref{sec:eleID}. The uncertainties assigned to the points are statistical only.}
        \label{fig:chargeIDeff}
\end{figure*}

When two or three out of the three measurements agree on the sign of the charge (majority method), it is assigned as the default electron charge sign. A very high probability of correct charge sign assignment can be obtained by requiring all three  measurements to agree (selective method).
While the former method is 100\% efficient by construction, the latter has some efficiency loss.
The fraction of electrons passing the loose identification requirements (as described in Section~\ref{sec:eleID}) with all
three charge sign estimations in agreement is shown in Fig.~\ref{fig:chargeIDeff}, as a function of \pt for electrons
from $\PZ$ boson decays in the barrel and in the endcap regions.
The efficiency of the selective method for the electron charge sign measurement is better than 90 (75)\% in the barrel (endcap).

The measurement of the correct charge identification probability uses the expected number of same-sign events ($ \text{N}_{\text{SS}}^{\text{expected}} $), which in a given $\pt{-}\eta$ bin is defined as:
\begin{equation}
N_{\text{SS}}^{\text{expected}}(i,j) = p_i(1-p_j)N(i,j)+ p_j(1-p_i)N(i,j)
\end{equation}
where $p_{i,j}$ is the probability of correctly determining the electron charge in the $(i,j)$-th $\pt{-}\eta$ bin and $N$ is the number of selected electron pairs. By performing a global fit (in all the bins simultaneously) of the $ \text{N}_{\text{SS}}^{\text{expected}} $ to the observed number, the probability $p$ for each bin can be obtained in both data and simulation.
The electrons are required to pass the loose identification requirements, as described in Section~\ref{sec:eleIDCB}. Tighter identification requirements, specifically those requiring no ``missing hits'' for the track, have different efficiencies and correct charge sign identification probabilities. In this procedure no background subtraction is applied.

\begin{figure}[hbtp]
\centering
        \includegraphics[ width=\cmsFigWidth]{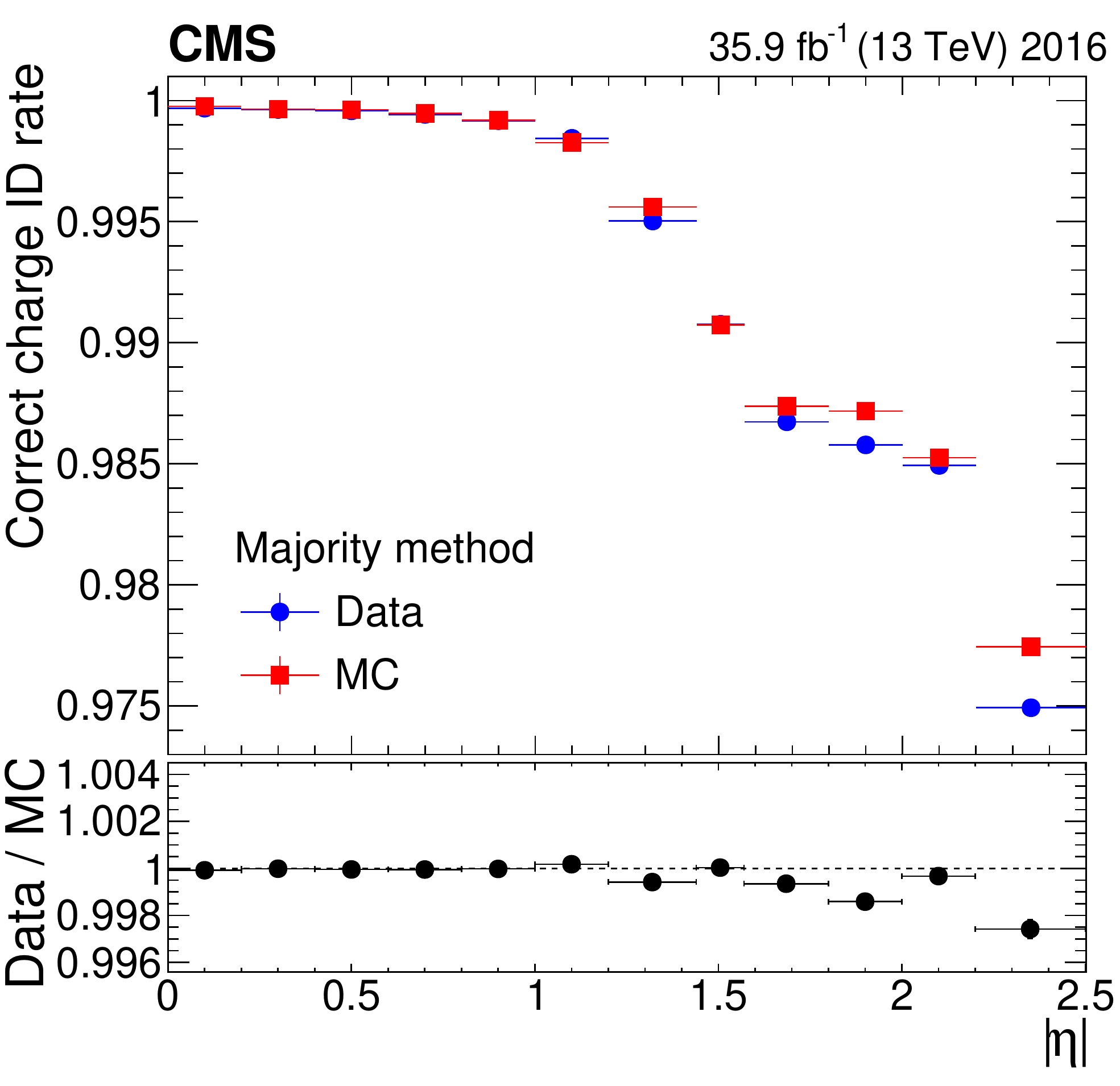}
        \includegraphics[ width=\cmsFigWidth]{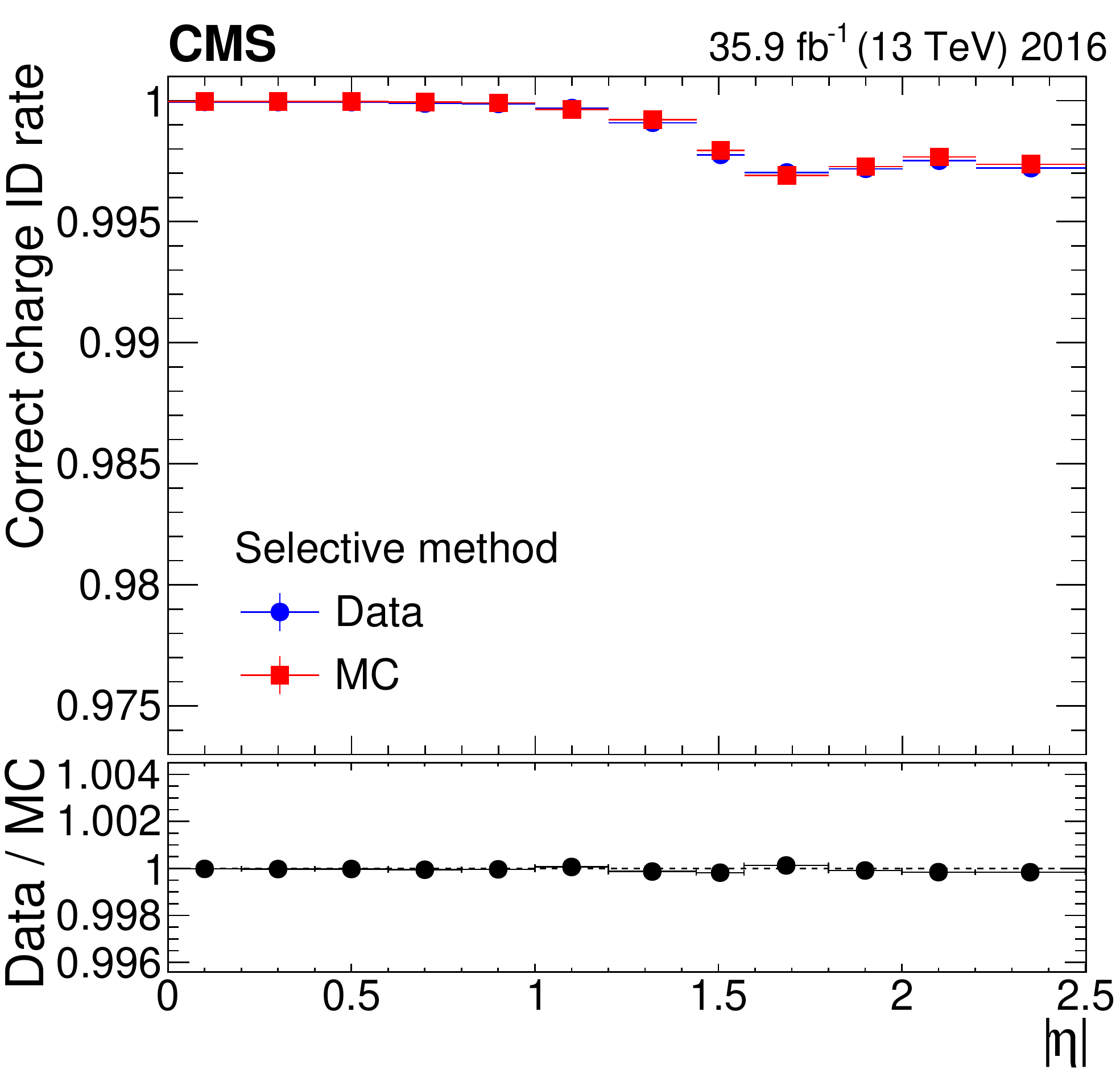}
        \caption{Correct charge identification probability for electrons using the majority method (\cmsLeft) and the selective method (\cmsRight), as measured in the 2016 data set and in simulated $\PZ\to \Pe\Pe$ events. The electrons are required to satisfy the loose identification requirements described in Section~\ref{sec:eleID}. The uncertainties assigned to the points are statistical only.}
        \label{fig:chargeID_vsAbsEta}
\end{figure}

Figure~\ref{fig:chargeID_vsAbsEta} shows the probability of correct charge assignment of the majority (left) and selective (right) methods, as a function of the electron's $\abs{\eta}$.
The charge identification rate using the 2016 data set is compared with the correct charge assignment probability obtained in $\PZ\to \Pe\Pe$ simulated events.

From the data-to-simulation comparison, the systematic uncertainty in the charge sign assignment probability for electrons is less than 0.1\% in the barrel and 0.3\% in the endcap regions.

\section{Online electron and photon reconstruction}\label{sec:sec5}
Electron and photon candidates at L1 are based on ECAL trigger towers defined by arrays of 5$\times$5 crystals in the barrel and by a more complicated pattern in the endcaps, because of the different layout of the crystals~\cite{CERN-LHCC-97-033}.
The central trigger tower with the largest transverse energy above a fixed threshold ($\ET> 2\GeV$) is designated as the seed tower.
To recover energy losses because of bremsstrahlung, clusters are built from surrounding towers with $\ET$ above 1\GeV to form the L1 candidates.
The sum of the $\ET$ of all towers in the cluster is the raw cluster $\ET^\text{L1}$. To obtain better identification of L1 e/$\gamma$ candidates,
requirements are set on: (i) the energy distribution between the central and neighboring towers; (ii) the amount of energy deposited in the HCAL downstream of the central tower, the $\ET^\text{L1}$ of the candidate; and (iii) variables sensitive to the spatial extent of the electromagnetic shower~\cite{CMS-PAS-TRG-17-001}. No tracker information is available at L1, so electrons and photons are indistinguishable at this stage.

The HLT electron and photon candidates are reconstructed from  energy deposits in ECAL crystals grouped into clusters around the corresponding L1 candidate (called the L1 seed). For a given L1 seed, the ECAL clustering algorithm is performed by the HLT from the readout channels overlapping a matrix of crystals centered on the L1 candidate. The HLT processing time is kept short by clustering only around the L1 seed. Based on this L1 seed, superclusters are built using offline reconstruction algorithms, and the HLT requirements are applied as follows.
For electron candidates, the ECAL SC is associated with a reconstructed track whose direction is compatible with its location. Electron and photon selection at the HLT relies on the identification and isolation criteria, together with minimal thresholds on the SC $\ET^\text{HLT}$ (i.e., the energy measured by the HLT using only the ECAL information, without any final calibration applied). The identification criteria are based on the transverse profile of the cluster energy deposited in the ECAL, the amount of energy in the HCAL downstream from the ECAL SC, and (for electrons) the degree of association between the track and the ECAL SC. The isolation criteria make use of the energy deposits that surround the HLT electron candidate in the ECAL, HCAL, and tracker detectors.

\subsection{Differences between online and offline reconstruction}
The HLT must ensure a large acceptance for physics signals, while keeping the CPU time and output rate under control. This is achieved by exploiting the same software used for the off\-line analysis that ensures high reconstruction efficiency and reduces the trigger rate by applying stringent identification criteria and quality selections. The differences between the HLT and offline reconstruction code are minimal and are mainly driven by: (i) the limited CPU time available at the HLT, fixed at about 260\unit{ms} by the number of processing CPUs and the L1 input rate; (ii) the lack of final calibrations, which are not yet computed during the data-taking period; and (iii) more conservative selection criteria to avoid rejecting potentially interesting events.

To keep the processing time short, all trigger paths have a modular structure and are characterized by a sequence of reconstruction and filtering blocks of increasing complexity. Thus faster algorithms are run first, and their products are immediately filtered, allowing the remaining algorithms to be skipped when the event fails any given filter. Another important time-saving optimization is to restrict the detector readout and reconstruction to regions of interest around the L1 candidates.
Moreover, HLT SCs, which are more robust against possible background contamination, have a simpler energy correction than the offline reconstruction.

The main difference between the online and offline reconstruction occurs in the tracking algorithms. Every electron candidate reconstructed at the HLT is ECAL-driven; the algorithm starts by finding a supercluster and then looks for a matching track reconstructed in the pixel detector.
The association is performed geometrically, matching the SC trajectory to pixel detector hits. Since 2017, the online pixel matching algorithm requires three pixel hits rather than two, as in the offline algorithm, to maximize early background rejection. Two pixel detector hits are accepted only if the trajectory passes through a maximum of three active modules. Once the SC is associated with the pixel detector seeds, the electron track is reconstructed using the same GSF algorithm as employed offline. Since this algorithm is used only when the pixel matching succeeds, the processing time is considerably reduced. Moreover, not all electron paths lead to reconstructed tracks; some of them can achieve significant rate reduction from pixel detector matching alone. For isolated electrons, all the nearby tracks must be reconstructed to build the track isolation variables. This is accomplished at the end of the path by using an iterative tracking algorithm similar to that applied offline, but specifically customized for the HLT and with fewer iterations of the tracking procedure.

Offline tracker-driven electron reconstruction is advantageous only for low energy or nonisolated electrons, neither of which is easy to trigger on. The use of only ECAL-driven electrons at the HLT is thus a reasonable simplification with respect to the offline reconstruction.

Other differences that exist with respect to the offline reconstruction concern calorimetry. At HLT the timing selection requirement applied offline to reject out-of-time hits (e.g., pileup, anomalous signals in ECAL from the interaction of particles in photodetectors, cosmic and beam halo events) is removed, since it does not significantly reduce the rate and risks losing rare signatures, such as the detection of long-lived particles. Moreover, the ECAL online calibration is also different; the response corrections for the crystal transparency loss that are applied at HLT during the data-taking period and updated twice per week are not as accurate as the ones used by the offline reconstruction.

Finally, some online variables are defined differently with respect to offline. The $\ET$ is computed with respect to the origin of the CMS reference system, instead of the actual position of the collision primary vertex, and it is measured using only calorimeter information, without any track-based corrections or final calibrations. The online particle isolation is defined by exploiting energy clusters built in the ECAL and HCAL and tracks reconstructed in the tracker, instead of using the more complete PF information, which is available offline. Some other variables, such as $H/E$, are defined in the same way both offline and online, although with slightly different parameters, that ensure the online selection is always looser than offline.

\subsection{Electron trigger requirements and performance}
The electron triggers correspond to the first selection step of most offline analyses using electrons, which requires the presence of at least one, two or three HLT electron candidates. Because of bandwidth limitations, both L1 seeds and HLT paths may be prescaled, i.e., they  may record only a fraction of the events, to reduce the trigger rate. Tables~\ref{tb:unprescale_L1} and~\ref{tb:unprescale_HLT} show the lowest unprescaled L1 and HLT $\ET$ thresholds and the corresponding L1 seed and the HLT path names of Run 2~\cite{CMS-PAS-TRG-17-001}.

\begin{table}[hbp]
\centering
\topcaption{Lowest unprescaled $\ET^\text{L1}$ thresholds and the corresponding seed names, for the three years of Run 2.}
\begin{tabular}{lll} \hline
& $\ET^\text{L1}$ threshold [\GeVns]& $\abs{\eta}$ range \\ \hline
2016&&\\ 
Single electron/photon & 40  & $<$3\\
Single electron/photon (isolated) & 30& $<$3 \\
Double electron/photon & 23, 10 &$<$3\\ 
[\cmsTabSkip]
2017&&\\ 
Single electron/photon & 40  &$<$3\\
Single electron/photon (isolated) & 32  &$<$3\\
Single electron/photon (isolated) & 30 &$<$2.1\\
Double electron/photon & 25, 14 &$<$3\\ 
[\cmsTabSkip]
2018&&\\ 
Single electron/photon & 40  &$<$2.5\\
Single electron/photon (isolated) & 32  &$<$2.5\\
Single electron/photon (isolated) & 30 & $<$2.1\\
Double electron/photon & 25, 14 &$<$2.5\\ \hline
\end{tabular}
\label{tb:unprescale_L1}
\end{table}

\begin{table}[hbp]
\centering
\topcaption{Lowest unprescaled $\ET^\text{HLT}$ thresholds and the corresponding path names, for the three years of Run 2 data-taking. The electrons are always required to be within the L1 $\abs{\eta}$ requirement and always within $\abs{\eta} <2.65$.}
\begin{tabular}{lll} \hline
& $\ET^\text{HLT}$ threshold [\GeVns]\\ \hline
2016&&\\ 
Single electron & 27 \\
Double electron (isolated)& 23, 12 \\
Double electron (nonisolated) &33\\
Triple electron & 16, 12, 8 \\ 
[\cmsTabSkip]
2017&&\\ 
Single electron & 32 \\
Double electron (isolated)& 23, 12 \\
Double electron (nonisolated) &33\\
Triple electron & 16, 12, 8 \\ 
[\cmsTabSkip]
2018&&\\ 
Single electron & 32 \\
Double electron (isolated)& 23, 12 \\
Double electron (nonisolated) &25\\
Triple electron & 16, 12, 8 \\ \hline
\end{tabular}
\label{tb:unprescale_HLT}
\end{table}

The single- and double-electron trigger performance is reported, using the full Run 2 data sample corresponding to an integrated luminosity of 136\fbinv.
Efficiencies are obtained using a data-driven method based on the tag-and-probe technique (described in detail in Section~\ref{sec:tnp}), which exploits $\PZ\to \Pe\Pe$ events and requires one electron candidate, called the tag, to satisfy tight selection requirements, while leaving the other electron of the pair, called the probe, unbiased to measure the efficiency.

For the results presented in this section, tag electrons are required to pass the criteria described in Section~\ref{sec:tnp}.
Moreover, they have to satisfy an unprescaled single-electron trigger, with $\ET^\text{HLT} > 27$ and 32\GeV for the 2016 and 2017--2018 data-taking periods, respectively.
Probe electrons must have $\abs{\eta} <  2.5$ and $\ET^\text{ECAL} >5\GeV$, with $\ET^\text{ECAL} = E^\text{ECAL} \sin{\theta_{\text{SC}}}$, where $E^\text{ECAL}$ is the best estimate of the electron energy measured by ECAL and $\theta_{\text{SC}}$ is the angle with respect to the beam axis of the electron SC. No additional identification criteria are applied to the probes. To measure the trigger efficiency, probes are then required to pass the HLT path under study. The electron triggers analyzed in this paper are the following:
\begin{itemize}
        \item HLT\_Ele(27)32\_WPTight\_Gsf: standard single-electron trigger with tight identification and isolation requirements. The electron $\ET^\text{HLT}$ is required to be above 27\GeV in 2016 and above 32\GeV in 2017-2018.
        \item HLT\_Ele23\_Ele12\_CaloIdL\_TrackIdL\_IsoVL: standard double-electron trigger with loose identification (CaloIdL, TrackIdL) and isolation (IsoVL) requirements. The $\ET^\text{HLT}$ thresholds of the two electrons are 23 and 12\GeV, respectively.
\end{itemize}

Photon triggers are not included in this paper, since they are very similar to electron triggers, except for the absence of the requirement on the presence of matching tracks. Moreover, photon triggers are usually designed for specific analyses and are not used as extensively as the electron triggers described above.

\begin{figure}[hbtp]
\centering
    \includegraphics[ width=\cmsFigWidth]{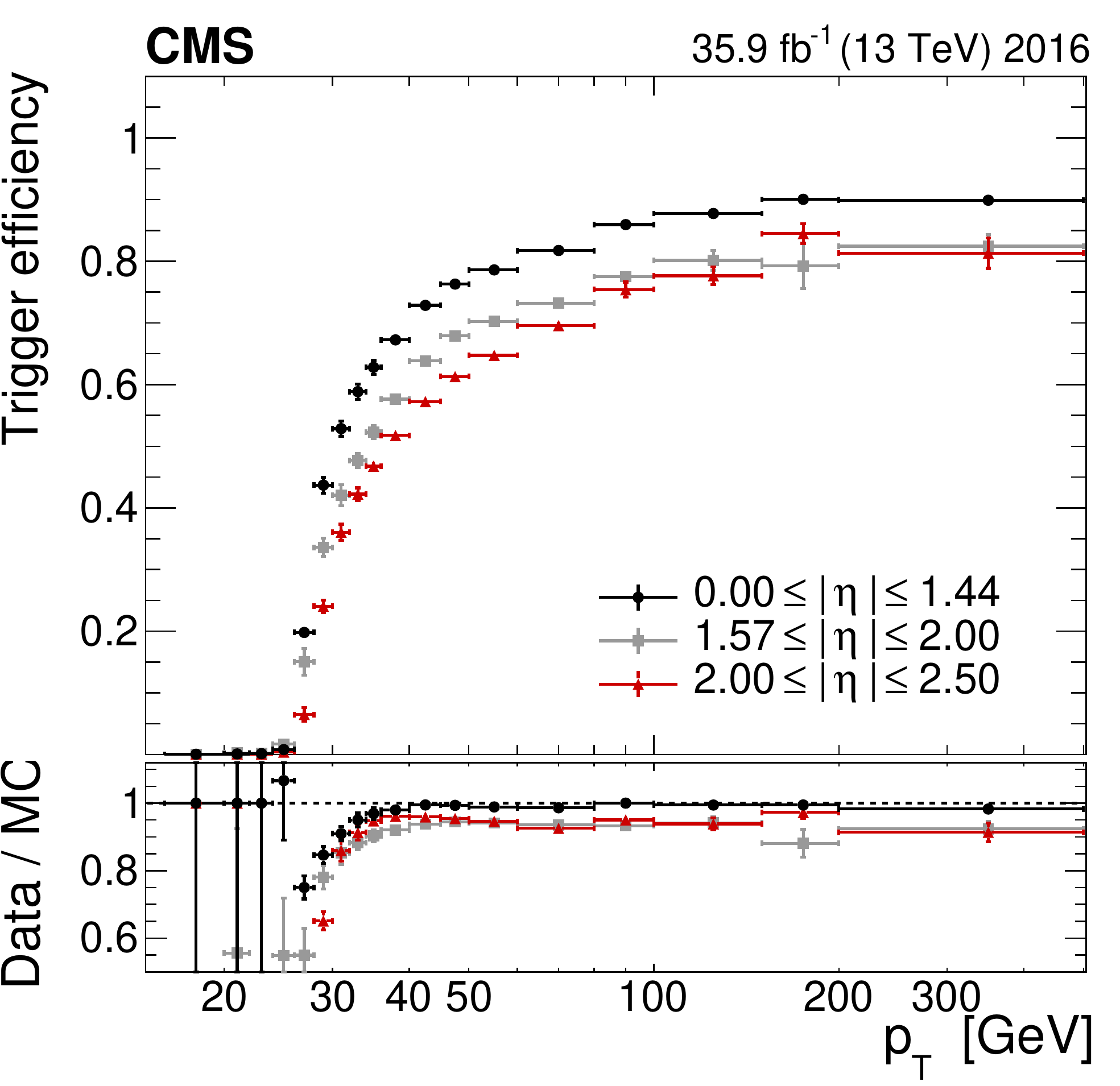}
    \includegraphics[ width=\cmsFigWidth]{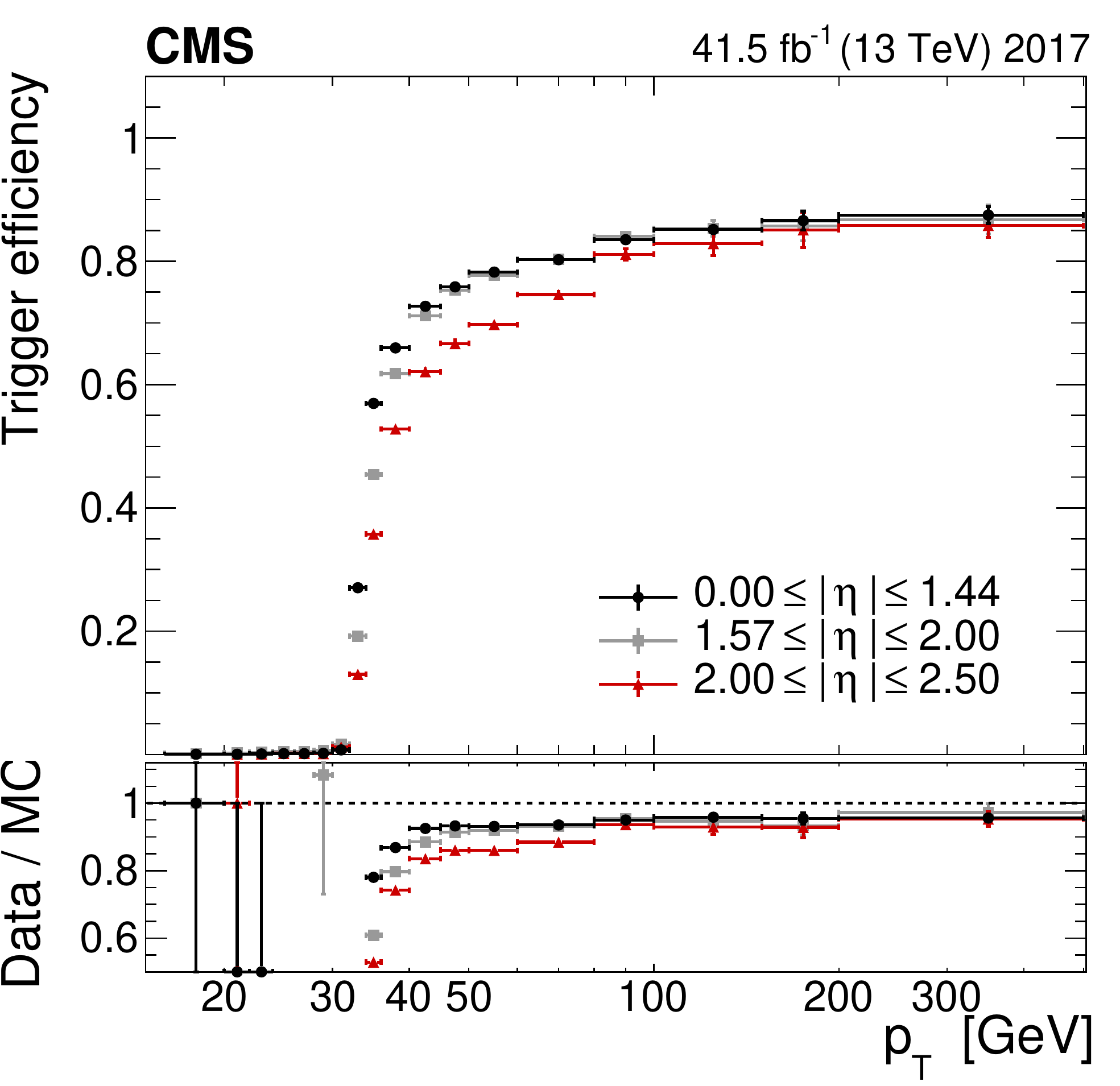}
    \includegraphics[ width=\cmsFigWidth]{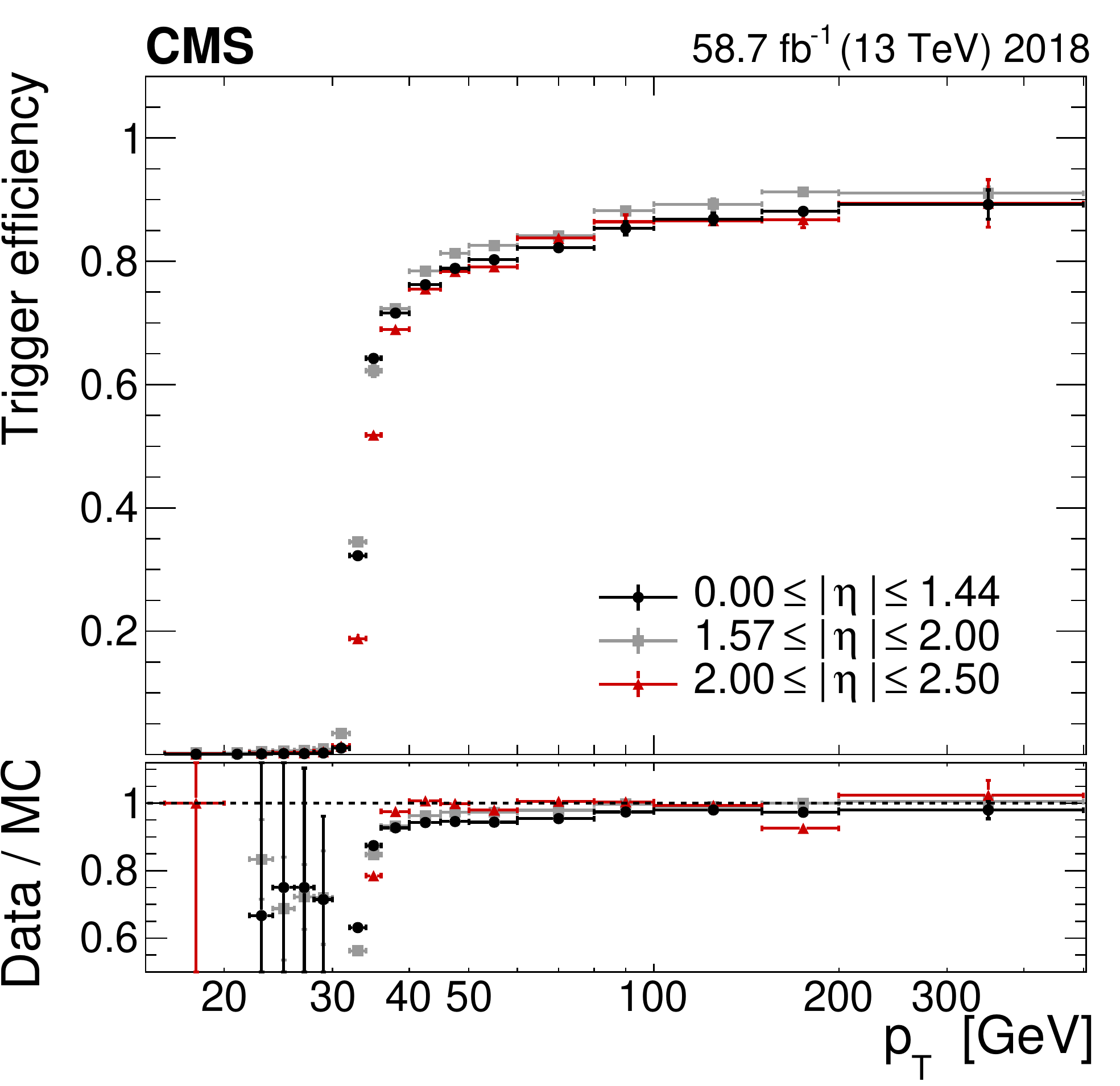}
\caption{Efficiency of the unprescaled single-electron HLT path with the lowest $\ET^\text{HLT}$ requirement (HLT\_Ele27\_WPTight\_Gsf in 2016, HLT\_Ele32\_WPTight\_Gsf in 2017--2018), with respect to the offline reconstruction, as a function of the electron \pt, for different $\eta$ regions using the 2016 (upper left), 2017 (upper right) and 2018 (lower) data sets.  The bottom panel shows the data-to-simulation ratio. The efficiency measurements combine the effects of the L1 and HLT triggers. The vertical bars on the markers represent combined statistical and systematic uncertainties.}
\label{fig:SingleEle_pt}
\end{figure}

\begin{figure}[hbtp]
\centering
    \includegraphics[ width=\cmsFigWidth]{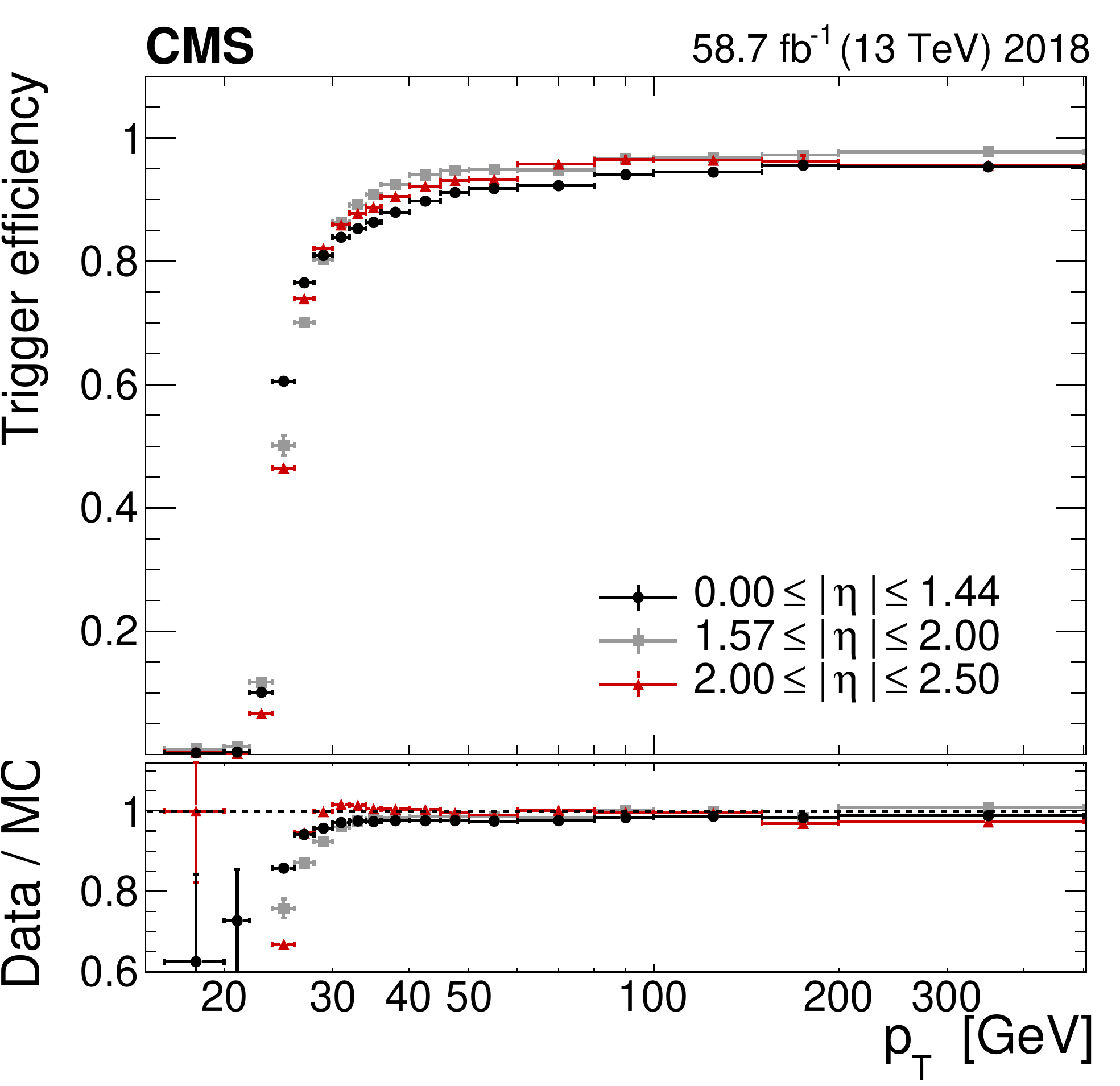}
    \includegraphics[ width=\cmsFigWidth]{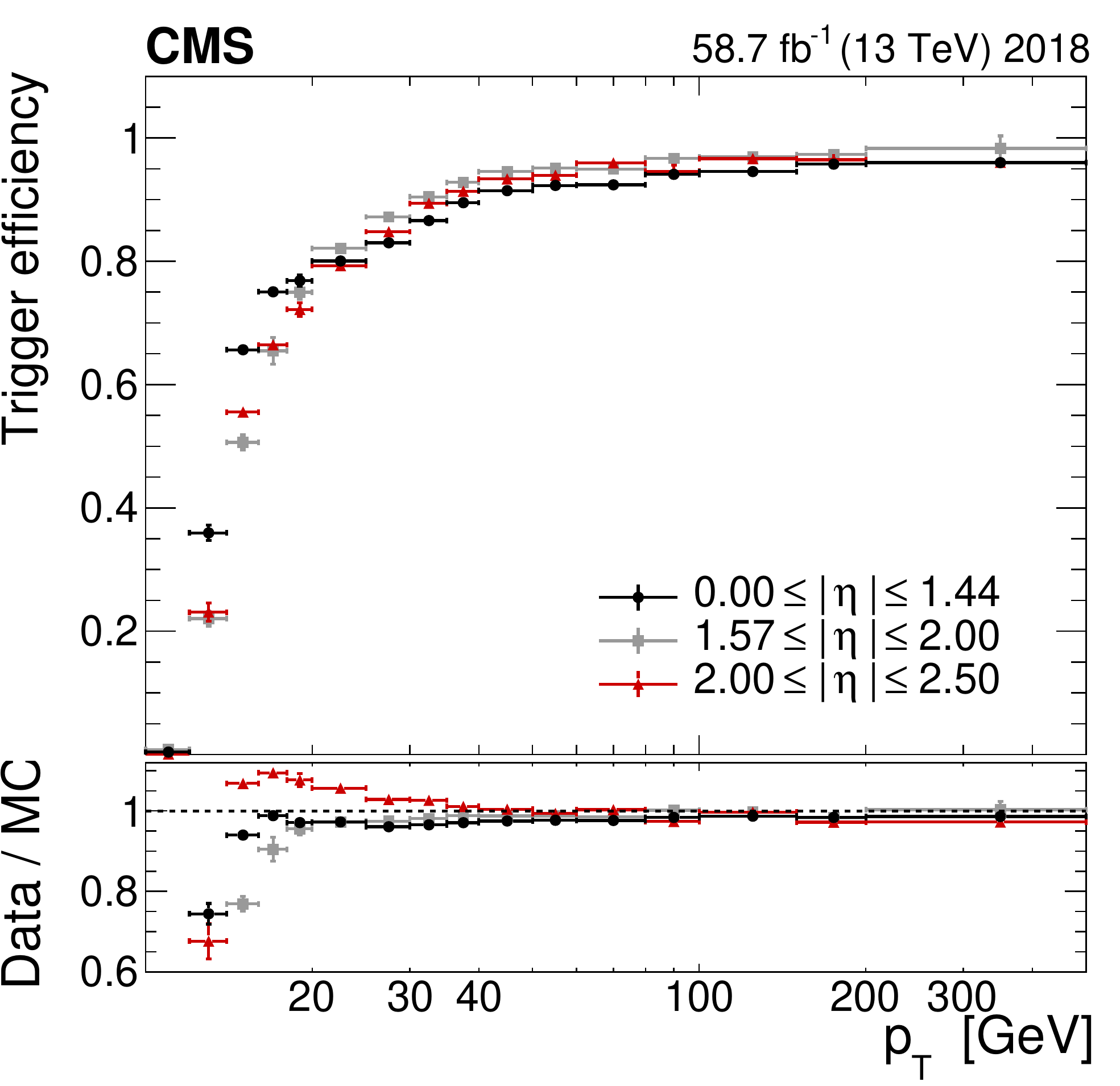}
    \caption{Efficiency of the Ele23 object (left) and Ele12 object (right) of the HLT\_Ele23\_Ele12 trigger, with respect to an offline reconstructed electron, as a function of the electron \pt, obtained for different $\eta$ regions using the 2018 data set. The Ele23 efficiency includes the requirement of passing the leading electron threshold of the asymmetric L1\_DoubleEG seed. The bottom panel shows the data-to-simulation ratio. The efficiency measurements combine the effects of the L1 and HLT triggers. The vertical bars on the markers represent combined statistical and systematic uncertainties.}
\label{fig:Ele23Ele12_pt}
\end{figure}

The efficiency of the two analyzed electron triggers in different SC $\eta$ regions is shown, with respect to an offline reconstructed electron, in Figs.~\ref{fig:SingleEle_pt} and~\ref{fig:Ele23Ele12_pt}, as a function of the electron \pt.
The region $1.44 < \abs{\eta} < 1.57$ is not included since it corresponds to the transition between the barrel and endcap regions of the ECAL where the quality of reconstruction, calibration and identification are not as good as in the rest of the ECAL~(see Fig.~\ref{fig:eleRecoEff2017}).
The DY + jets simulated event samples produced with MadGraph5~\cite{Alwall:2014hca} are used for comparison. The measurement combines both the L1 and HLT efficiencies.

{\tolerance=800 At the HLT level, both objects required by the double-electron path must correspond to an L1 seed, which can require either a single-electron with a higher momentum threshold (L1\_SingleEG) or two electrons (L1\_DoubleEG) with lower momentum thresholds (as shown in Table~\ref{tb:unprescale_L1}). This requirement also needs to be applied offline when performing the tag-and-probe measurement. 
Since the tag needs to pass a single-electron HLT path, it must pass an L1\_SingleEG seed. As a consequence, it will also satisfy the requirements of the $\ET$-leading object of the lowest unprescaled L1\_DoubleEG, lowering the L1 requirement on the probe to be only above the subleading threshold of the lowest unprescaled L1\_DoubleEG. When the $\ET$-leading object (Ele23) of the double-electron path is tested, the probe is thus specifically requested to pass the leading threshold of the path's L1\_DoubleEG seed. As reported in Table~\ref{tb:unprescale_L1}, the $\ET^\text{L1}$ thresholds of the lowest unprescaled L1\_DoubleEG seed increased across the years, leading to a larger efficiency at low \pt for the double-electron trigger in 2016 than in 2017 and 2018. \par}

The single-electron trigger analyzed in this paper is characterized by a sequence of strict identification and isolation selections, known as ``tight working point'' (WPTight). This selection was retuned in 2017 to ensure better performance.  As a consequence, the single-electron trigger efficiency is higher in 2017--2018 than in 2016.

\begin{figure}[hbtp]
\centering
    \includegraphics[ width=\cmsFigWidth]{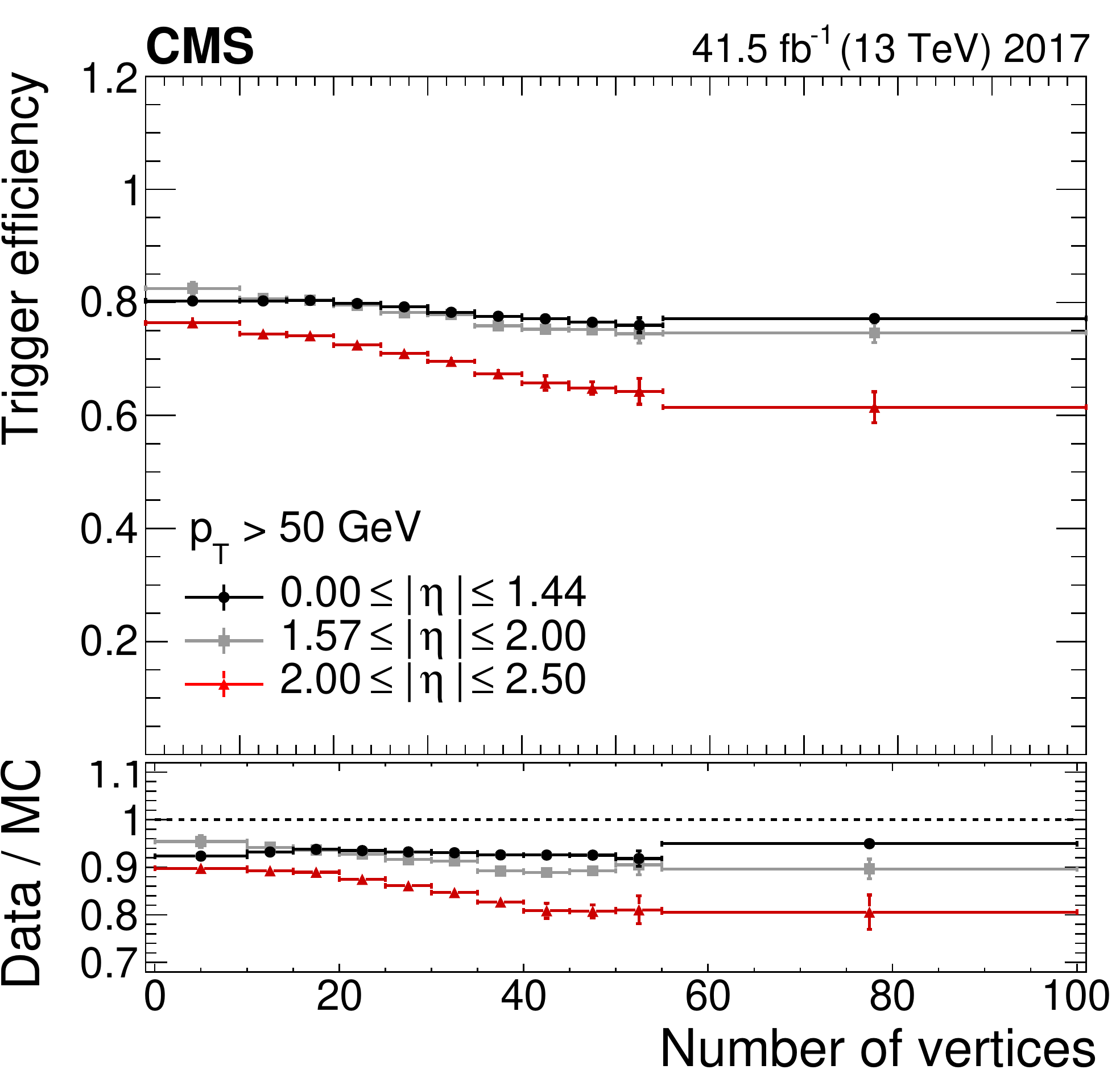}
    \includegraphics[ width=\cmsFigWidth]{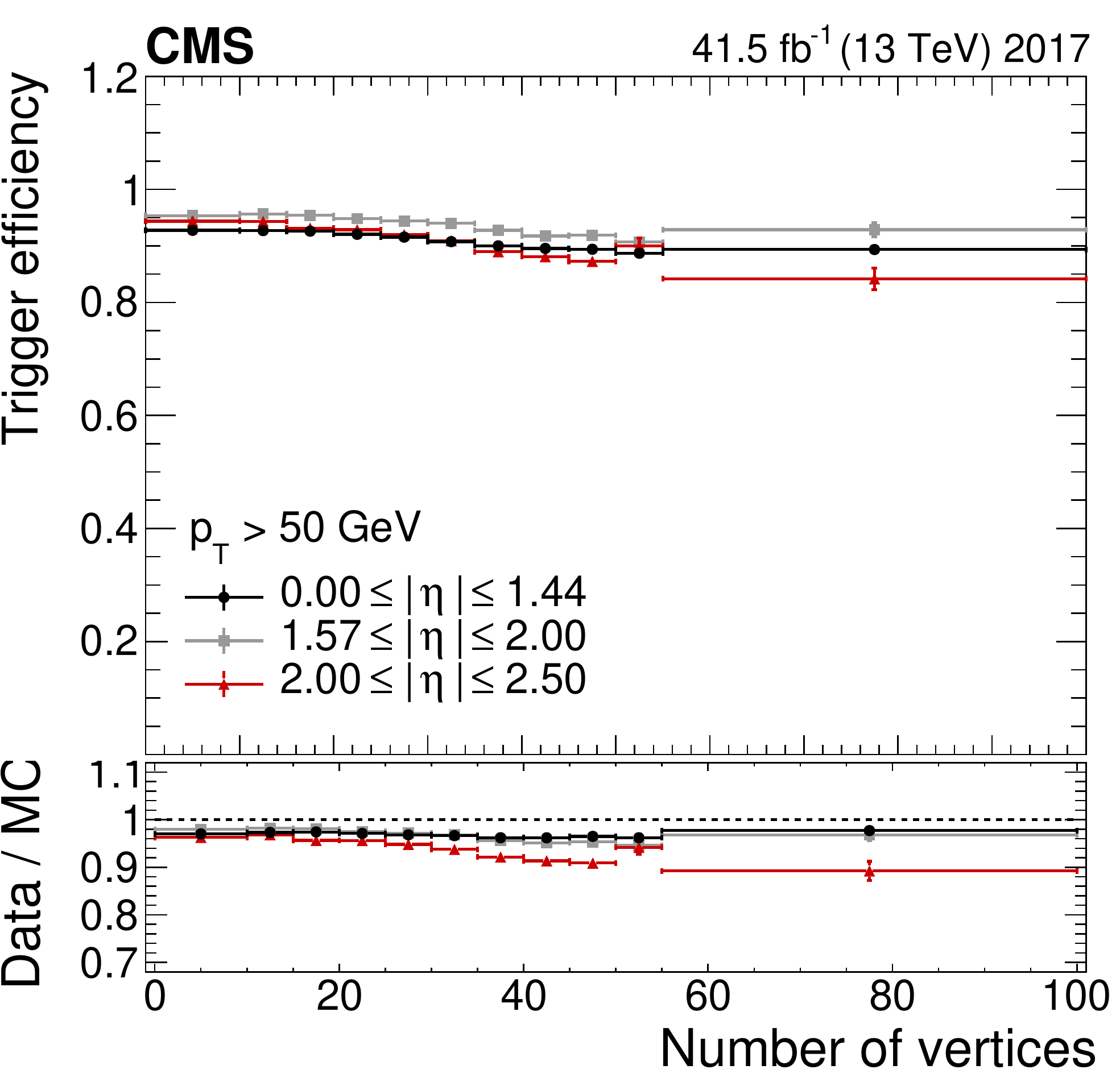}
   \caption{Efficiency of HLT\_Ele32\_WPTight\_Gsf (left) and HLT\_Ele23\_Ele12 (right) trigger,  with respect to an offline reconstructed electron, as a function of the number of reconstructed primary vertices, obtained for different $\eta$ regions using the 2017 data set. Electron $\ET$ is required to be $>$50\GeV. The bottom panel shows the data-to-simulation ratio. The efficiency measurements combine the effects of the L1 and HLT triggers. The vertical bars on the markers represent combined statistical and systematic uncertainties. In 2017 the majority of the high-pileup data was recorded at the end of the year, the same time the pixel DC-DC convertors exhibited efficiency losses. Thus the efficiency loss versus number of vertices in the event is not solely due to the pileup. However the efficiency loss is significant only for $2.0 < \abs{\eta} < 2.5$.}
\label{fig:2017_npv}
\end{figure}

As previously described, electron candidates at the HLT are built by associating a track reconstructed in the pixel detector with an ECAL SC. In 2017, the CMS pixel detector was upgraded by introducing extra layers in the barrel and forward regions. At the beginning of that year, a commissioning period of the pixel detector led to a slightly reduced efficiency, which mostly affected barrel electrons.
Moreover, as a consequence of the new detector, the algorithm used to reconstruct electrons, by matching ECAL superclusters to pixel tracks, was revised. Since the beginning of 2017 data taking, the algorithm requires two hits in the pixel detector when the particle trajectory passes through three or less active modules and three hits otherwise, whereas in 2016 only two hits were demanded in all cases.
This change produced a significant rate reductions with minimal efficiency losses. To operate with the new pixel detector, DC-DC converters were installed.
After a few months of smooth  operation, some  converters started to fail once the luminosity of the accelerator was increased, at the beginning of October 2017, leading to a  decreasing efficiency toward the end of the year. For these reasons related to the pixel detector, 2017 trigger performance is slightly worse than for the other years, in particular for the double-electron trigger, where the retuning of the tight working point does not have any effect. In Fig.~\ref{fig:2017_npv}, the 2017 efficiencies of the single- and double-electron HLT paths are reported as a function of the number of reconstructed primary vertices.
In 2017 the majority of the high-pileup data was recorded at the end of the year, the same time the pixel DC-DC convertors exhibited efficiency losses. Thus the efficiency loss versus number of vertices in the event  is not solely due to the pileup. However, as Fig.~\ref{fig:2017_npv} shows, the efficiency loss is significant only for $2.0 < \abs{\eta} < 2.5$.

The combined L1 and HLT trigger efficiency for the lowest unprescaled single-electron trigger path is about 80\% at the \pt plateau, with slightly lower values in the endcaps in 2016--2017. Because of the looser selection applied, the double-electron trigger has an efficiency close to unity for both objects.
The increase in dead regions in the pixel tracker arising from DC-DC convertor failure is difficult to simulate, and is one of the main causes of disagreement between data and simulation, in particular in 2017, especially at high \pt. The discrepancy in the turn-on at low \pt, seen for all years and $\eta$ values, is mainly because of the small differences that exist between the online and offline ECAL response corrections, as described above.

\section{Energy corrections}\label{sec:sec6}
The energy deposited by electrons and photons in the ECAL and collected by the superclustering algorithm is subject to losses for several reasons. Electromagnetic shower energy in the ECAL can be lost through lateral and longitudinal shower leakage, or in intermodule gaps or dead crystals;
the shower energy can also be smaller than the initial electron energy because of the energy lost in the tracker.

These losses result in systematic variations of the energy measured in the ECAL.
Without any corrections, this would lead to a degradation of the energy resolution for reconstructed electrons and photons.
To improve the resolution, a multivariate technique is used to correct the energy estimation for these effects,
as discussed below. The regression technique described in Section~\ref{sec:EnergyReg} uses simulation events only, whereas the energy scale and spreading corrections detailed in Section~\ref{sec:SS} are based on the comparison between data and simulation.

\subsection{Energy corrections with multivariate regressions}
\label{sec:EnergyReg}
A set of regression fits based on BDTs are applied to correct the energy of
$\Pe$/$\gamma$~\cite{Khachatryan:2014ira}. The minimum $\ET$ for electrons (photons) considered
for the BDT training is 1 (5)\GeV at the simulation level.
Each of these energy regressions is built as follows. The regression
target $y$ is the ratio between the true energy of an $\Pe$/$\gamma$ and its reconstructed
energy, thus the regression prediction for the target is the correction factor to be applied
to the measured energy to obtain the best estimate of the true energy. The regression input variables, represented by the vector $\vec{x}$,
includes the object and event parameters most strongly correlated with the target. The regression is implemented as a gradient-boosted decision tree, and a log-likelihood function is employed~\cite{Khachatryan:2014ira}:

\begin{equation}
\mathcal{L} = - \sum_{\text{MC}\;\; \Pe/\gamma\;\; \text{objects}} \ln{p(y|\vec{x})},
\end{equation}
\noindent
where $p(y|\vec{x})$ is the estimated probability for an object to have the observed value $y$, given the input variables $\vec{x}$, and the sum runs over all
objects in a simulated sample in which the true values of the object energies are known.
The probability density function used in this regression algorithm is a double-sided Crystal Ball (DSCB) function~\cite{Oreglia:1980cs} that has
a Gaussian core with power law tails on both sides.
The definition of the DSCB function is as follows:

\begin{equation}
\text{DSCB}(y; \mu, \sigma, \alpha_{\text{L}}, n_{\text{L}}, \alpha_{\text{R}}, n_{\text{R}}) =
\begin{cases}
  N\;  \re^{-\frac{\xi(y)^2}{2}}, & \text{if } -\alpha_{\text{L}}\le \xi(y)\le \alpha_{\text{R}} \\
  N\;  \re^{-\frac{\alpha_{\text{L}}^2}{2}}
     \left(\frac{\alpha_{\text{L}}}{n_{\text{L}}}
       \left(\frac{n_{\text{L}}}{\alpha_{\text{L}}}-\alpha_{\text{L}}-\xi(y)\right)
       \right)^{-n_{\text{L}}},
     & \text{if } \xi(y) < -\alpha_{\text{L}} \\
  N\;  \re^{-\frac{\alpha_{\text{R}}^2}{2}}
     \left(\frac{\alpha_{\text{R}}}{n_{\text{R}}}
       \left(\frac{n_{\text{R}}}{\alpha_{\text{R}}}-\alpha_{\text{R}}+\xi(y)\right)
       \right)^{-n_{\text{R}}},
     & \text{if } \xi(y) > \alpha_{\text{R}} \\
\end{cases}
,
\end{equation}
where $N$ is the normalization constant, $\xi(y)=(y-\mu)/\sigma$, the
variables $\mu$ and $\sigma$ are the parameters of the Gaussian core,
and the $\alpha_{\text{R}}$ ($\alpha_{\text{L}}$) and $n_{\text{R}}$ ($n_{\text{L}}$) parameters control the
right (left) tails of the function.  Through the training phase, the regression algorithm
performs an estimate of the parameters of the double Crystal
Ball probability density as a function of the input vector of the
object and event characteristics $\vec{x}$:

\begin{equation}
    p(y|\vec{x}) = \text{DSCB}( y; \mu(\vec{x}), \sigma(\vec{x}),
    \alpha_{\text{L}}(\vec{x}), n_{\text{L}}(\vec{x}),
    \alpha_{\text{R}}(\vec{x}), n_{\text{R}}(\vec{x})
      ).
\end{equation}
Subsequently, for an $\Pe$/$\gamma$ candidate, the most probable value $\mu$ is the estimate of the correction to the object's energy, and
the width of the Gaussian core $\sigma$ is the estimate of the
per-object energy resolution. Both $\mu$ and $\sigma$ are predicted
by the regression, as functions of the object and event parameter
vector $\vec{x}$.

The electron energy is corrected via the sequential application of three
regressions: the first regression (step 1) provides the correction to the SC
energy, the second regression (step 2) provides an estimate of  the SC energy resolution, taking into account the additional spread in data due to real detector conditions, and the third regression (step 3)
yields the final energy value, correcting the combined energy estimate from the SC and the electron track information. The photon energy is corrected using the same method, except that step 3 is omitted.

The electron and photon regressions are trained on samples of simulated events with two electrons or photons in each event,
generated with a flat transverse momentum spectrum, where the true value of the $\Pe$/$\gamma$  energy is known
and the geometric condition $\Delta R < 0.1$ is used to find a match of the reconstructed $\Pe$/$\gamma$ to the true ones.
The ECAL crystals exhibit slight
variations in the light output for a given electromagnetic shower. This effect is corrected by the intercalibration of the crystals~\cite{Chatrchyan:2013dga}, and a corresponding modeling of this variation is applied in the simulation.
In addition the knowledge of the crystal intercalibrations is affected by random deviations~\cite{Khachatryan:2015hwa},
which impact the energy resolution. This effect is usually simulated by applying
a random spreading of the crystal intercalibrations within the expected inaccuracy.
To avoid the random spreading of the simulation, the regression fit corrects the data using two MC samples: a sample without the intercalibration spreading (called ideal IC samples) is used to train the
energy regression, and a sample with the intercalibration spreading (called real IC samples) is used for the energy resolution estimation and for the SC and track combination.

The workflow for the electron and photon energy regressions is summarized
in Table~\ref{tab:regression-steps}. Each
subsequent step depends on the output of the previous step.

\begin{table}[hbp]
\centering
\topcaption{Details of the three energy regression steps used in electron and photon energy reconstruction.}
\begin{tabular}{lllll}
\hline
Regression & $\Pe$/$\gamma$  & High level & Simulated   & Quantity  \\
 index     &                   &   object   & sample type & corrected \\
\hline
step 1 & supercluster & electrons/photons & ideal IC & energy \\
step 2 & supercluster & electrons/photons & real IC  & energy resolution\\
step 3 & supercluster and track     & electrons only    & real IC  & energy \\
\hline
\end{tabular}
\label{tab:regression-steps}
\end{table}

The step 1 regression primarily corrects for the energy that is lost in the tracker
material or in modular gaps in the ECAL.
The regression inputs include the energy and position of the SC, and the variable $R_9$, which is defined as the energy sum of the $3{\times}3$ crystal array centered around the
most energetic crystal in the SC divided by the energy of the SC. Other quantities, including lateral shower shapes in $\eta$ and $\phi$, number of saturated
crystals, and other SC shape parameters, as well as an estimate of the pileup
transverse energy density in the calorimeter are also included.

Step 2 is performed to
obtain an estimate of the per-object resolution.
It uses the same inputs as in step 1, but the SC energy is scaled by the correction factor obtained from the step 1
regression, and the target of the step 2 regression is the ratio of
the true energy of the particle to the
measured energy corrected by step 1.
Since imperfect intercalibration affects the spread of the energy response
between crystals and not the mean value of the average response, in step 2 the mean $\mu(\vec{x})$ of the DSCB probability density function is
fixed to that obtained from step 1. The primary result of the step 2
regression is the estimated value of the energy resolution
$\sigma(\vec{x})$.

\begin{figure*}[h!]
\centering
\includegraphics[ width=\cmsFigWidth]{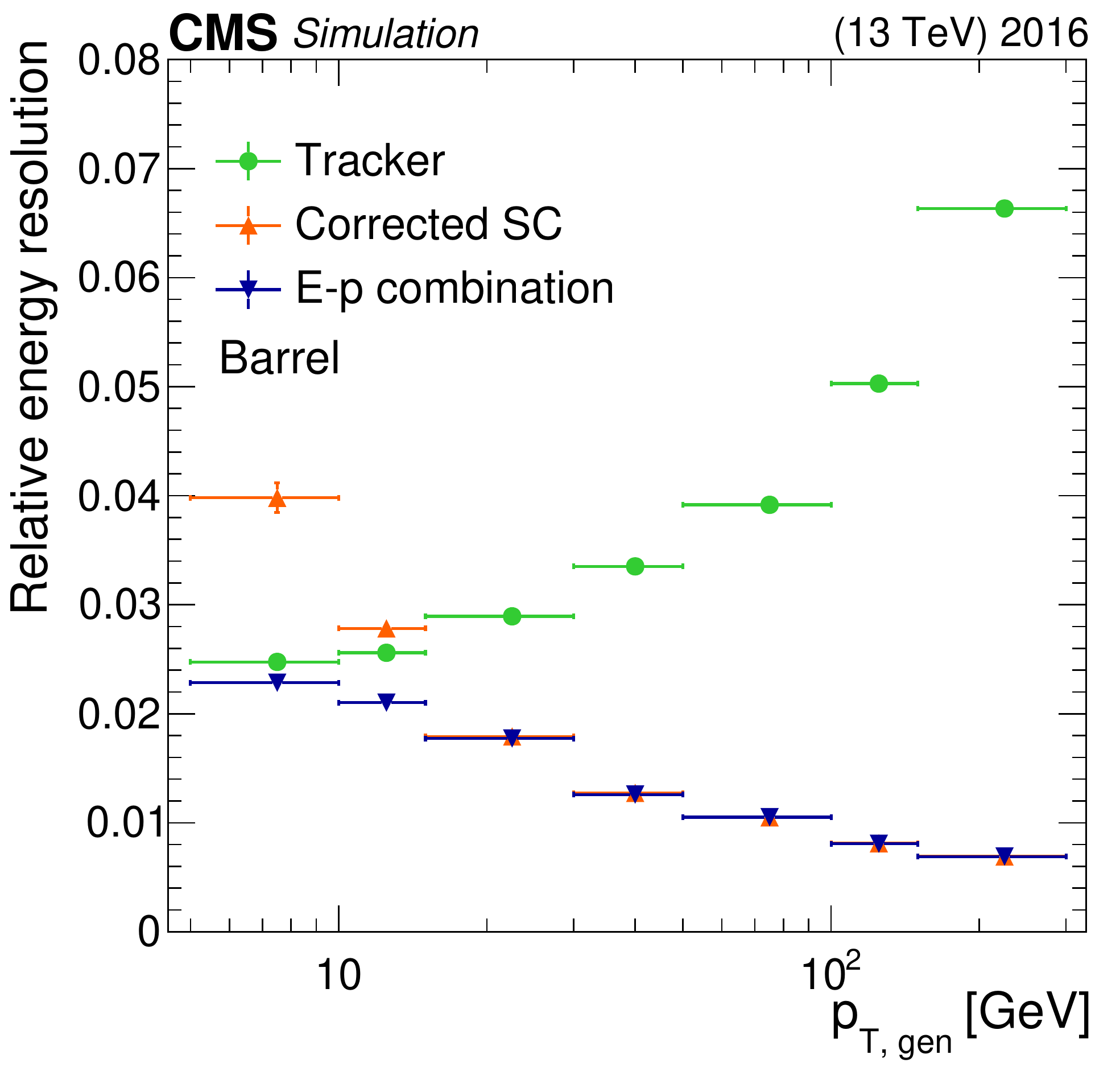}
\includegraphics[ width=\cmsFigWidth]{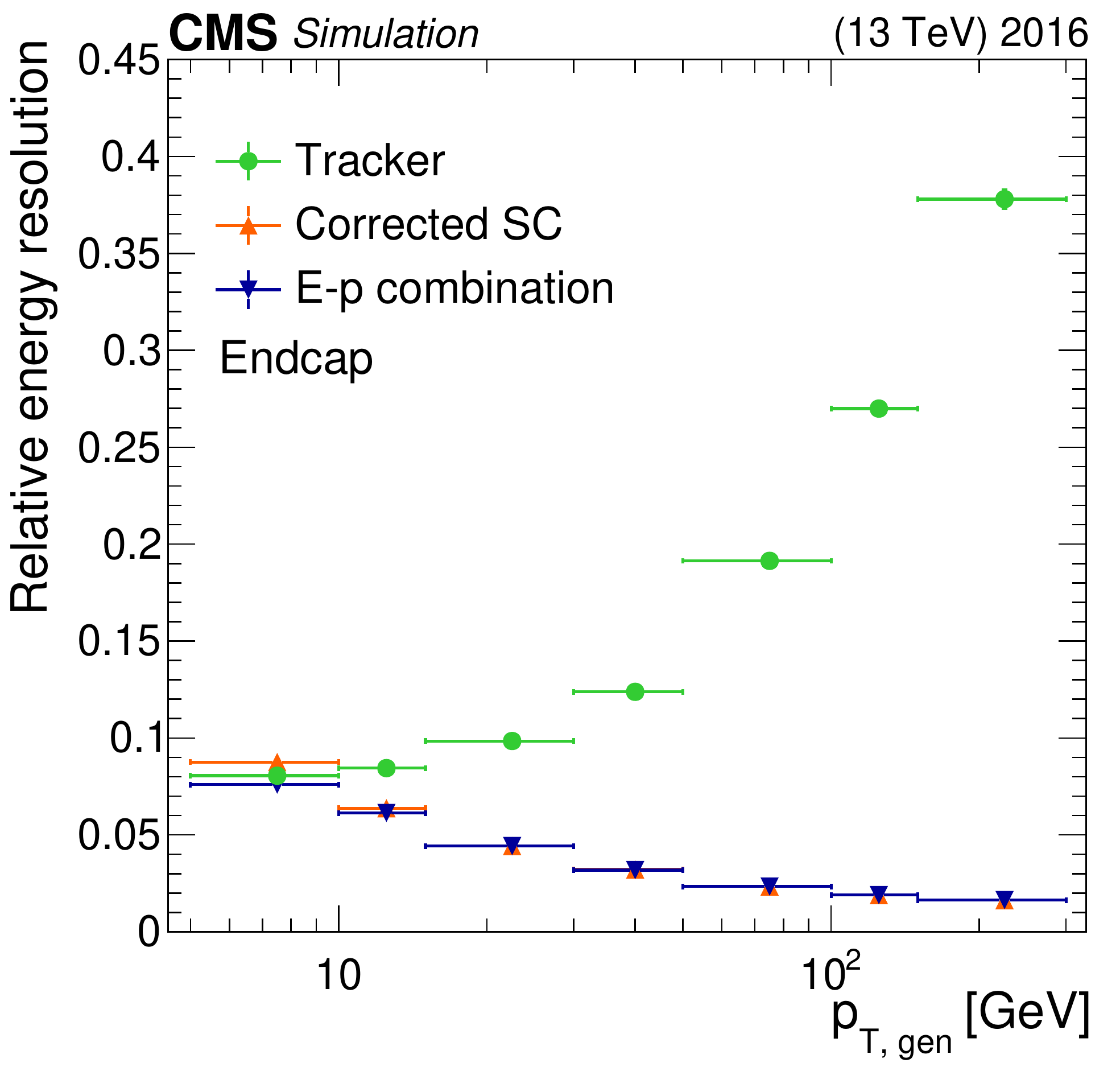}\\
\caption{Relative electron resolution versus electron $\pt$, as measured by the ECAL (``corrected SC''), by
  the tracker, and seen in the $E$-$p$ combination after the step 3 regression, as found in 2016 MC samples for
  barrel (left) and endcap (right) electrons. Vertical bars on the markers represent the uncertainties coming from the fit procedure.}
\label{fig:EcalAndTrackerResolution}
\end{figure*}

\begin{figure*}[h!]
\centering
\includegraphics[ width=\cmsFigWidth]{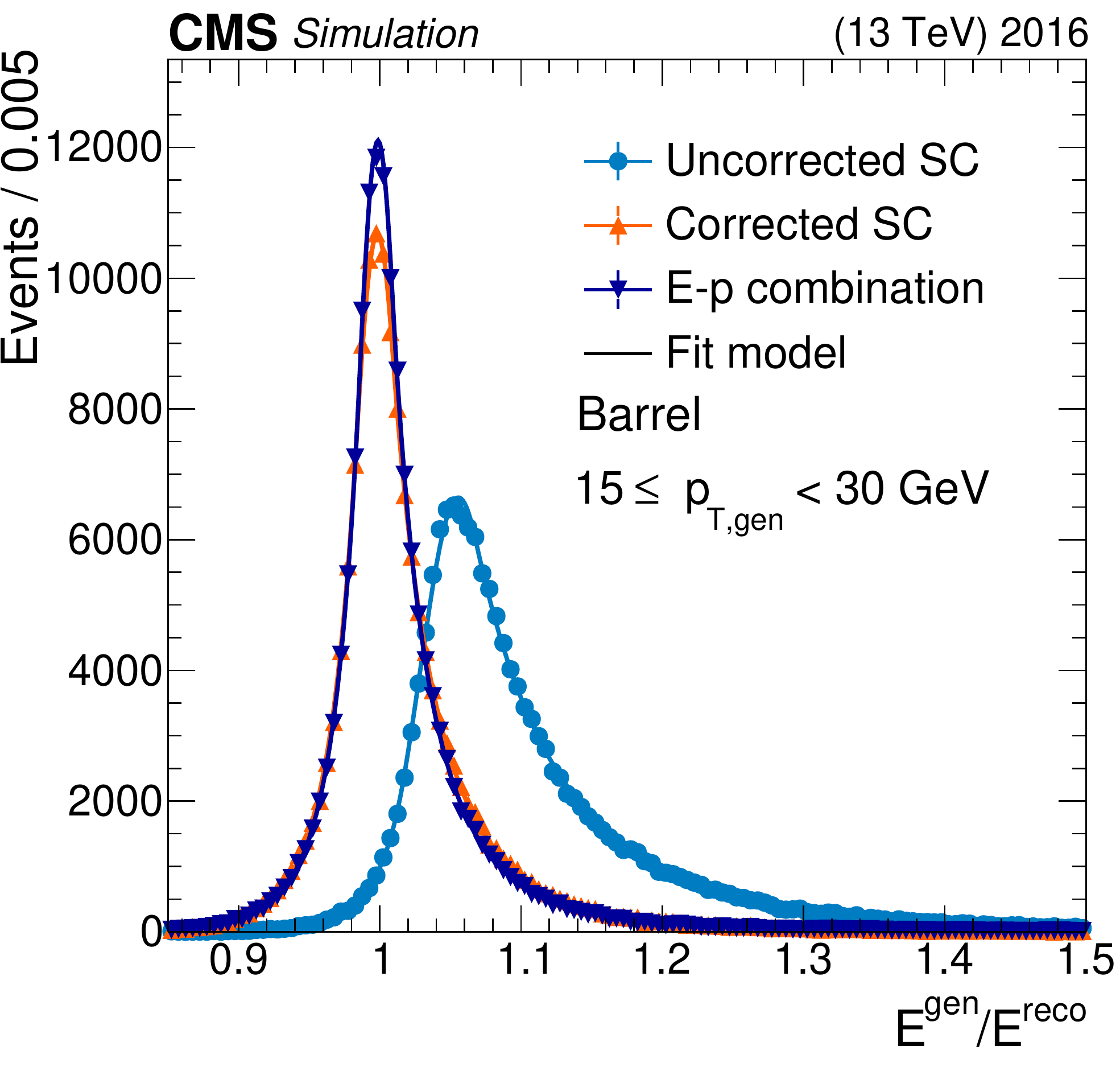}
\includegraphics[ width=\cmsFigWidth]{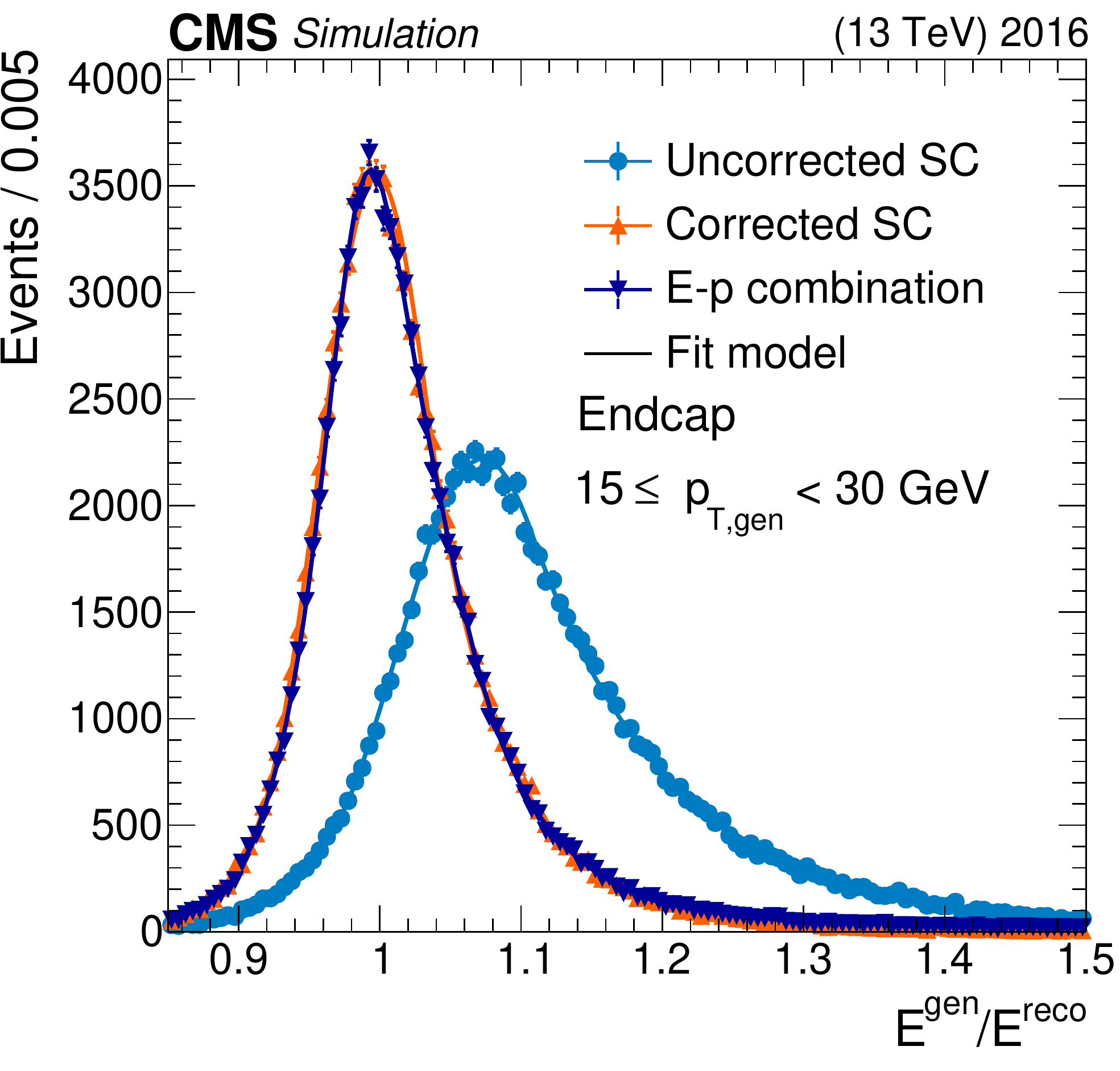}\\
\caption{Ratio of the true to the reconstructed electron energy in
  the \pt range 15--30\GeV with and without regression corrections, with
  a DSCB function fit overlaid, in 2016 MC samples for barrel (left) and endcap (right) electrons. Vertical bars on the markers represent the statistical uncertainties of the MC samples.}
\label{fig:regressionFitExamples}
\end{figure*}

\begin{figure*}[h!]
\centering
\includegraphics[ width=\cmsFigWidth]{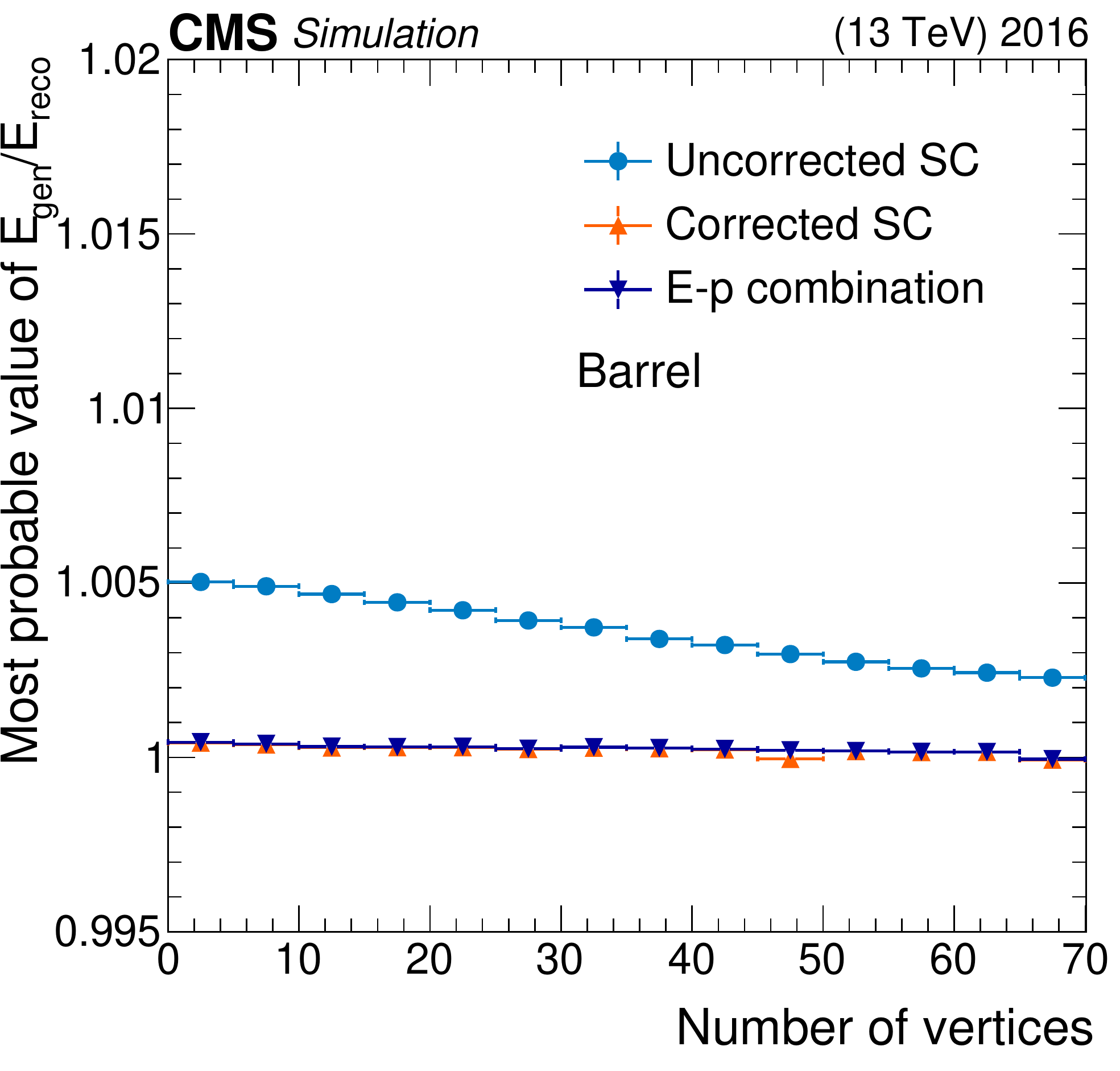}
\includegraphics[ width=\cmsFigWidth]{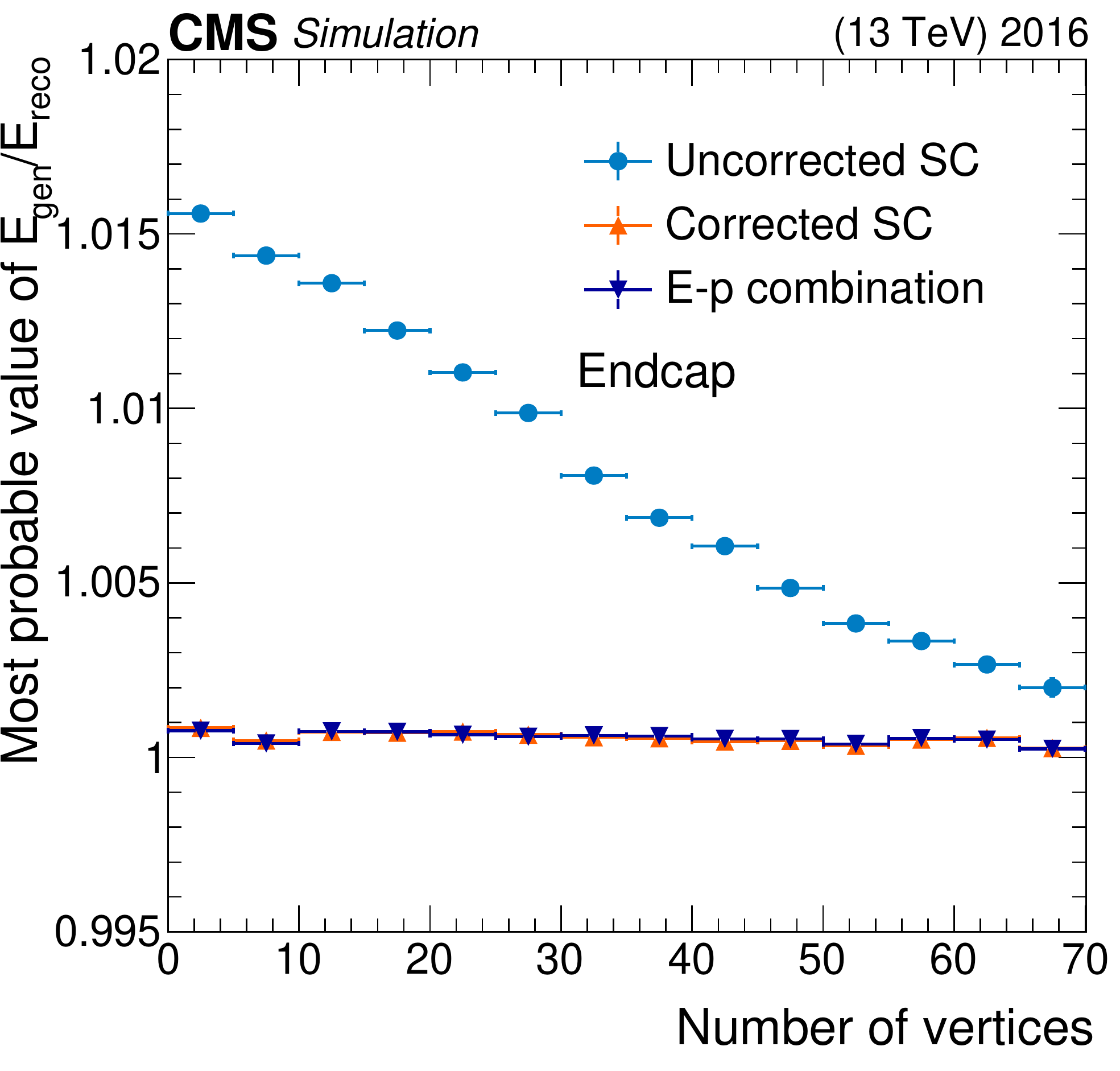}\\
\caption{Most probable value of the ratio of true to reconstructed electron energy, as a function of pileup, with and without regression corrections in 2016 MC samples for  barrel (left) and endcap (right) electrons. Vertical bars on the markers represent the uncertainties coming from the fit procedure and are too small to be observed from the plot.}
\label{fig:energyVSPU}
\end{figure*}

\begin{figure*}[h!]
\centering
\includegraphics[ width=\cmsFigWidth]{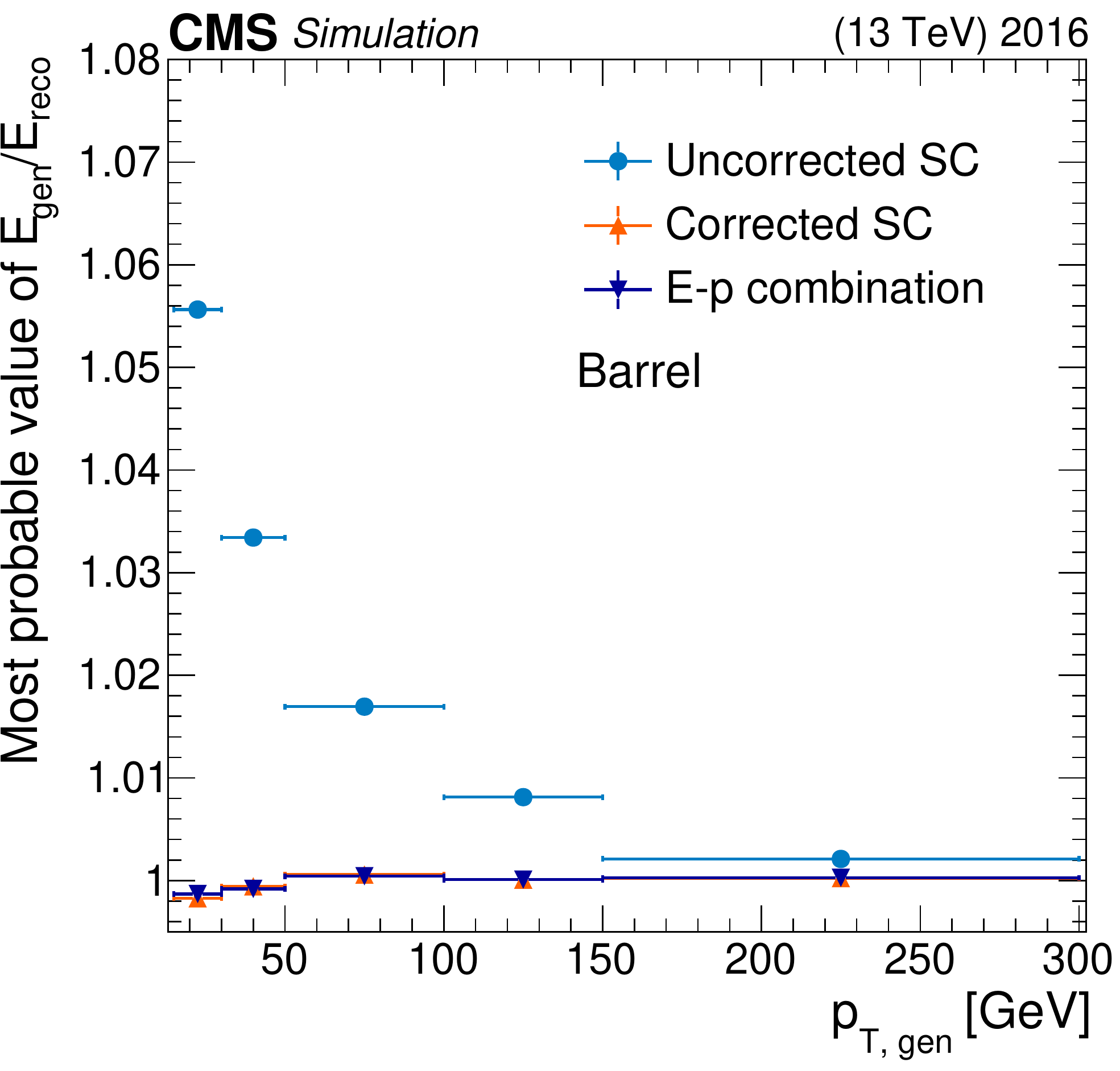}
\includegraphics[ width=\cmsFigWidth]{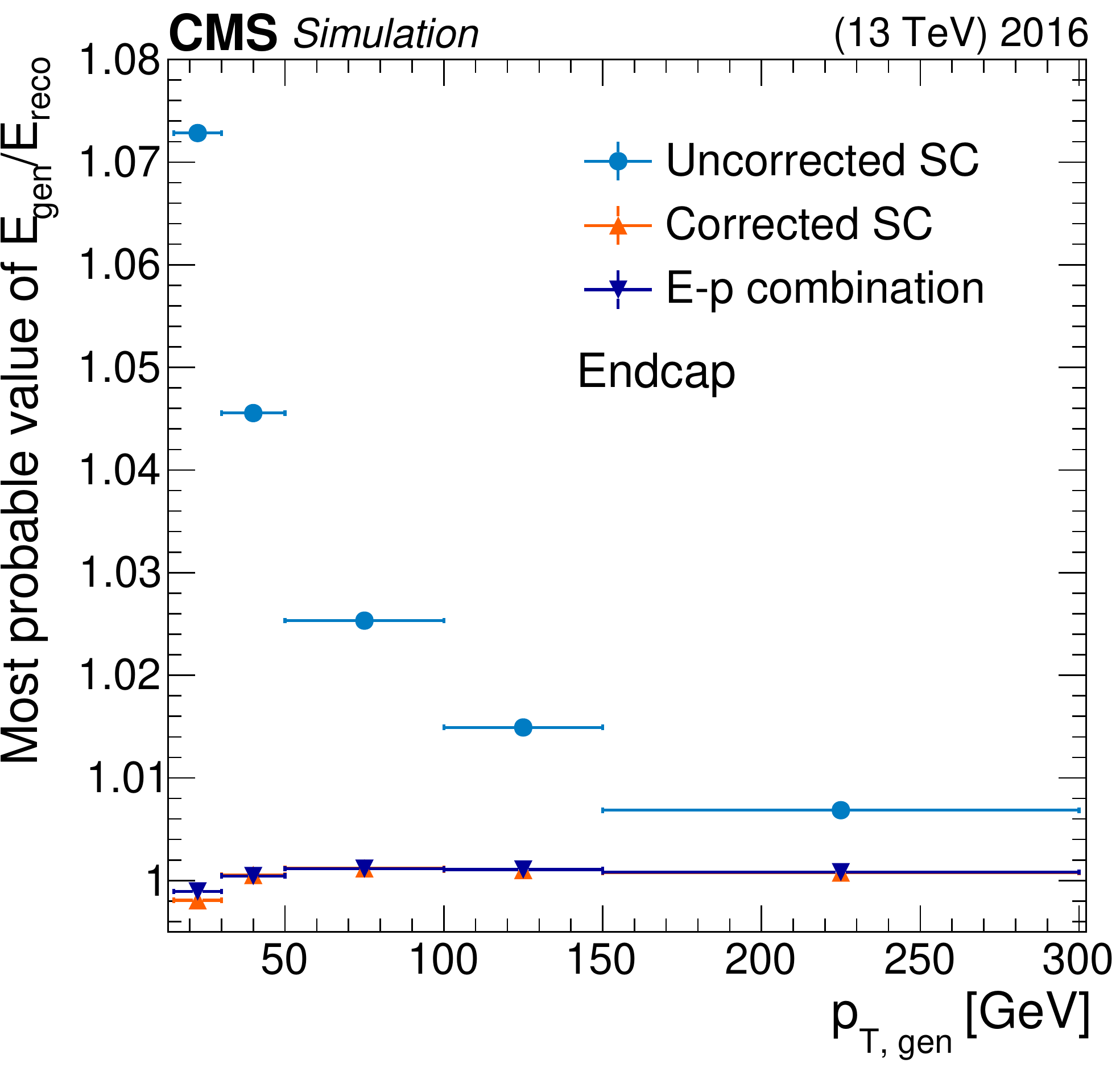}\\
\caption{Most probable value of the ratio of true to reconstructed electron energy, as a function of the electron \pt with and without the regression corrections, evaluated using 2016 MC samples for barrel (left) and endcap (right) electrons. Vertical bars on the markers represent the uncertainties coming from the fit procedure and are too small to be observed from the plot.}
\label{fig:energyVSET}
\end{figure*}

For electrons, since the energy measurement is performed independently in
the ECAL and the tracker, an additional step
 combining the ECAL energy and momentum estimate from the
tracker is performed.
A weighted combination of the two independent measurements can be
formed as:

\begin{equation}
E_{\text{combined}}^{\text{reco}}=\frac{ E_{\text{ECAL}}/\sigma_{E}^2 + p_{\text{tracker}}/\sigma_p^2 }
{ 1/\sigma_{E}^2 + 1/\sigma_p^2 },
\end{equation}
\noindent
where $E_{\text{ECAL}}$ and $\sigma_E$ are the ECAL measurements of the
energy and the energy resolution of the SC of the electron
corrected with the step 1 and 2 regressions, respectively, and
$p_{\text{tracker}}$ with $\sigma_p$ are the momentum magnitude and momentum
resolution measured by the electron tracking algorithm (as described in
Section~\ref{sec:electron-track-reconstruction}).
This improves the predicted electron energy at low $\ET$,
especially where the momentum measurement from the tracker has a better resolution than the corresponding ECAL measurement.
The average relative momentum resolution of the tracker ($\sigma_p$) and energy resolution of the ECAL ($\sigma_E$) are
shown in Fig.~\ref{fig:EcalAndTrackerResolution}. The momentum resolution of the tracker is better than
the ECAL energy resolution for transverse momenta below 10--15\GeV and deteriorates at higher energies. The $E$-$p$ combination
in CMS is only performed for electrons with energies less than 200\GeV. For higher-energy electrons
 only the SC energy is used, corrected by the above described regression steps.
The step 3 regression uses as a target the ratio of the true electron
energy and $E_{\text{combined}}^{\text{reco}}$ computed as the $E$-$p$ combination discussed
above. The inputs for the regression include all quantities that enter
the $E_{\text{combined}}^{\text{reco}}$ expression, plus several additional tracker quantities including
the fractional amount of energy lost by the electron in the tracker, whether
the electron was reconstructed as ECAL-driven or tracker-driven (as discussed in Section~\ref{sec:electron-track-reconstruction}), and a few other tracker-related parameters. In Figs.~\ref{fig:EcalAndTrackerResolution}-\ref{fig:energyVSET} the outcome of the step 3 regression is referred to as the $E$-$p$ combination.
 
These regressions lead to significantly
improved measurements of electron and photon energies and energy resolutions as seen in Fig.~\ref{fig:regressionFitExamples}. The primary improvement occurs in the regressions applied to the energy of the SC (steps 1 and 2). Correcting the $E$-$p$ combination, which already uses the improved
SC energy, has a smaller impact. The effects of the regression corrections for the various steps of the correction procedure are illustrated in Fig.~\ref{fig:regressionFitExamples}
for low-$\pt$ electrons.
The regressions
are robust, and the performance is stable for electrons and photons in a wide energy range
in all regions of ECAL, and as a function of pileup, as shown in Figs.~\ref{fig:energyVSPU}~and~\ref{fig:energyVSET}.

\subsection{Energy scale and spreading corrections}
\label{sec:SS}
After applying the corrections described in Section~\ref{sec:EnergyReg}, small differences remain between data and simulation in both the electron and photon energy scales and resolutions. In particular, the resolution in simulation is
better than that in data.

An additional spreading needs to be applied to the photon and electron energy resolutions in simulation to match that observed in data.
The electron and photon energy scales are corrected by varying the scale in the data to match that observed in simulated events.
The magnitude of the final correction is up to 1.5\% with a total uncertainty estimated to be smaller than 0.1 (0.3)\% in the barrel (endcap).

Two dedicated methods, the ``fit method'' and the
``spreading method''~\cite{Khachatryan:2015hwa}, were developed in  Run 1 to estimate these corrections from $\PZ \to \Pe\Pe$ events.
In the fit method, an analytic fit is performed to the invariant mass distribution of the $\PZ$ boson ($m_{\Pe\Pe}$), with a  convolving of a BW
and a OSCB function.
The invariant mass distributions obtained from data and from simulated events are fitted separately and the results are compared to extract a scale offset.
The BW width is fixed to that of the $\PZ$ boson: $\Gamma_{\PZ} = 2.495\GeV$~\cite{Zyla2020}.  The parameters of the OSCB function, which describes  calorimeter resolution effects and bremsstrahlung losses in the detector material upstream of the ECAL, are free parameters of the fit. The spreading method, on the other hand, utilizes the simulated
$\PZ$ boson invariant mass distribution as a probability density
function in a maximum likelihood fit to the data. The simulation already accounts for all known detector effects, reconstruction inefficiencies, and the $\PZ$ boson kinematic properties. The residual discrepancy between data and simulation is described by an energy spreading function, which is applied to the simulation.  A Gaussian spreading, which ranges from 0.1 to 1.5\%, is applied to the simulated energy response; it is adequate to describe the data in all the examined categories of events.
Compared with the fit method, the spreading method can accommodate a larger number of electron categories in which these corrections are derived.

A multistep procedure is implemented, based on the fit and spreading methods, to fine-tune the electron and photon energy scales. To derive the corrections to the
photon energy scale,
electrons from $\PZ$ boson decays are used, reconstructed using information exclusively from the ECAL.

In the first step, any residual long-term drifts in the energy
scale in data are corrected by using the fit method, in approximately 18-hour intervals
(corresponding approximately to one LHC fill). Further subcategories are defined based on various $\eta$ regions,
owing to the different levels of radiation damage and of the amount of material budget upstream of the ECAL.
There are two $\eta$ regions in the barrel, $\abs{\eta}< 1.00$ and
$1.00 < \abs{\eta}< 1.44$. In the endcap, the two $\eta$ categories are defined by
$1.57 < \abs{\eta}< 2.00$ and $2.00 < \abs{\eta}< 2.50$.
After applying these time-dependent residual scale corrections, the energy scale in data is stable with time.

In the second step,
corrections to both the
energy resolution in the simulation and the scale for the
data are derived simultaneously in bins of $\abs{\eta}$ and $R_9$ for electrons, using the spreading method.
The energy scale corrections are derived in 50 electron categories: 5 in $\abs{\eta}$ and 10 in $R_9$.
This is a significant improvement in granularity compared with Run 1~\cite{Khachatryan:2015hwa},
 where only 8 electron categories were used (4 in $\abs{\eta}$ and 2 in $R_9$),
thus leading to an improvement in the precision of the derived scale corrections.
The $R_9$ value of each electron or photon SC is used
to select electrons that interact or photons that undergo a conversion in the material upstream of the ECAL.
The energy
deposited by photons that convert before reaching the ECAL tends to
have a wider transverse profile and thus lower $R_9$ values than those for
unconverted photons. The same is true for electrons that radiate upstream of the ECAL.

\begin{figure*}[hbtp]
\centering
\includegraphics[ width=\cmsFigWidth]{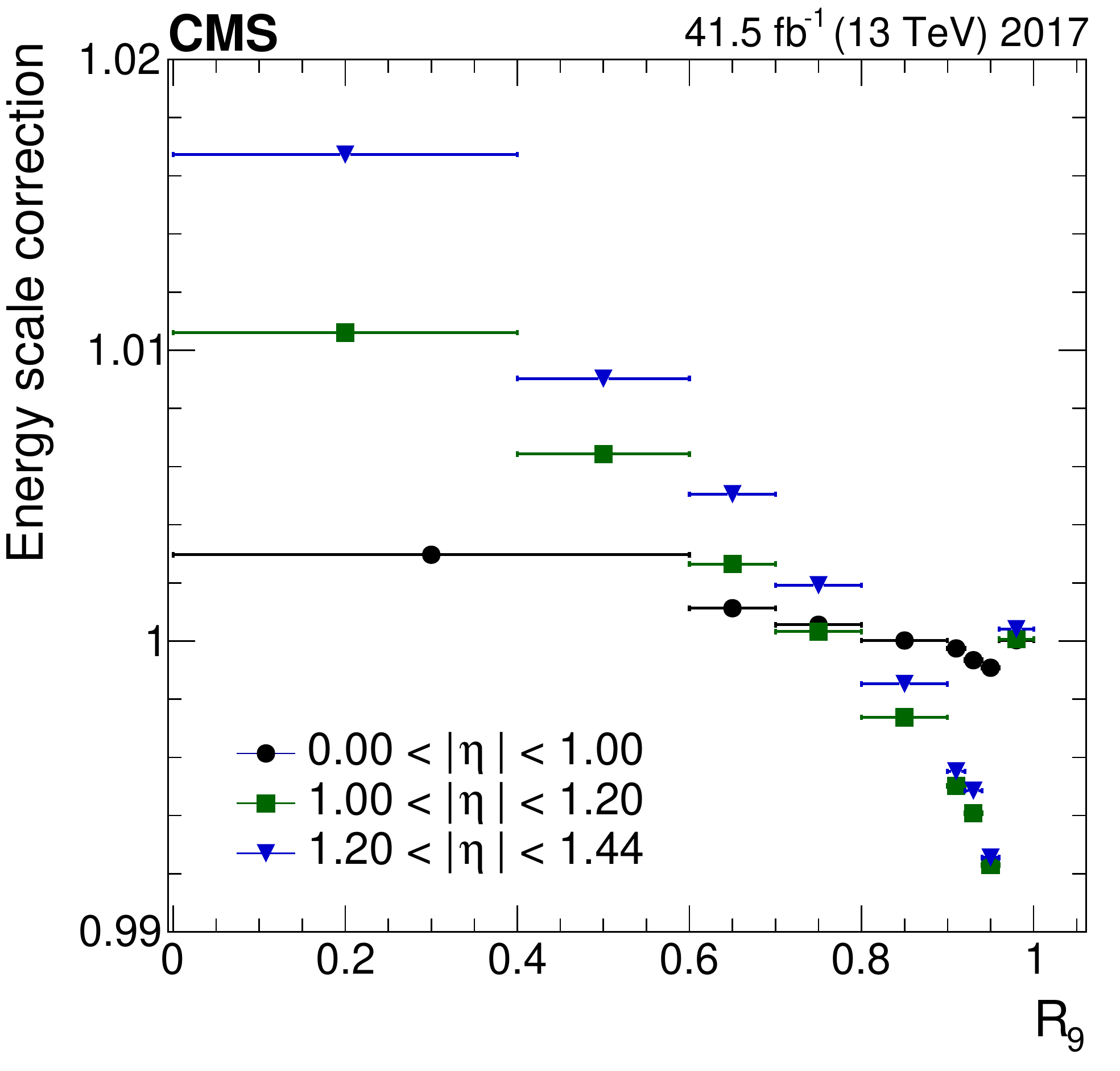}
\includegraphics[ width=\cmsFigWidth]{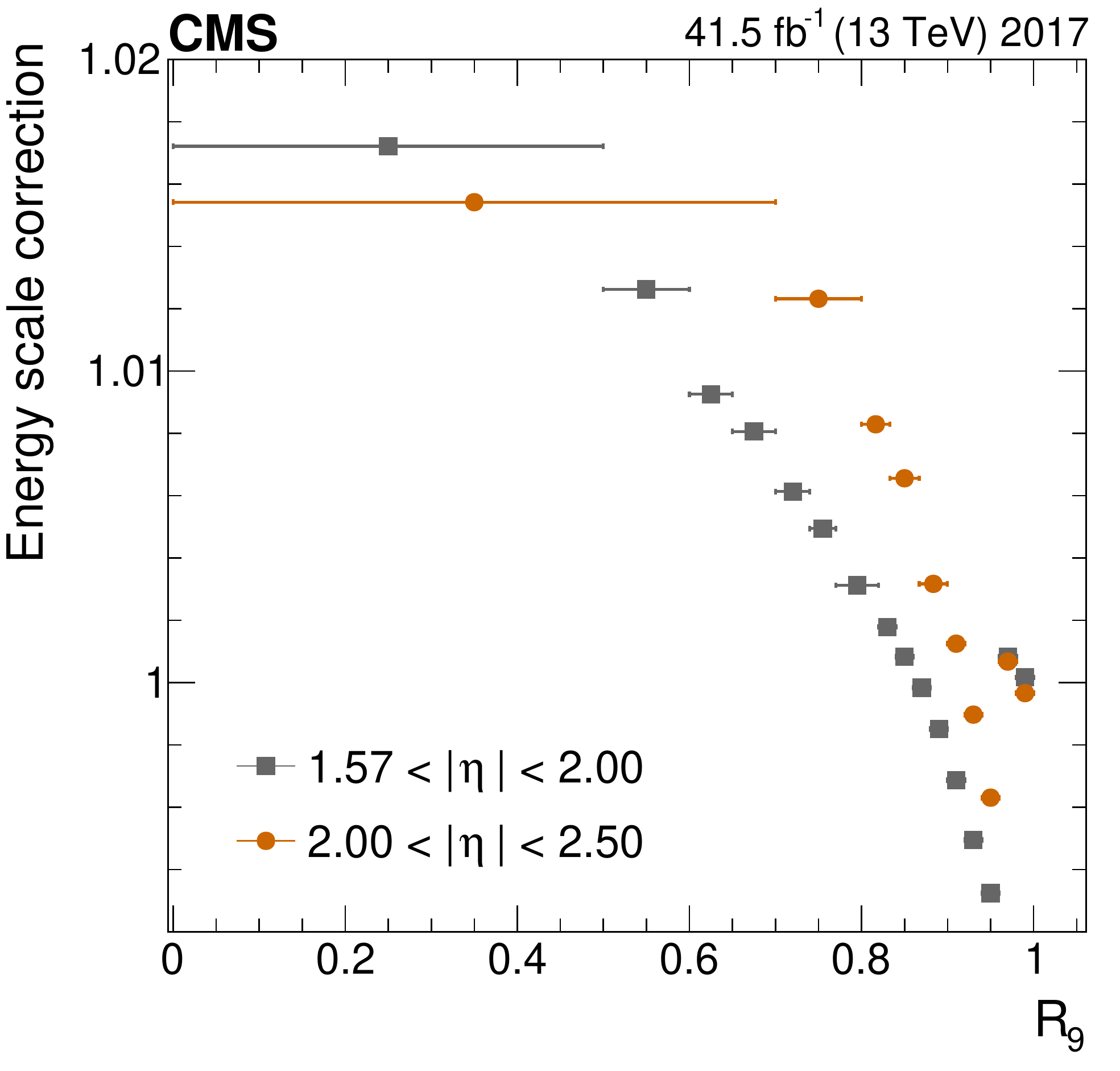}\\
\caption{Energy scale corrections for the 2017 data-taking period, as function of $R_9$, for different $\abs{\eta}$ ranges, in the barrel (left) and for the endcap (right). The error bars represent the statistical uncertainties in the derived correction and are too small to be observed from the plot.}
\label{fig:scaleVSr9}
\end{figure*}

The energy scale corrections obtained from this step in fine bins of $R_9$ are shown in Fig.~\ref{fig:scaleVSr9} for the 2017 data-taking period.
The uncertainties shown are statistical only.

The ECAL electronics operate with three gains: 1, 6, and 12, depending on the energy recorded in a single readout channel.
Most events are reconstructed with gain 12, whereas events with the highest energies are
reconstructed with gains 6 or 1.  The gain switch from 12 to 6 (6 to 1)
typically happens for electron/photon energies above 150 (300)\GeV in the barrel, and higher values in the endcaps.
A residual scale offset of nearly 1\% is measured for gain~6 both in the EB and EE and of 2 (3)\% for gain~1 in the EB (EE).
Thus an additional gain-dependent residual correction is derived and applied.

The systematic uncertainties
in the electron energy scale and resolution corrections are derived using
$\PZ\to \Pe\Pe$ events by varying the distribution of $R_9$, the electron selections used, and the $\ET$ thresholds on
the electron pairs used in the derivation of the corrections.
The contributions of these individual sources are added in quadrature to obtain the total uncertainty.
This uncertainty in the energy scale
is 0.05--0.1 (0.1--0.3)\% for electrons in the EB (EE), where the range corresponds to the variation in the  $R_9$ bins.

\begin{figure*}[h!]
\centering
\includegraphics[ width=\cmsFigWidth]{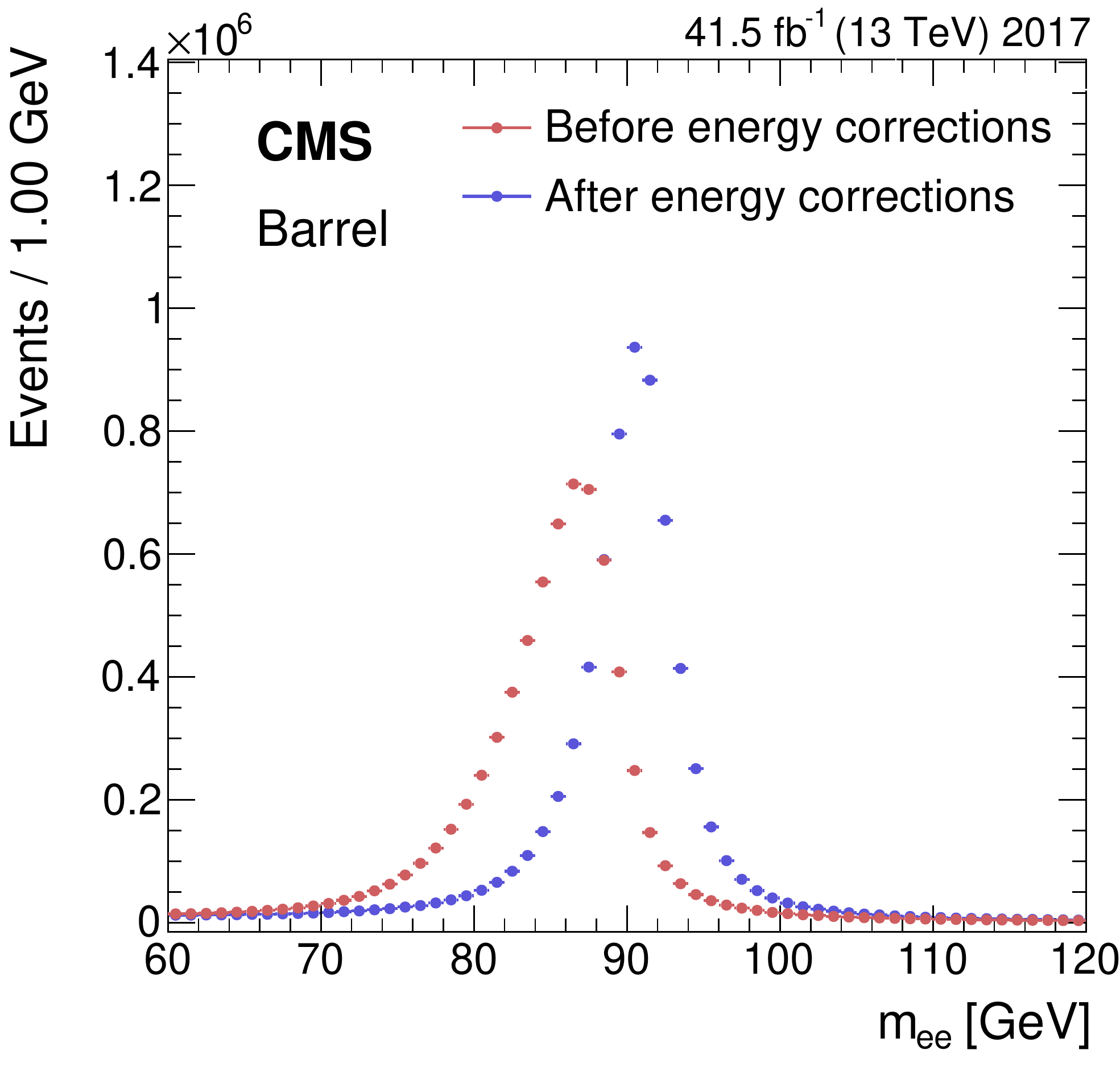}
\includegraphics[ width=\cmsFigWidth]{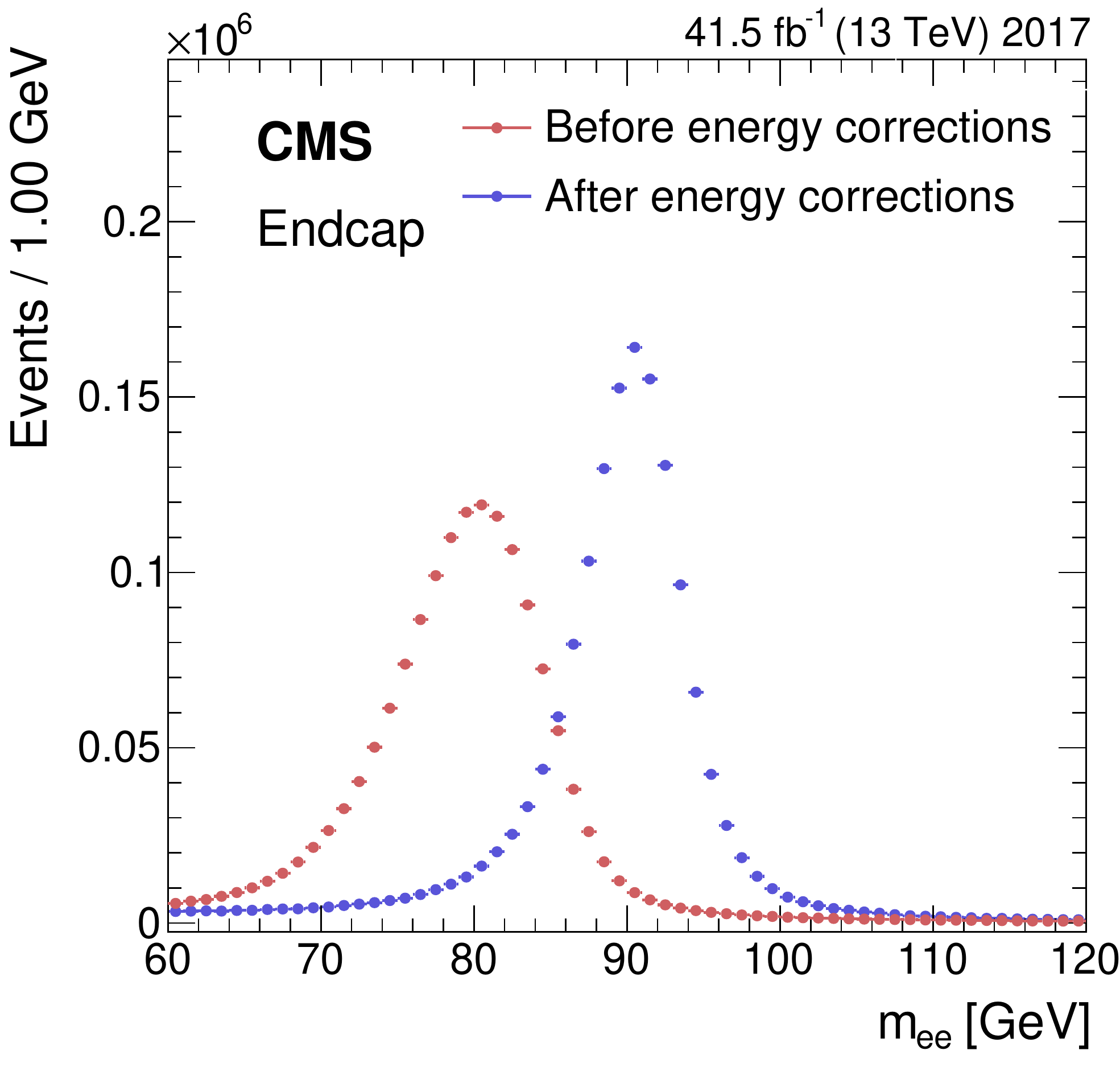}\\
\caption{Dielectron invariant mass distribution before and after all the energy corrections (regression and scale corrections) for barrel (left) and endcap (right) electrons for $\PZ\to \Pe\Pe$ events. The error bars represent the statistical uncertainties in data and are too small to be observed from the plot.}
\label{fig:effectRegression}
\end{figure*}

The performance of energy corrections in data, including the ones described in Section~\ref{sec:EnergyReg}, is illustrated by the reconstructed $\PZ\to \Pe\Pe$ mass distribution before and after corrections, as shown in Fig.~\ref{fig:effectRegression}.
The regression clearly improves the mass resolution for electrons from $\PZ$ boson decays,
both in the barrel and endcaps, and the absolute energy scale correction shifts
the dielectron mass distribution peak closer to the world-average $\PZ$ boson mass value.

\begin{figure*}[hbtp]
\centering
\includegraphics[ width=\cmsFigWidth]{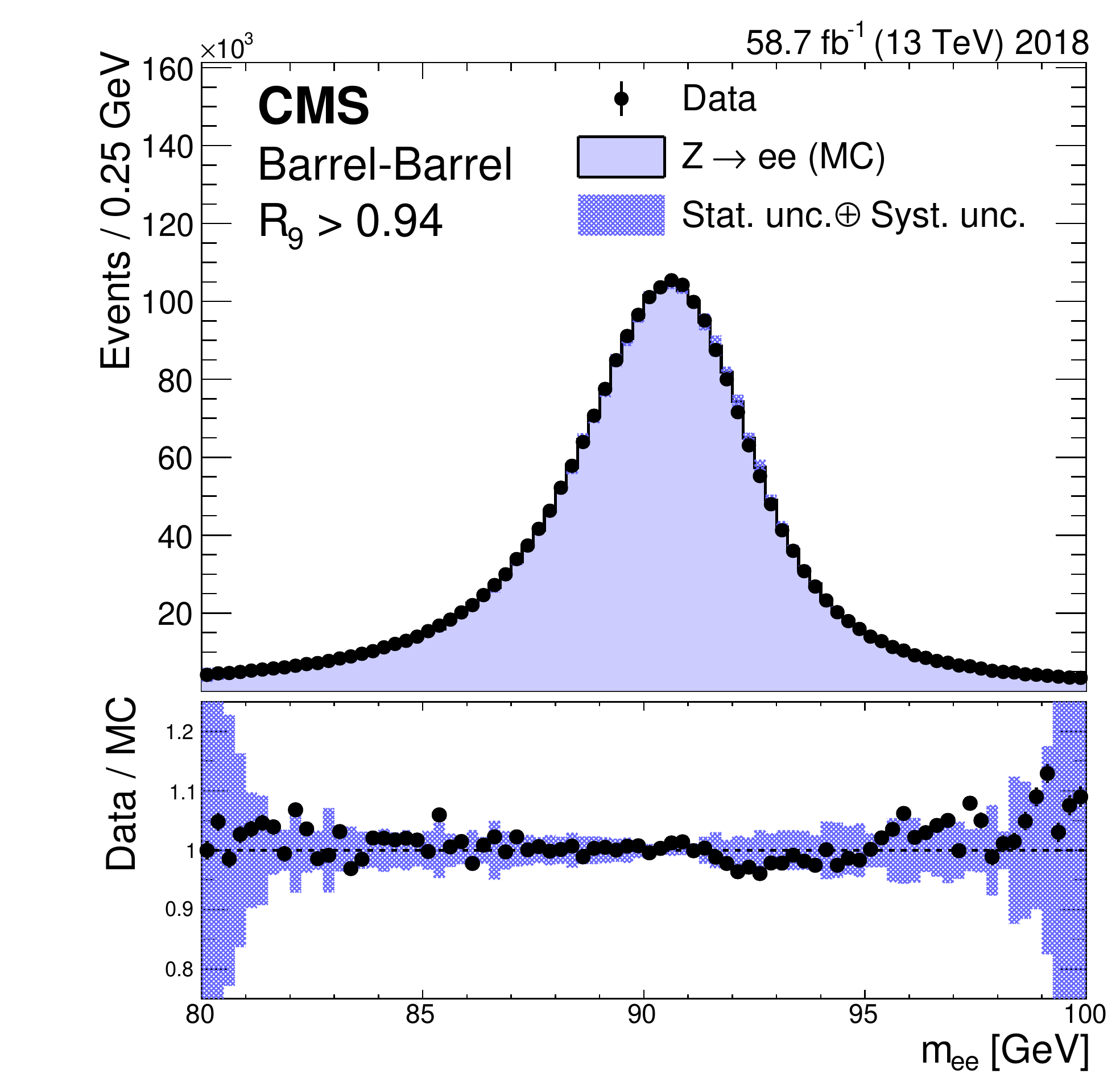}
\includegraphics[ width=\cmsFigWidth]{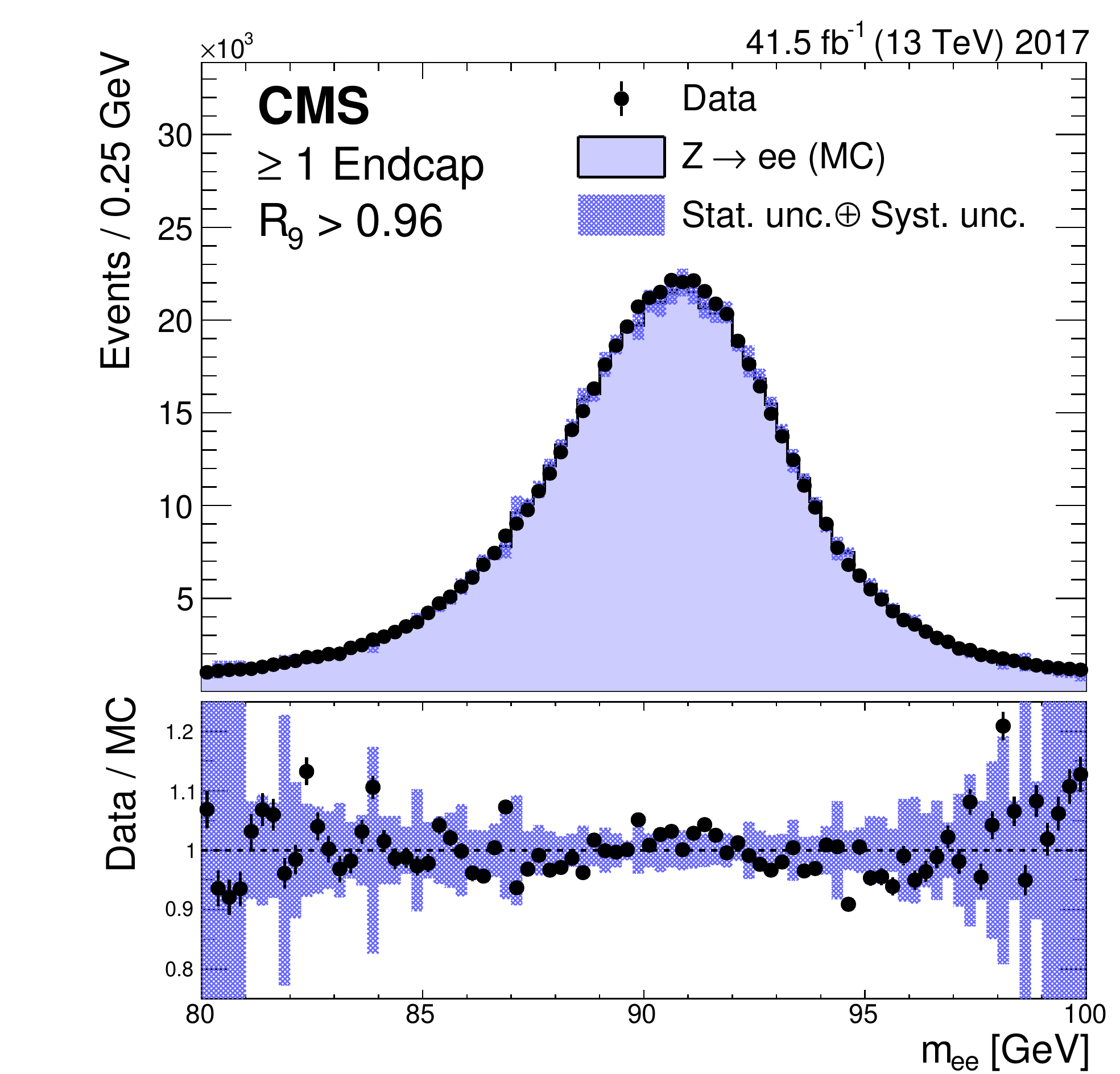}\\
\caption{Invariant mass distribution of $\PZ\to\Pe\Pe$ events, after spreading is applied to simulation and scale corrections to data. The results are shown for barrel (left) and endcap (right) electrons. The simulation is shown with the filled histograms and data are shown by the markers. The vertical bars on the markers represent the statistical uncertainties in data. The hatched regions show the combined statistical and systematic uncertainties in the simulation. The lower panels display the ratio of the data to the simulation with the bands representing the uncertainties in the simulation.}
\label{fig:ZpeakAfterSmearing}
\end{figure*}

The data-to-simulation agreement, after the
application of residual scales to data and spreadings to simulated events, is
shown in Fig.~\ref{fig:ZpeakAfterSmearing} for two representative categories.

The ultimate energy resolution after all the corrections (regression and scale corrections) ranges from 2 to 5\%, depending on electron pseudorapidity and energy loss through bremsstrahlung in the detector material.

\subsection{Performance and validation with data}
\label{sec:PerformanceAndValidationWithData}
Energy scale and spreading corrections, derived with electrons from $\PZ$ boson decays with a mean $\ET$ of around 45\GeV, are applied also to electrons and photons over a wide range of $\ET$ up to several hundreds of GeV. Therefore, it is important to validate the performance of the residual energy corrections on a sample of unbiased photons and on high-energy $\Pe$/$\gamma$.

\begin{figure*}[hbtp]
\centering
\includegraphics[ width=\cmsFigWidth]{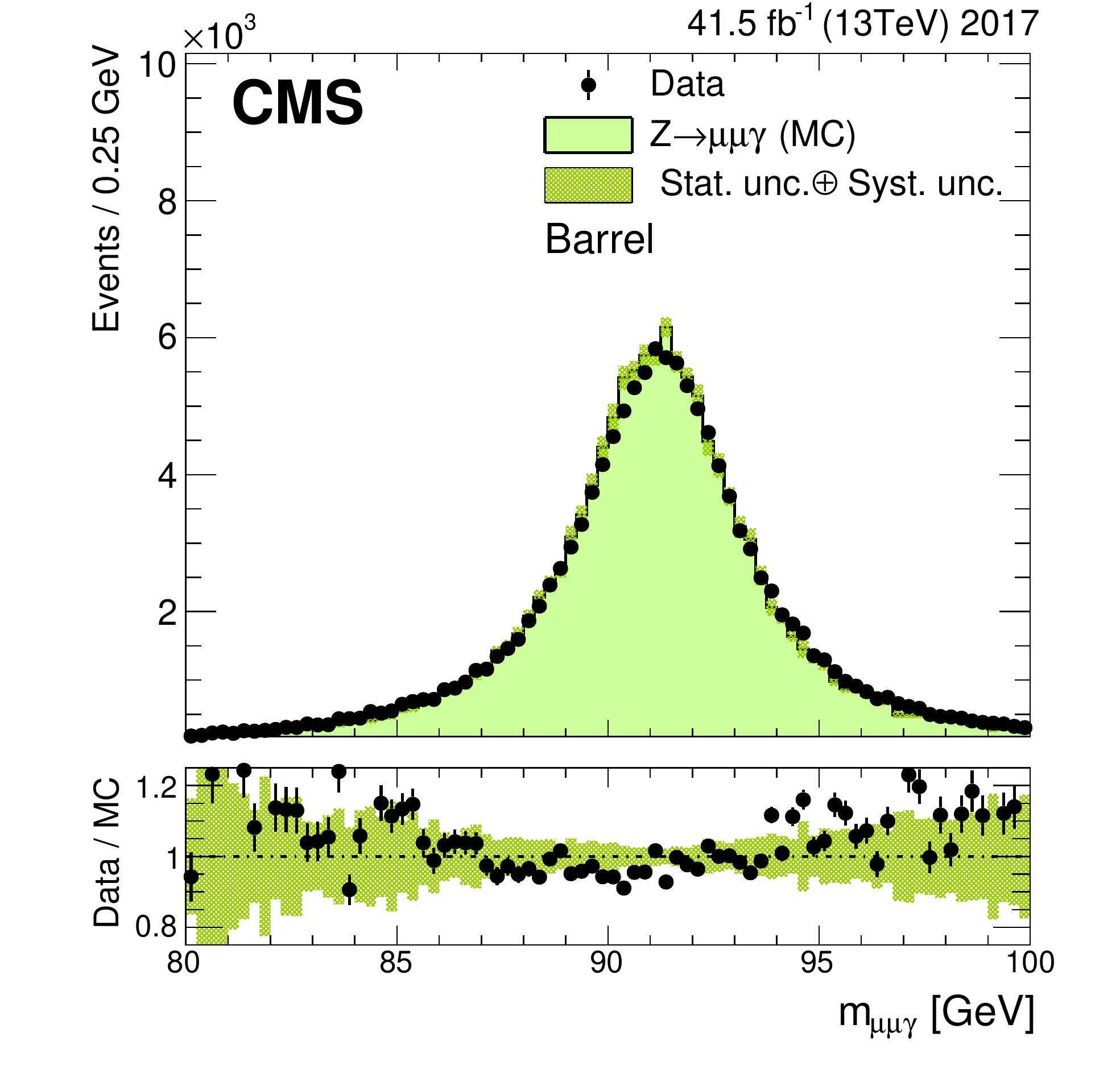}
\includegraphics[ width=\cmsFigWidth]{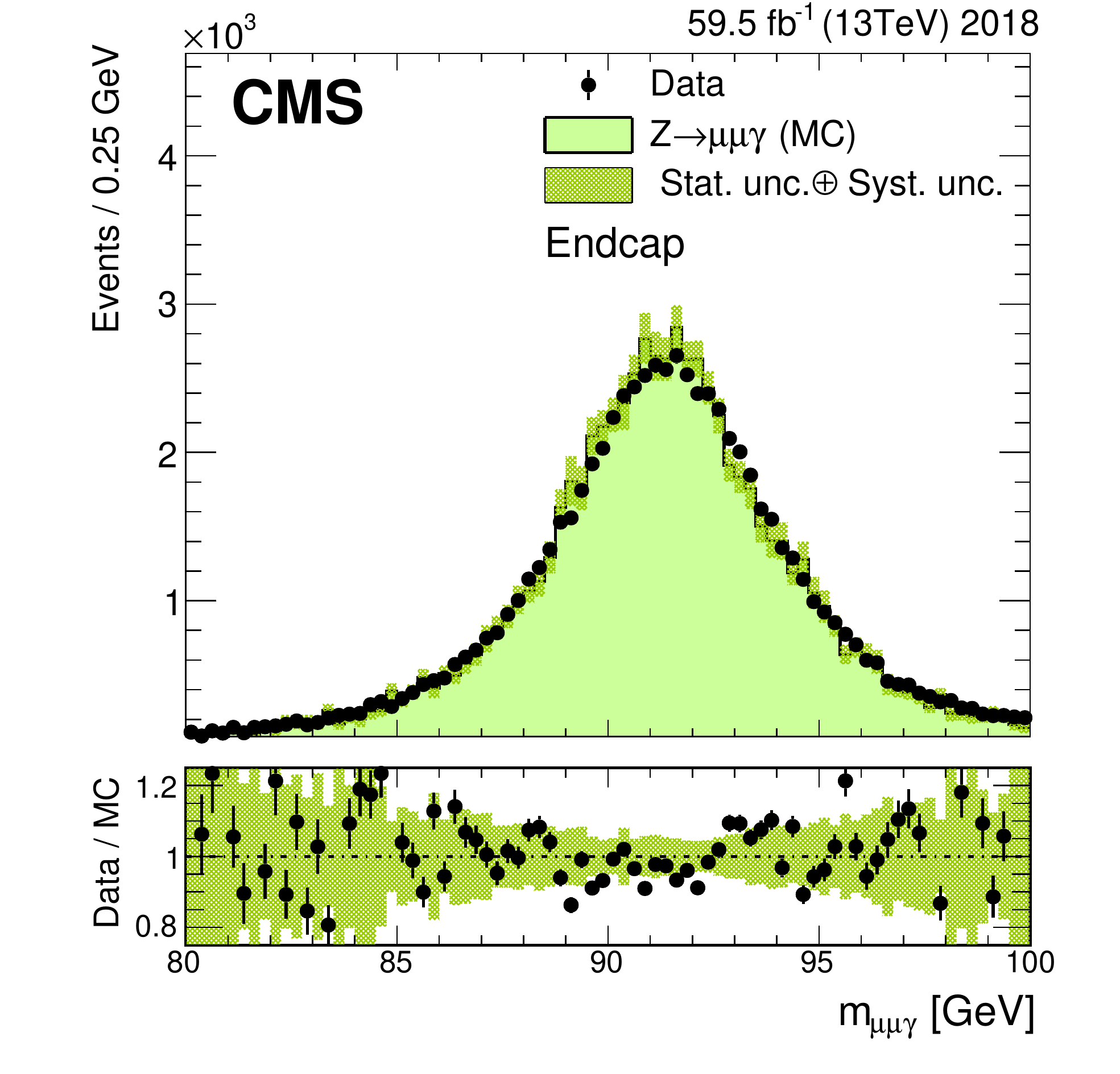}\\
\caption{Invariant mass distributions of $\PZ\to\mu\mu\gamma$ shown for barrel (left) and endcap (right) photons selected from 2017 data and simulation (left), 2018 data and simulation (right). The simulation is shown with the filled histograms and data are shown by the markers. The vertical bars on the markers represent the statistical uncertainties in data. The hatched regions show the sum of statistical and systematic uncertainties in the simulation. The lower panels display the ratio of the data to the simulation with bands representing the uncertainties in the simulation.}
\label{fig:ZFSRpeak}
\end{figure*}

To validate the unbiased photons, a sample of $\PZ \to \mu\mu\gamma$ events selected from data with 99\% photon purity is used.
Events in both data and simulation are required to satisfy standard dimuon trigger requirements.
An event is kept if there are at least two muons passing the tight muon identification requirements~\cite{Sirunyan:2018fpa}, with $\pt>30$ and $10\GeV$ and $\abs{\eta} < 2.4$.
The two muons must have opposite charges and an invariant mass
($m_{\mu\mu}$) greater than 35\GeV.
Once the dimuon system is identified,
a photon in the event is
required to have
$\abs{\eta}<2.5$, be reconstructed outside
the barrel-endcap transition region, and have
$\ET>20\GeV$.
The $\mu \mu \gamma$ system is then selected
by requiring that the photon is within $\Delta R = 0.8 $ of at least one of the muons.
After applying these criteria, roughly $140\times 10^3$ ($230\times 10^3$) of events have been selected with a photon in the EB for 2016--2017 (2018) and roughly $40\times 10^3$ ($80\times 10^3$) with a photon in the EE for 2016--2017 (2018) data sets.
Figure \ref{fig:ZFSRpeak} shows the invariant mass distribution of the  $\PZ\to\mu\mu\gamma$ system ($m_{\mu\mu\gamma}$) obtained after
applying the scale and spreading corrections derived with electrons from $\PZ$ boson decays to the photons, shown separately for barrel and endcap photons in 2017 data and simulation, and in 2018 data and simulation.
The photon energy scale is extracted for data and simulation from the mean of the distribution of a per-event estimator~\cite{Chatrchyan:2013dga} defined as $s=(m_{\mu\mu\gamma}^2-m_{\PZ}^2)/(m_{\PZ}^2-m_{\mu\mu}^2)$, where $m_{\PZ}$ denotes the Particle Data Group world-average $\PZ$ boson mass~\cite{Zyla2020}.
The energy scale difference between data, corrected with the energy scale corrections derived with electrons from $\PZ$ boson decays, and simulation, both from $\PZ\to\mu\mu\gamma$ events, is smaller than 0.1\% for photons both in the barrel or in the endcaps. This correction is within the quadratic sum of the statistical and systematic uncertainties associated with this scale extraction process, which include scale and spreading systematic uncertainty, as well as systematic uncertainties due to corrections applied to muon momenta~\cite{Sirunyan:2018fpa}.

The performance of the energy corrections on high-energy $\Pe$/$\gamma$ is validated by using $\PZ\to\Pe\Pe$ data and MC samples, with scale and spreading corrections applied. The residual corrections for $\ET$ between 120 and 300\GeV are better than 0.8 (1.1)\% in the barrel (endcaps). These values are used to derive the systematic uncertainties in the energy correction extrapolation above 300\GeV, where the statistics are very low. In this $\ET$ range, the systematic uncertainty is conservatively assumed to be 2 (3) times the systematic uncertainty in EB (EE) of the $\ET$ range 120--300\GeV.

\subsubsection{Impact of residual corrections in $\PH\to \gamma\gamma$ channel}
\label{sec:rescorr_Hgg}

\begin{figure}[t]
\centering
\includegraphics[ width=\cmsFigWidth]{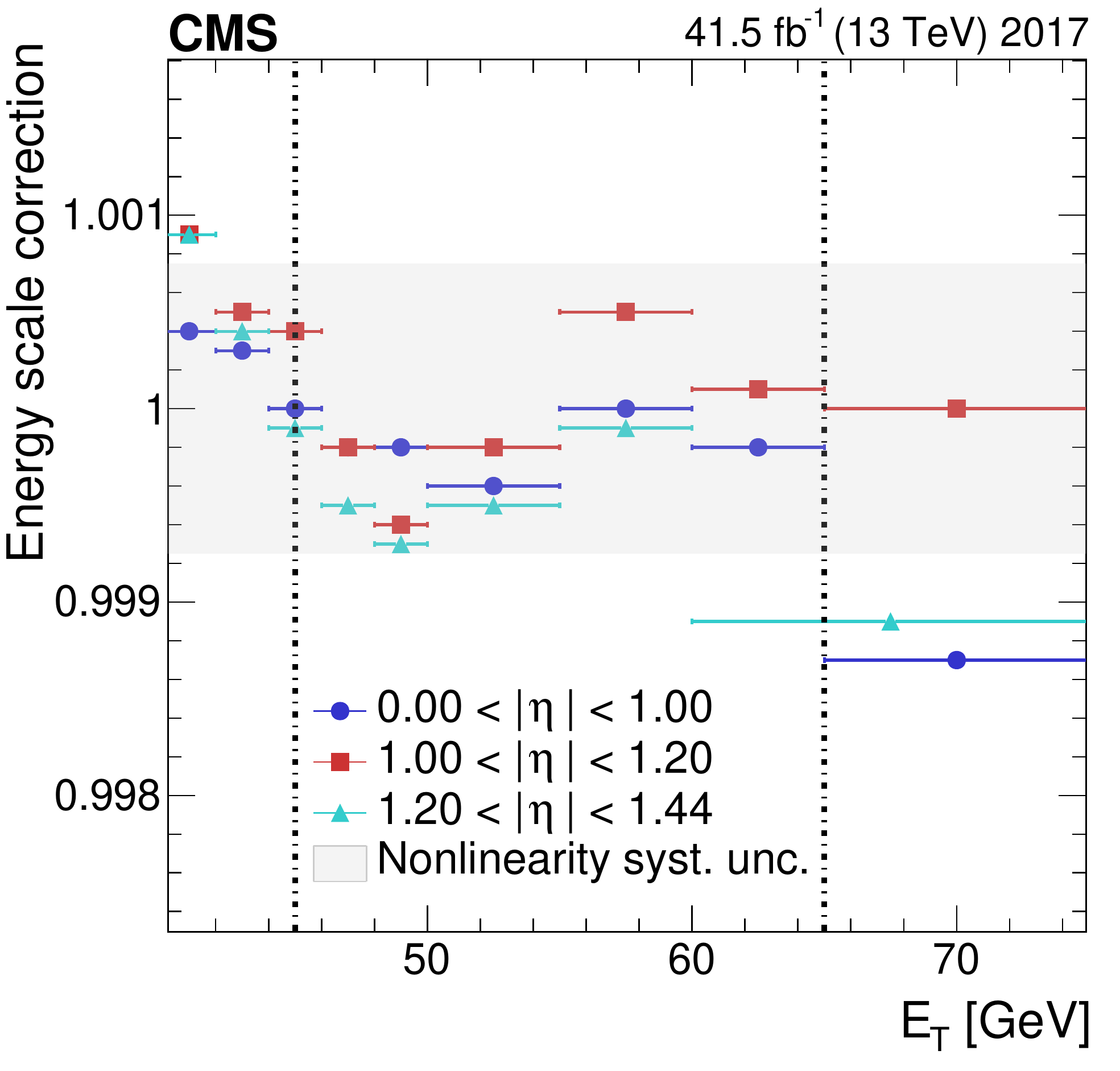}
\caption{Energy scale corrections as a function of the photon $\ET$. The
horizontal bars represent the variable bin width. The
systematic uncertainty associated with this correction is approximately the
maximum deviation observed in the $\ET$ range between 45 and 65\GeV for
electrons in the barrel.}
\label{fig:scale_vs_et}
\end{figure}

The mass of the Higgs boson in the diphoton channel has been recently measured exploiting $\Pp\Pp$ collision data collected at a center-of-mass energy of 13\TeV by the CMS experiment during 2016, corresponding to an integrated luminosity of 35.9\fbinv~\cite{Sirunyan:2020xwk}.
This result benefits from a refined calibration of the ECAL, while exploiting new analysis techniques to constrain the uncertainty in the Higgs boson mass to $m_{\PH} = 125.78 \pm 0.26\GeV$~\cite{Sirunyan:2020xwk}.
A key requirement for this measurement was to
measure and correct for nonlinear discrepancies between data and simulation in the energy scale, as a function of $\ET$, using electrons from $\PZ$ boson decays.
Additional energy scale corrections were derived in bins of $\abs{\eta}$ and $\ET$ to account for any
nonlinear response of the ECAL with energy for the purpose of this high-precision measurement.
The corrections obtained from this step are shown in
Fig.~\ref{fig:scale_vs_et} for electrons, as functions of $\ET$, in three
bins of $\abs{\eta}$ in the EB. This improves the precision of the $m_{\PH}$ measurement, since the energy spectrum of the electrons from the $\PZ$ boson decays ($\langle \ET \rangle\approx 45\GeV$) used to derive the scale corrections, is different from the energy spectrum of photons from the Higgs boson
decay ($\langle \ET \rangle\approx 60\GeV$).

The accuracy of the energy scale correction extrapolation in the energy range of interest of the $\PH\to \gamma\gamma$ search (between 45 and 65\GeV in $\ET$) is 0.05--0.1 (0.1--0.3)\% for photons in the EB (EE)~\cite{Sirunyan:2020xwk}. The quality of the energy resolution achieved provides the highest precision to the Higgs boson mass measurement in the diphoton channel.

\section{Electron and photon selection}\label{sec:sec8}
Many physics processes under study at the LHC are
characterized by the presence of electrons or photons
in the final state. The performance of the identification
algorithms for electrons and photons is therefore crucial
for the physics reach of the CMS experiment.
Two different techniques are used in CMS  for the identification of electrons and photons. One is based on sequential requirements (cut-based), and the other is based on a multivariate discriminant. Although the latter is more suited for precision measurements and physics analyses with well-established final states, the former is largely used for model independent searches of nonconventional signatures.
We describe below in detail the main strategies of electron and photon identification and the performance through the full Run 2.

\subsection{Electron and photon identification variables}

Different strategies are used to identify prompt (produced at the primary vertex) and isolated electrons and photons, and separate them from
background sources. For prompt electrons, background sources can originate from photon conversions, hadrons misidentified as electrons, and secondary electrons from semileptonic decays of \cPqb or \cPqc quarks.
The most important background to prompt photons
arises from jets fragmenting mainly
into light neutral mesons
$\pi^0$ or $\eta$, which subsequently decay promptly to two photons. For the energy range of interest, the $\pi^0$ or $\eta$ are
significantly boosted, such that the two photons from the decay are nearly collinear and
are difficult to distinguish from a single-photon incident on the calorimeter.
Different working points are defined to identify either electrons or photons, corresponding to identification efficiencies of approximately
70, 80, and 90\%, respectively.
In all cases data and simulation efficiencies are compatible within 1--5\% over the full $\eta$ and $\ET$ ranges for electrons and photons.

\subsubsection{Isolation criteria}\label{sec:isos}
One of the most efficient ways to reject electron and photon backgrounds is the use of isolation energy sums,
a generic class of discriminating variables that are constructed from the sum of the reconstructed energy in a cone
around electrons or photons in different subdetectors. For this purpose, it is convenient to define cones in
terms of an $\eta$--$\phi$ metric; the distance with respect to the reconstructed electron or photon direction is
defined by $\Delta R$.
To ensure that the energy from the electron or photon itself
is not included in this sum, it is necessary to define a veto region inside the isolation cone, which is excluded from the isolation sum.

Electron and photon isolation exploits the information provided by the PF event reconstruction~\cite{Sirunyan:2017ulk}.
The isolation variables are obtained by summing the transverse momenta of charged hadrons ($I_{\text{ch}}$), photons ($I_{\gamma}$), and neutral hadrons ($I_{\text{n}}$), inside an isolation cone of $\Delta R = 0.3$ with respect to the electron or photon direction. The larger the energy of the incoming electrons or photons, the larger the amount of energy spread around its direction in the various subdetectors. For this reason, the thresholds applied on the isolation quantities are frequently parametrized as a function of the particle $\ET$, as indicated in Tables~\ref{tab:barr_endcap_cutpho}
and~\ref{tab:cbarr_endcap_cutele}.

The isolation variables are corrected to mitigate the contribution from pileup. 
This contribution in the isolation region is estimated as
$\rho A_{\text{eff}}$, where $\rho$ is the
median of the transverse energy density per unit area in the event
and $A_{\text{eff}}$ is the area of the isolation region weighted by a factor that accounts for the dependence of the pileup transverse energy density on the object $\eta$~\cite{Khachatryan:2015hwa}. The quantity $\rho A_{\text{eff}}$ is subtracted from the isolation quantities.

\begin{figure*}[hbtp]
\centering
\includegraphics[ width=\cmsFigWidth]{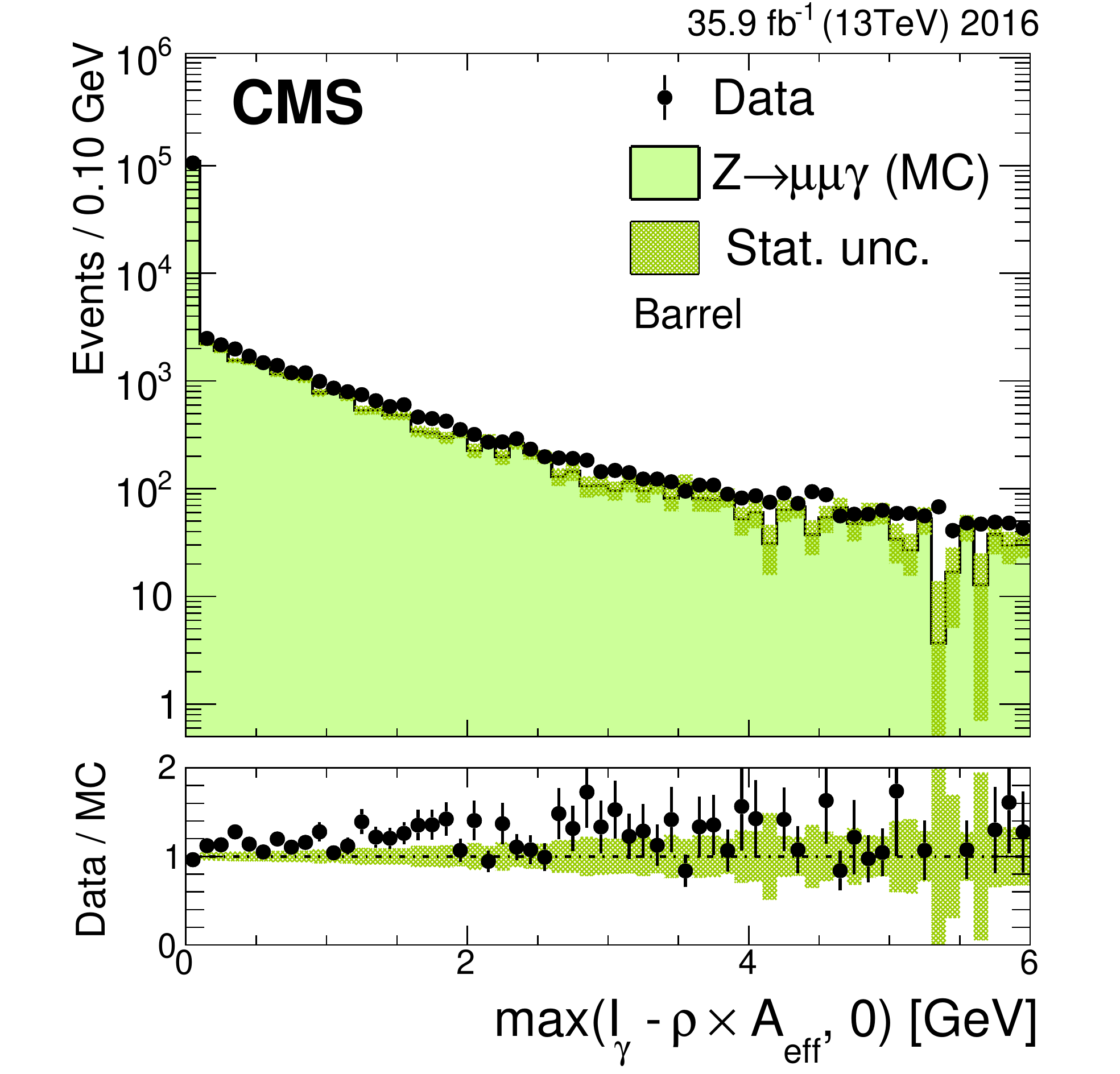}
\includegraphics[ width=\cmsFigWidth]{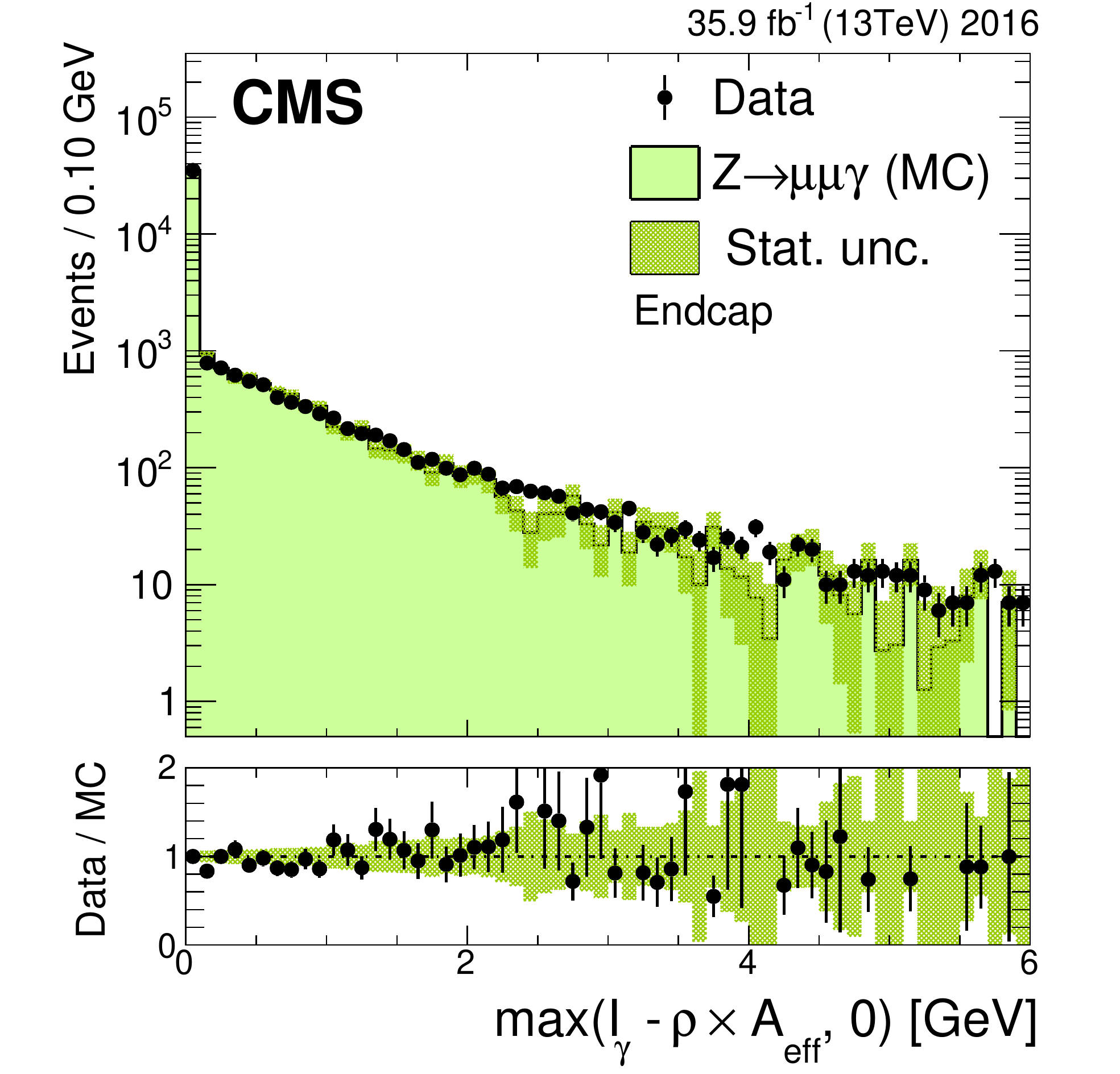}\\
\caption{The $\rho$-corrected PF photon isolation in a cone defined by $\Delta R = 0.3$ for photons in $\PZ\to\mu\mu\gamma$ events in the EB (left) and in the EE (right). Photons are selected from 2016 data and simulation. The simulation is shown with the filled histograms and data are represented by the markers. The vertical bars on the markers represent the statistical uncertainties in data. The hatched regions show the statistical uncertainty in the simulation. The lower panels display the ratio of the data to the simulation, with the bands representing the uncertainties in the simulation.}
\label{fig:ZFSR_pfPhotonISO}
\end{figure*}

The distributions of $I_{\gamma}$ for photons after the $\rho$ corrections are shown in Fig.~\ref{fig:ZFSR_pfPhotonISO} for photons in the EB and EE.

\subsubsection{Shower shape criteria}
Another method to reject jets with high electromagnetic content exploits the shape of the electromagnetic shower in the ECAL.
Even if the two photons from neutral hadron decays inside a jet cannot be fully resolved, a wider shower profile is expected, on average, compared with a single incident electron or photon. This is particularly true along the $\eta$ axis of the cluster, since the presence of the material combined with the effect of the magnetic field reduce the discriminating power resulting from the $\phi$ profile of the shower. This can elongate the electromagnetic cluster in
the $\phi$ direction for both converted photons as well as pairs of photons from
neutral hadron decays where at least one of the photons has converted.
Several shower-shape variables are constructed to parameterize the differences between the geometrical shape of energy deposits from prompt photons or electrons compared with those caused by hadrons from jets. The following are two of the most relevant variables used for photon and electron identification.

\begin{itemize}

\item Hadronic over electromagnetic energy ratio ($H/E$):
The $H/E$ ratio is defined as the ratio between
the energy deposited in the HCAL in a cone of radius $\Delta R = 0.15$ around the SC
direction and the energy of the photon or electron candidate.
There are three sources that significantly contribute to the measured hadronic energy (H) of a genuine electromagnetic object: HCAL noise, pileup, and leakage of electrons or photons through the inter-module gaps. For low-energy
electrons and photons, the first two sources are the primary contributors, whereas for high-energy electrons, the last contribution dominates.
Therefore, to cover both low- and high-energy regions, the $H/E$ selection requirement is of the form
 $H < X + Y \rho + J E$,
where $X$ and $Y$ represent the noise and pileup terms, respectively, and $J$ is a scaling term for high-energy electrons and photons.
An example of the $H/E$ distribution in data and simulation is shown in Fig.~\ref{fig:hoeEle} for electrons in the barrel and endcap regions.
The data-to-simulation ratio in Fig.~\ref{fig:hoeEle} is mostly consistent with one except for $H/E > 0.16$ where background from events with nonprompt electrons starts to contribute.

\begin{figure*}[hbtp]
\centering
\includegraphics[ width=\cmsFigWidth]{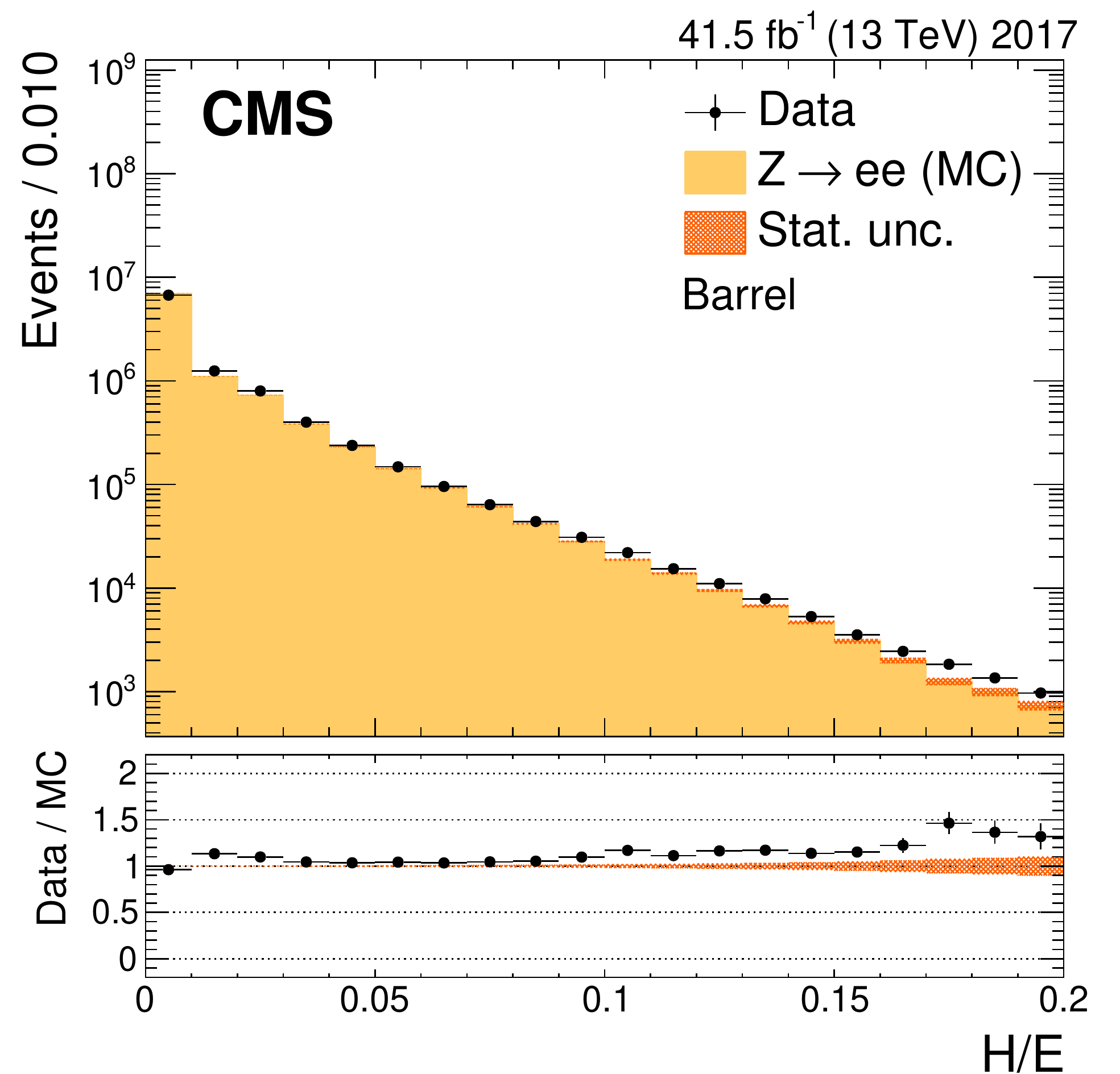}
\includegraphics[ width=\cmsFigWidth]{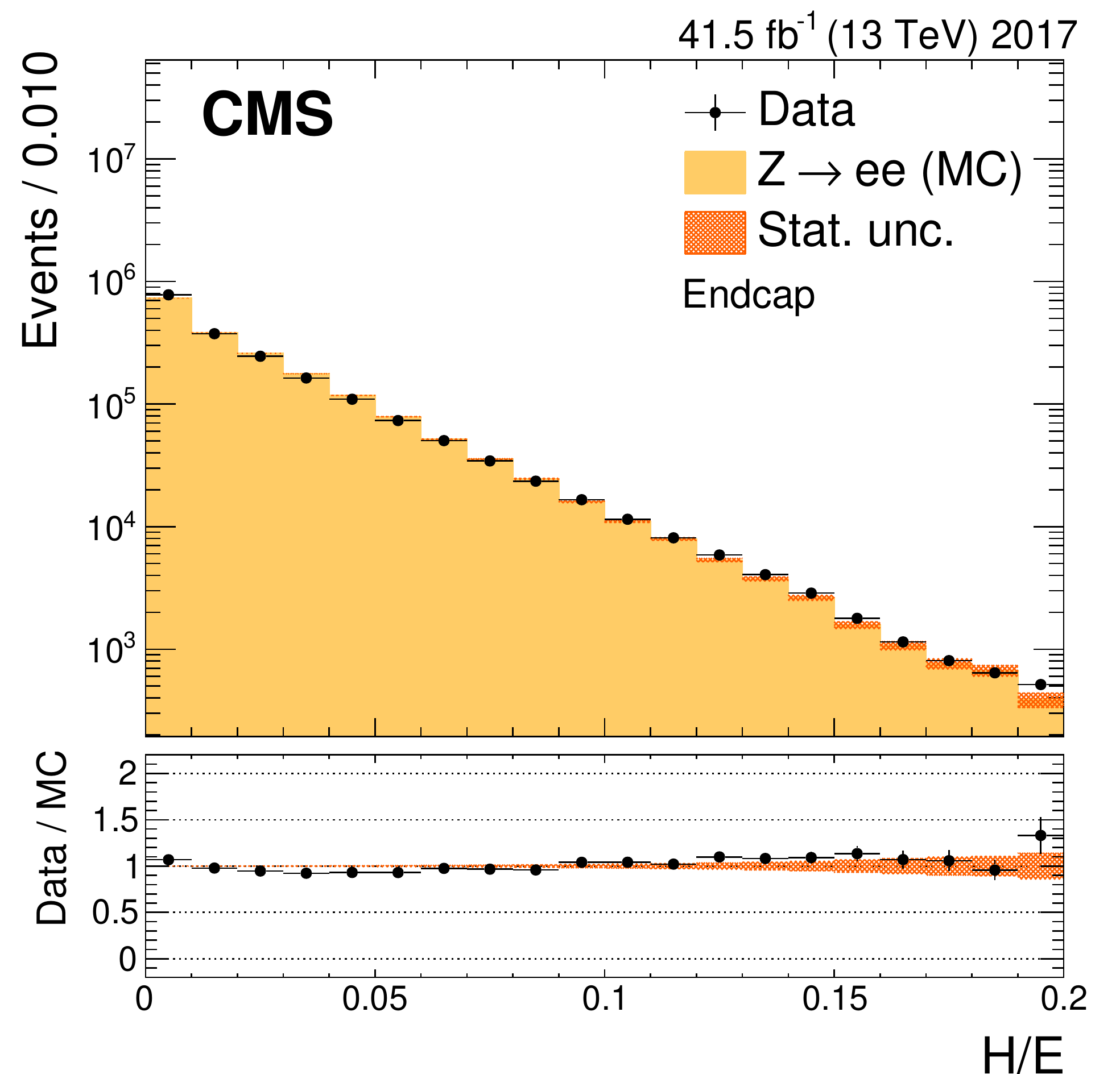}\\
\caption{Distribution of $H/E$ for electrons in the barrel (left) and in the endcaps (right). The simulation is shown with the filled histograms and data are represented by the markers. The vertical bars on the markers represent the statistical uncertainties in data. The hatched regions show the statistical uncertainty in the simulation. The lower panels display the ratio of the data to the simulation, with the bands representing the uncertainties in the simulation.
}
\label{fig:hoeEle}
\end{figure*}
\item $\sigma_{i \eta i \eta}$: the second moment of the log-weighted distribution
of crystal energies in $\eta$, calculated in the $5{\times}5$ matrix around the most
energetic crystal in the SC and rescaled to units of crystal size. The mathematical expression is given below:

\begin{equation}\label{eq:sieie}
\sigma_{i \eta i \eta} = \sqrt{ \frac{ \sum_{i}^{5{\times}5} w_i (\eta_i - \overline{\eta}_{5{\times}5})^2 }{\sum_{i}^{5{\times}5} w_i}  }.
\end{equation}
Here, $\eta_i$ is the pseudorapidity of the $i$th crystal, $\overline{\eta}_{5{\times}5}$ denotes the pseudorapidity mean position, and $w_i$
is a weight factor which is defined as: $w_i = \max (0 , 4.7 + \ln (E_i / E_{5{\times}5}))$, and is nonzero if $\ln (E_i / E_{5{\times}5}) > -4.7$, i.e., $E_i > 0.9\% $ of $E_{5{\times}5}$. This selection requirement is intended to reject ECAL noise, by
ensuring that crystals with energy deposits of at least 0.9\% of $E_{5{\times}5}$, the energy deposited in a $5{\times}5$ crystal matrix around the most energetic crystal,
will contribute to the definition of $\sigma_{i \eta i \eta}$.
Because of the presence of upstream material and the magnetic field, the shower from an electron or a photon spreads into more than one crystal. The size of the crystal in $\eta$ in the EB is 0.0175 and in the EE it varies from 0.0175 to 0.05. Following Eq.~\ref{eq:sieie}, the $\sigma_{i \eta i \eta}$ variable essentially depends on the distance between two crystals in $\eta$. Thus, the spread of $\sigma_{i \eta i \eta}$ in EE is twice larger than in the EB.
The distribution of $\sigma_{i \eta i \eta}$ is expected to be narrow for single-photon or electron showers, and broad for
two-photon showers that arise from neutral meson decays.
An example of the $\sigma_{i \eta i \eta}$ distribution in data and MC is shown in Fig.~\ref{fig:ZFSR_sigamIeIe} for photons in the barrel and in the endcap regions.

\begin{figure*}[hbtp]
\centering
\includegraphics[ width=\cmsFigWidth]{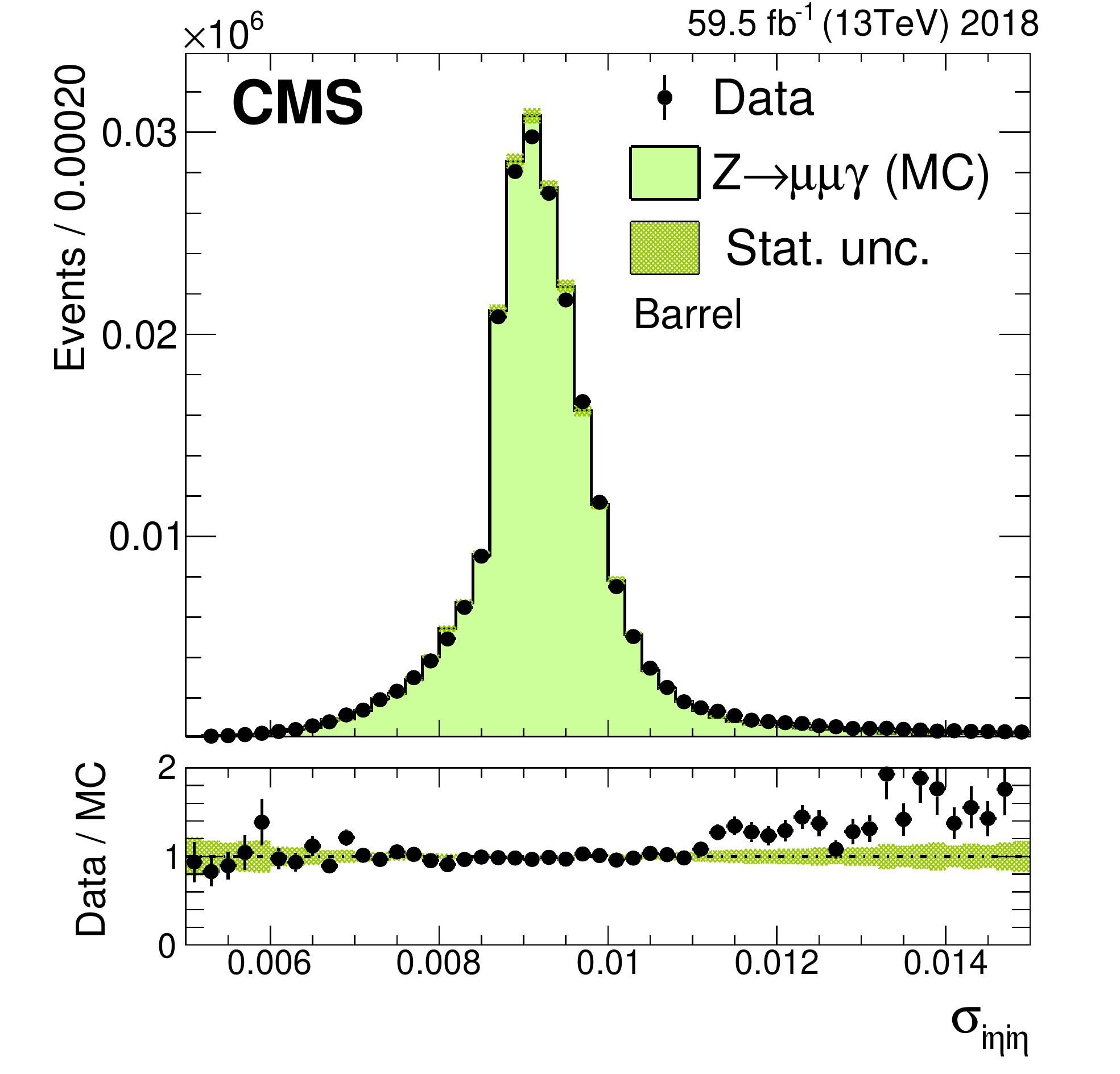}
\includegraphics[ width=\cmsFigWidth]{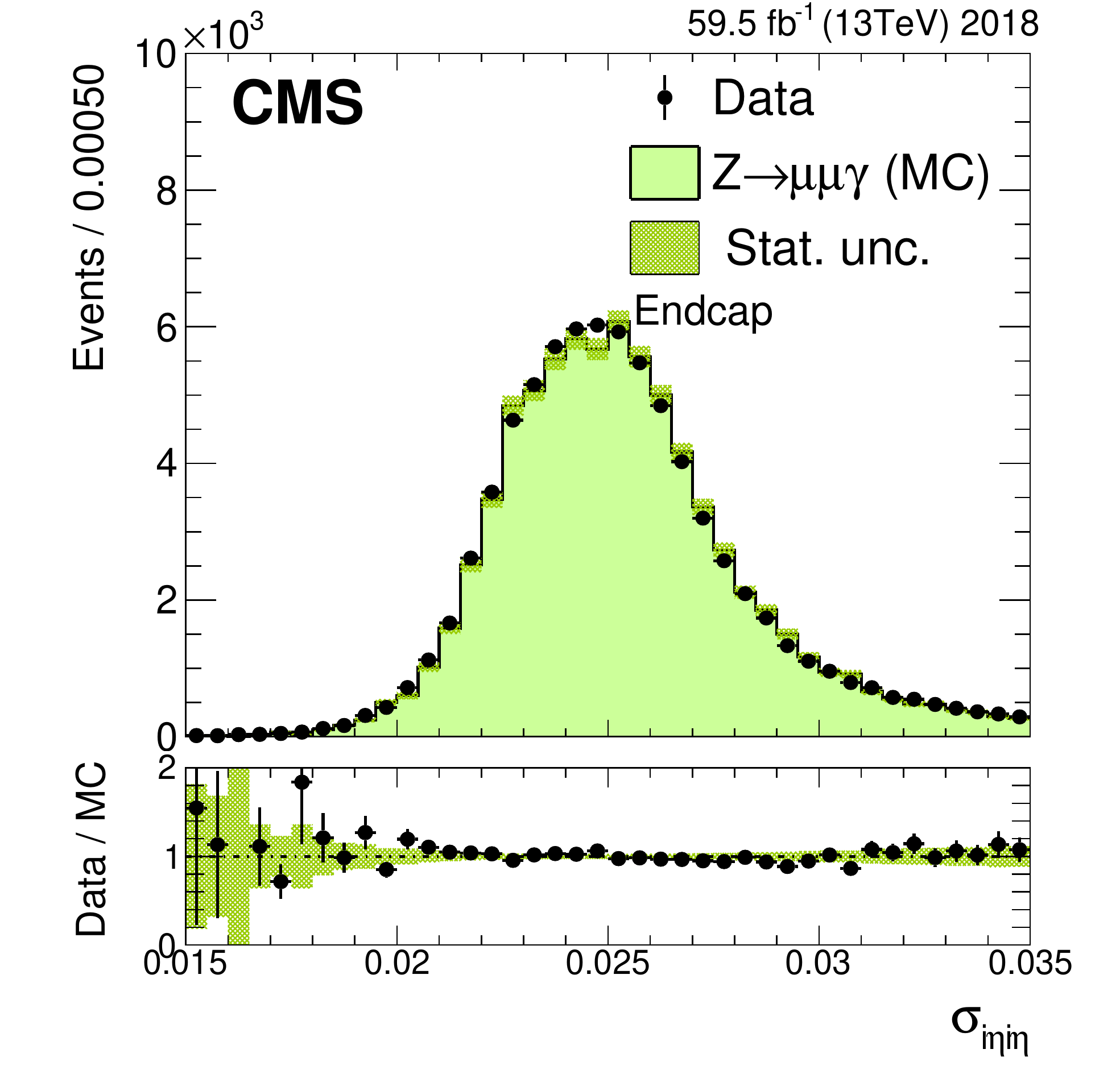}\\
\caption{ Distribution of $\sigma_{i \eta i \eta}$ for photons from $\PZ\to\mu\mu\gamma$ events in the barrel (left) and in the endcaps (right). Photons are selected from 2018 data and simulation. The simulation is shown with the filled histograms and data are represented by the markers. The vertical bars on the markers represent the statistical uncertainties in data. The hatched regions show the statistical uncertainty in the simulation. The lower panels display the ratio of the data to the simulation, with the bands representing the uncertainties in the simulation.}
\label{fig:ZFSR_sigamIeIe}
\end{figure*}
\end{itemize}

Another important variable is $R_9$. Showers of photons
that convert before reaching the calorimeter have wider transverse profiles and lower values
of $R_9$ than those of unconverted photons. The energy weighted $\eta$-width and $\phi$-width of the SC provide further information of
the lateral spread of the shower. In the endcaps, where CMS is
equipped with a preshower detector, the variable $\sigma_{\text{RR}} = \sqrt{\smash[b]{\sigma_{xx}^2 + \sigma_{yy}^2}}$ is also used, where $\sigma_{xx}$ and $\sigma_{yy}$
measure the lateral spread in the two orthogonal directions of the sensor planes of the preshower detector.

\subsubsection{Additional electron identification variables}\label{sec:AddVar}
\begin{figure*}[hbtp]
\centering
\includegraphics[ width=\cmsFigWidth]{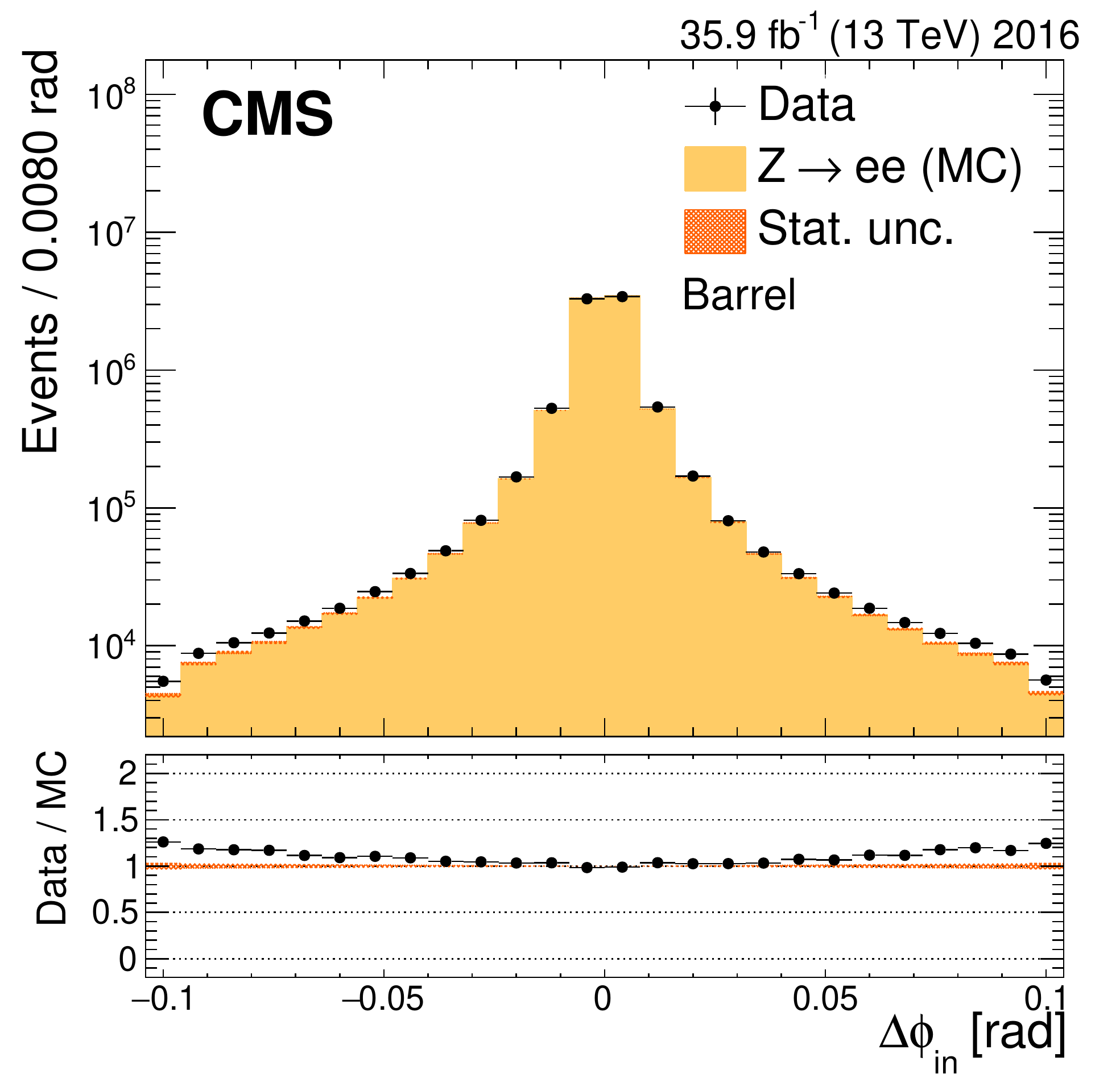}
\includegraphics[ width=\cmsFigWidth]{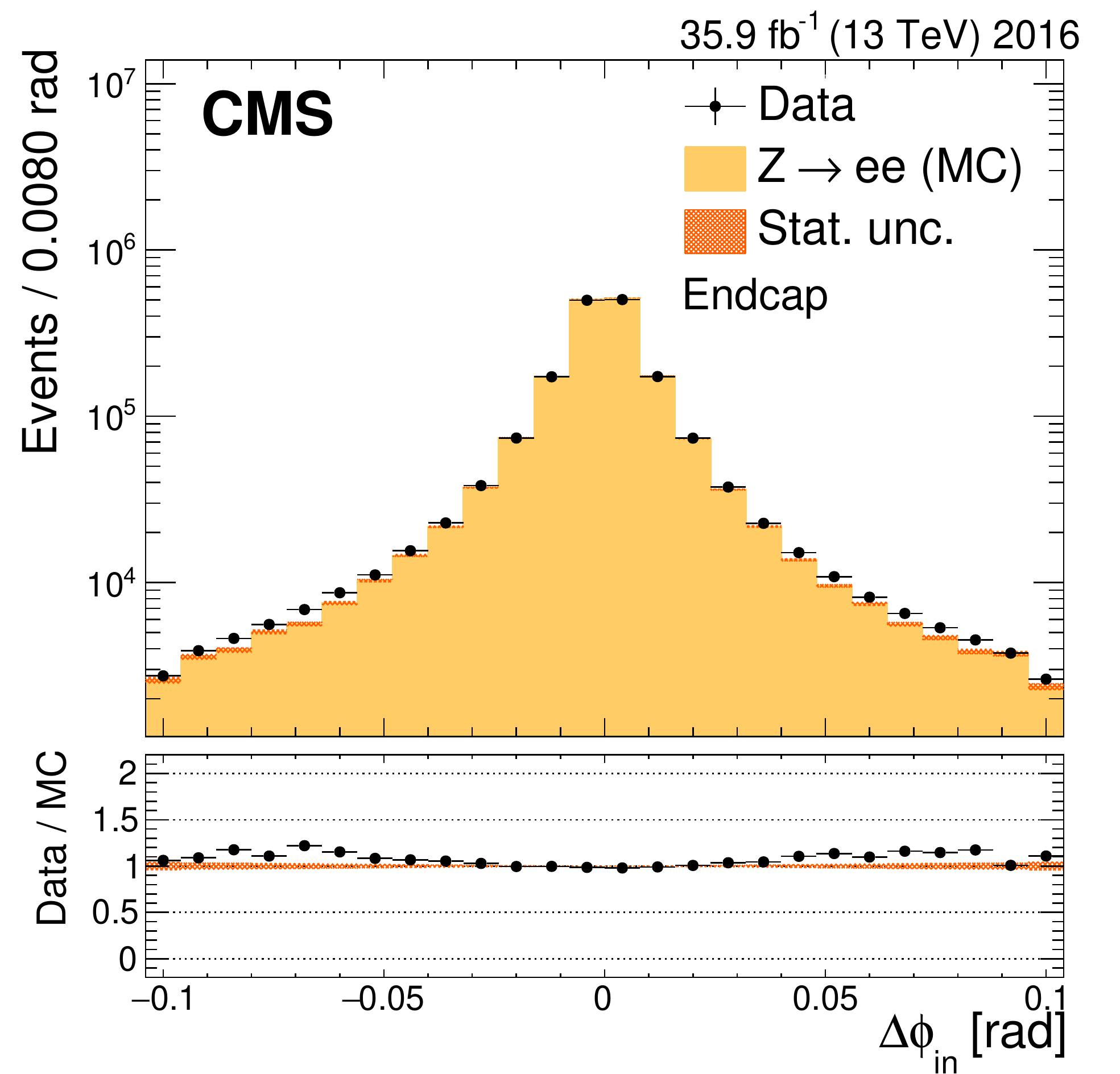}\\
\caption{Distribution of $\Delta\phi_{\text{in}}$ for electrons in the barrel (left) and endcaps (right). For the cut-based identification, selection requirements on this variable are listed in Table~\ref{tab:barr_endcap_cutpho}. The simulation is shown with the filled histograms and data are represented by the markers. The vertical bars on the markers represent the statistical uncertainties in data. The hatched regions show the statistical uncertainty in the simulation. The lower panels display the ratio of the data to the simulation, with the bands representing the uncertainties in the simulation.}
\label{fig:dphiin}
\end{figure*}

Additional tracker-related variables are used for the identification of electrons.
One such discriminating variable is $\abs{1/E - 1/p}$, where $E$ is the SC energy and $p$ is the track momentum at the point of closest approach
to the vertex.
Another important variable for the electron identification is $\abs{\Delta\eta^{\text{seed}}_{\text{in}}}$
 defined as
$\abs{\eta_{\text{seed}} - \eta_{\text{track}}}$, where $\eta_{\text{seed}}$ is the position of the seed cluster in $\eta$,
and $\eta_{\text{track}}$ is the track $\eta$ extrapolated from the innermost track position.
Similarly, $\abs{\Delta\phi_{\text{in}}} = \abs{\phi_{\text{SC}} - \phi_{\text{track}}}$ is another discriminating variable that uses the SC energy-weighted position in $\phi$ instead of the seed cluster $\phi$. An example of the $\Delta\phi_{\text{in}}$ distribution in data and simulation is shown in Fig.~\ref{fig:dphiin} for electrons in the barrel and endcap regions.

An important source of background to prompt electrons
arises from secondary electrons produced in conversions of photons in the tracker material.
To reject this background, CMS algorithms exploit the pattern of hits associated with the electron track.
When photon conversions take place inside the tracker volume, the first hit of the electron tracks from the
converted photons is not likely to be located in the innermost tracker layer, and missing hits
are therefore expected in the first tracker layers.
For prompt electrons, whose trajectories start from the
beamline, no missing hits are expected in the innermost layers.
Distributions of some of the identification variables for electrons are shown in Figs.~\ref{fig:hoeEle} and~\ref{fig:dphiin}, for electrons from $\PZ$ boson decays in data and simulation. The data-to-simulation ratio is close to unity, except at the very high end tail, where background from events with nonprompt electrons start to contribute.

\subsection{Photon identification}\label{sec:phoID}
A detailed description of photon identification strategies is given below.

\subsubsection{Cut-based photon identification}\label{sec:phoIDCB}
Requirements are made on $\sigma_{i \eta i \eta}$, $H/E$, and the isolation sums after correcting for pileup as detailed in Section~\ref{sec:isos}.
A summary of the standard identification requirements
for photons in the barrel and the endcaps is given in Table~\ref{tab:barr_endcap_cutpho}
for the tight
working point.
The selection requirements were tuned using a MC sample with 2017 data-taking conditions, but these identification criteria are suitable for use in all three years of Run 2.
The ``loose'' working point has an average signal efficiency of about 90\%, and is generally used when backgrounds are low.
The ``medium'' and ``tight'' working points have an average efficiency of about 80\% and 70\%, respectively, and are used in situations where the
background is expected to be larger.

\begin{table}[thb]
\centering
\topcaption{Cut-based photon identification requirements for the tight working point in the EB and EE. }
\begin{tabular}{ccc}
\hline
Variable  & Barrel (tight WP) & Endcap (tight WP) \\
\hline
$H/E$       &  $<$0.021  &  $<$0.032 \\
$\sigma_{i \eta i \eta}$ & $<$0.0099 & $<$0.027 \\
$I_{\text{ch}}$ &  $<$0.65\GeV  &  $<$0.52\GeV \\
$I_\text{n}$ & $<$$0.32\GeV + 0.015 \ET +$  & $<$$2.72\GeV + 0.012 \ET +$ \\
             &  $2.26 \times 10^{-5} \ET^2/\GeVns$ & $2.3 \times 10^{-5} \ET^2/\GeVns$\\
$I_\gamma$ & $<$$2.04\GeV + 0.0040 \ET$
& $<$$3.03\GeV + 0.0037 \ET$ \\
\hline
\end{tabular}
\label{tab:barr_endcap_cutpho}
\end{table}

\subsubsection{Electron rejection}\label{sec:CSEV}
Along with the cut-based photon identification criteria, a prescription is required to reject electrons in the photon identification scheme.
The most commonly used method is the conversion-safe electron veto~\cite{Khachatryan:2015iwa}.
This veto requires the absence of charged particle tracks, with a
hit in the innermost layer of the pixel detector not matched to a reconstructed conversion vertex,
pointing to the photon cluster in the ECAL.
A more efficient rejection of electrons can be achieved by rejecting any photon for which a pixel detector seed consisting of at least two hits in the pixel detector points to the ECAL within some window defined around the photon SC position.
The conversion-safe electron veto is appropriate in the cases where electrons do not constitute a major
background, whereas the pixel detector seed veto is used when electrons misidentified as photons are expected to be an important background.

\subsubsection{Photon identification using multivariate techniques}\label{sec:phoIDMVA}
A more sophisticated photon identification strategy is based on a multivariate technique, employing
a BDT implemented in the TMVA framework~\cite{Hocker:2007ht}. Here, a single discriminant variable is built
based on multiple input variables, and provides excellent separation between signal (prompt photons) and background from misidentified jets.
The signal is defined as reconstructed photons from a $\gamma$ + jets simulated sample that are matched at generator level with prompt photons within a cone of size $\Delta R = 0.1$, whereas the background is defined by reconstructed photons in the same sample that do not match with a generated photon within a cone of size $\Delta R = 0.1$.
Photon candidates with $\ET > 15\GeV$, $\abs{\eta} < 2.5$, and satisfying very loose preselection requirements are used for the training of the BDT.
The preselection requirements consist of very loose cuts on $H/E$, $\sigma_{i \eta i \eta}$, $R_9$, PF photon isolation and track isolation.

The variables used as input to the BDT include
shower-shape and isolation variables, already presented above. Three more quantities are used that improve the discrimination between signal and background by including the dependencies of the shower-shapes and isolation variables on the event pileup, $\eta$ and $\ET$ of the candidate photon: the median energy per unit area, $\rho$; the $\eta$; and the uncorrected energy of the SC corresponding to the photon candidate.

\begin{figure}[hbtp]
  \centering
    \includegraphics[width=0.5\textwidth]{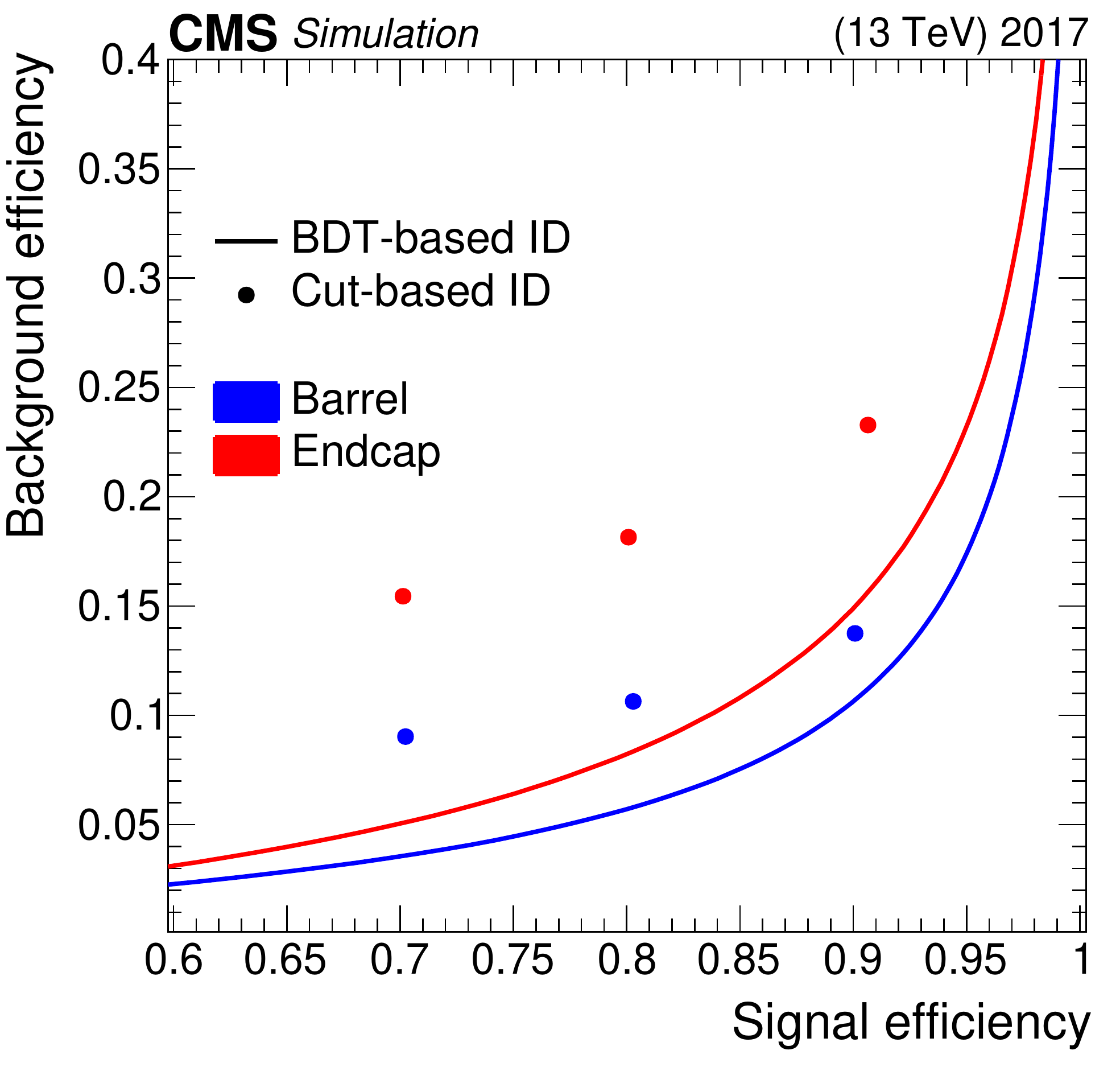}
     \caption{Performance of the photon BDT and cut-based identification algorithms in  2017. Cut-based identification is shown for three different working points, loose, medium and tight.}
    \label{fig:MVAvsCutBasedPhotons}
\end{figure}

A comparison of the performance between cut-based identification and BDT identification for photons is shown in Fig.~\ref{fig:MVAvsCutBasedPhotons}. The background efficiency as a function of the signal efficiency is reported for the multivariate identification (curves) and for the cut-based selection (discrete points).

\subsection{Electron identification}\label{sec:eleID}
A detailed description of electron identification strategies is given below.
\subsubsection{Cut-based electron identification}\label{sec:eleIDCB}

The sequential electron identification selection includes the requirements for seven identification variables, with thresholds as
listed in Table~\ref{tab:cbarr_endcap_cutele} for the representative tight working point.
The selection requirements were tuned using a MC sample with 2017 data-taking conditions, and this selection is suitable for use in all three years of Run 2.
The combined PF isolation is used, combining information from $I_{\text{ch}}$, $I_{\gamma}$ and $I_\text{n}$. It is defined as:
$I_{\text{combined}} = I_{\text{ch}} + \max( 0, I_\text{n}+I_{\gamma}-I_{\text{PU}} )$, where $I_{\text{PU}}$ is the correction related to the event pileup.
The isolation-related variables are very sensitive to the extra energy
from pileup interactions, which affects the isolation efficiency when there are many interactions per bunch crossing.
The contribution from pileup in the isolation cone, which must be subtracted, is computed assuming
$ I_{\text{PU}} = \rho A_{\text{eff}}$, where $\rho$ and $A_{\text{eff}}$ are defined before.
The variable $I_{\text{combined}}$ is divided by the electron $\ET$, and is called the relative combined PF isolation.
For the cut-based electron identification, four working points are generally used in CMS.
The ``veto'' working  point, which corresponds to an average signal efficiency of about 95\%, is used in analyses to reject events with more reconstructed electrons than expected from the signal topology.
The ``loose'' working point corresponds to a signal efficiency of around 90\%, and is used in analyses where backgrounds to electrons are low.
The ``medium'' working point can be used for generic measurements involving $\PW$ or $\PZ$ bosons, and corresponds to an average signal efficiency of around 80\%.
The ``tight'' working point is around 70\% efficient for genuine electrons, and is used when backgrounds are larger.
Requirements on the minimum number of missing hits, together with requirements on the pixel conversion veto described in Section~\ref{sec:CSEV}, are also applied.

\begin{table}[h]
\centering
\topcaption{Cut-based electron identification requirements for the tight working point in the barrel and in the endcaps.}
\begin{tabular}{ccc}
\hline
Variable  &  Barrel (tight WP) & Endcaps (tight WP) \\
\hline
$\sigma_{i \eta i \eta}$ & $<$0.010  &  $<$0.035 \\
$\abs{\Delta \eta_{\text{in}}^{\text{seed}}}$  & $<$0.0025 &  $<$0.005 \\
$\abs{\Delta\phi_{\text{in}}}$    & $<$0.022\unit{rad}  &  $<$0.024\unit{rad} \\
$H/E$                    & $<$$0.026 + 1.15\GeV/E_{\text{SC}}$  &  $<$$0.019 + 2.06\GeV/E_{\text{SC}}$ \\
                       & $ + 0.032 \rho /E_{\text{SC}}$   &  $ + 0.183 \rho /E_{\text{SC}}$ \\
$I_{\text{combined}}/\ET$     & $<$$0.029 + 0.51\GeV/\ET$ & $<$$0.0445 + 0.963\GeV/\ET$ \\
$\abs{ 1/E - 1/p }$        & $<$$0.16\GeV^{-1}$  & $<$$0.0197\GeV^{-1}$ \\
Number of missing hits & ${\le}1$  & ${\le}1$\\
Pass conversion veto   & Yes  & Yes \\
\hline
\end{tabular}
\label{tab:cbarr_endcap_cutele}
\end{table}

\subsubsection{Electron identification using multivariate techniques}\label{sec:eleIDMVA}

To further improve the performance of the electron identification, especially at $\ET$ less than 40\GeV, several variables are combined using a BDT.
The set of observables is extended relative to the simpler sequential selection: the track-cluster matching observables are computed both at the ECAL surface and at the vertex. More cluster-shape and track-quality variables are also used.
The fractional difference between the track momentum at the innermost tracker layer and at the outermost tracker layer, $f_{\text{brem}}$, is also included.
Similar sets of variables are used for electrons in the barrel and in the endcaps.
Electron candidates in DY + jets simulated events with $\ET$ greater than 5\GeV and $\abs{\eta} < 2.5$ are used to train several BDTs in bins of $\ET$ and $\eta$ with the XGBoost algorithm~\cite{Chen:2016btl}.
The splits in pseudorapidity are at the barrel-endcap transition and at $\abs{\eta} = 0.8$, because the tracker material budget steeply increases beyond this point. The split in $\ET$ is at 10\GeV, allowing for a dedicated training in a region where background composition and amount of background is different.
Signal electrons are defined as reconstructed electrons that match  generated prompt electrons within a cone of size $\Delta R = 0.1$.
Background electrons are defined as all reconstructed electrons that either match generated nonprompt electrons
(usually electrons from hard jets) or that don't match any generated electron.
Reconstructed electrons that match generated electrons from $\tau$ leptonic decays are not considered either for the signal or for the background.

\begin{figure*}[hbtp]
\centering
\includegraphics[width=0.5\textwidth]{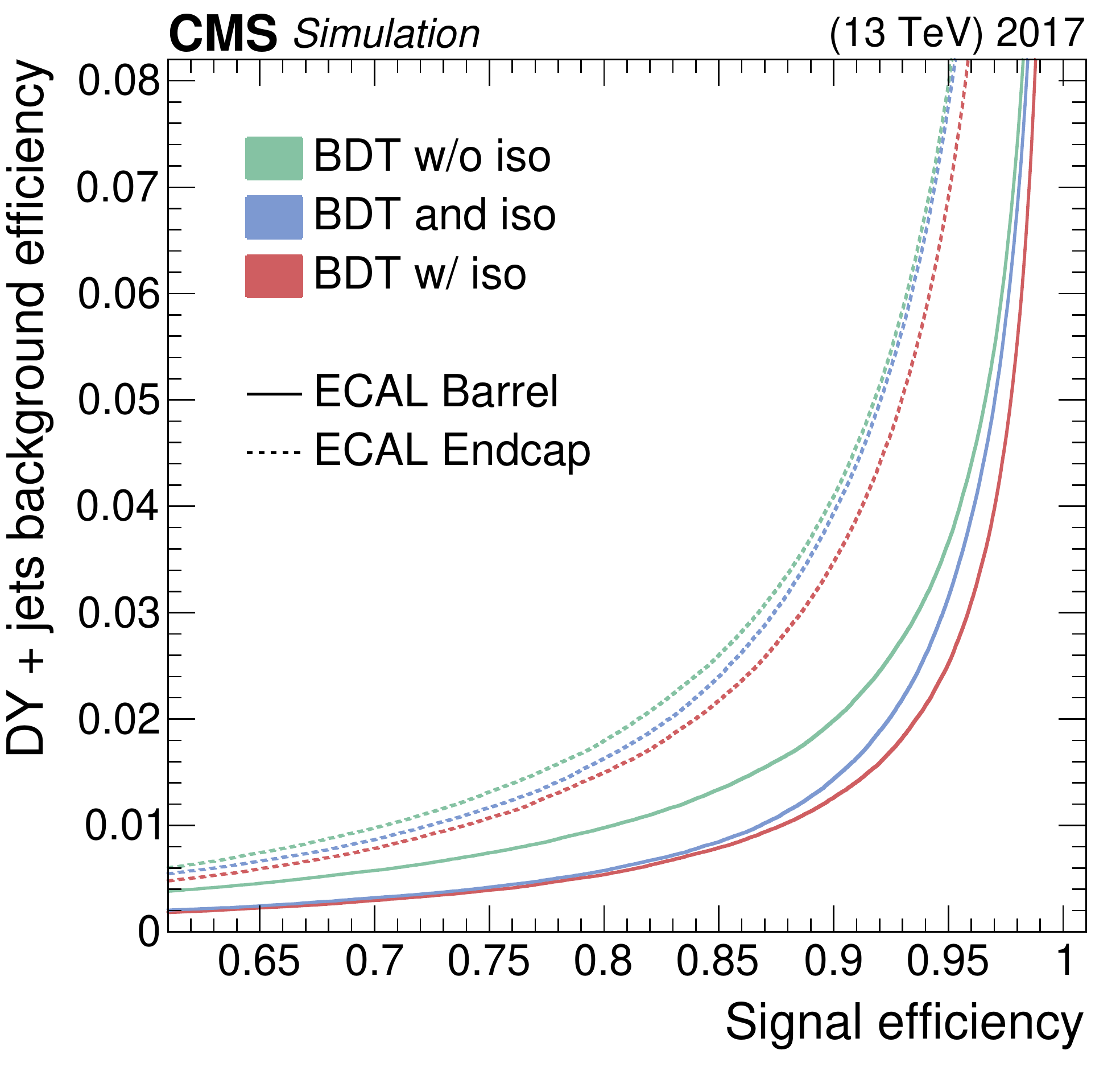}
    \caption{Performance of the electron BDT-based identification algorithm with (red) and without (green) the isolation variables, compared to an optimized sequential selection using the BDT without the isolations followed by a selection requirement on the
    combined isolation (blue). Electrons are selected for the BDT training with an $\ET$ of at least 20\GeV.}
\label{fig:MVAiso}
\end{figure*}

For a maximum flexibility at analysis level, the electron identification BDTs are trained with and
without including the isolation variables.
The performance of the BDT-based identification is reported in Fig.~\ref{fig:MVAiso}, and compared with a sequential selection in which the BDT and
isolation selection requirements are applied one after the other. The BDT trained with isolation variables shows a clear advantage over the sequential approach, especially for high selection efficiencies. Whereas the BDT is optimized on
background from DY + jets simulated events, the cut-based identification is optimized on background from
$\ttbar$ events, which contains a much higher fraction of nonprompt electrons.
This optimizes the performance of the identification algorithms for the analyses in which they are applied,
but prevents a like-for-like comparison between the two algorithms.

Although the BDT-based identifications have better background rejection for a given signal efficiency, compared with the cut-based identifications,
a significant fraction of physics analyses still prefer to use cut-based identifications because it is easy to flip or
undo a specific cut to perform a sidebands study. This is particularly true for searches focussing on high-$\ET$ ranges, where background is so low that an improved electron identification does not bring any sizeable improvement,
and these analyses profit from the simplicity and flexibility of the cut-based identifications.

\subsubsection{High-energy electron identification}\label{sec:HEEP}
The CMS experiment employs a dedicated cut-based identification method for the selection of high-$\ET$ electrons, known as high-energy electron pairs (HEEP).
Variables similar to those used for the cut-based general electron identification are
used to select high-$\ET$ electrons, starting at 35\GeV and extending up to about 2\TeV or more.
This selection requires that the lateral spread of energy deposits in the ECAL is consistent with that of a single electron and that the track is matched to the ECAL deposits and
is consistent with a particle originating from the nominal interaction point.
The associated energy in the HCAL around the electron direction must be less than 5\% of
the reconstructed
energy of the electron, once the noise and pileup contributions are included.

The main difference between the high-$\ET$ electron identification and the cut-based electron identification
is the use of subdetector-based isolation instead of PF isolation. Although the two algorithms are expected to provide
similar performance, the detector-based isolation behavior is
better suited for high-$\ET$ electrons.

Very high-energy electrons can lead to saturation of the ECAL electronics. In the presence of a saturated crystal, the shower-shape variables become biased and the requirements on lateral shower-shape variables, as for example the ratio of the energy collected in $n{\times}m$ arrays of crystals $E_{2 {\times}5}/E_{5{\times}5}$ and $\sigma_{i \eta i \eta}$ are disabled if
a saturated crystal occurs within a $5{\times}5$ crystal matrix around the central crystal.
The selection requires
that the electron be isolated in a cone of radius $\Delta R = 0.3$ in both the calorimeters ($I_{\text{ECAL}}$ and $I_{\text{HCAL}}$) and tracker ($I_{\text{tracker}}$).
The HCAL subdetector had two readout depths available in the endcap regions in Run 2.
Only the first longitudinal depth is used for the HCAL isolation, because the second one suffered from higher detector noise.

Only well-measured tracks that are consistent with originating from
the same vertex as the electron are included in the isolation sum.
Moreover, in the barrel, the $E_{2 {\times}5}/E_{5{\times}5}$ and $E_{1{\times}5}/E_{5{\times}5}$ variables are used, since they are very effective at high-$\ET$.

As mentioned in Sec.~\ref{sec:electron-track-reconstruction}, as \pt of the electrons increase,
they are less likely to be seeded by a tracker-driven approach.
Such electrons are rejected in the high-$\ET$ electron identification algorithm, which is mostly meant for high-energy electrons.
Requirements are applied on the minimum number of hits the electron leaves in the inner tracker and on the impact parameter relative to the center of the luminous region in the transverse 
plane ($d_{\mathrm{xy}}$).
This selection, which is about 90\% efficient for electrons with $\ET>50\GeV$, is used in many searches for exotic particles
published by CMS in Run 2~\cite{CMS-PAS-EXO-19-019}.
A summary of the requirements applied in the HEEP identification algorithm is shown in Table~\ref{tab:par1}. These selection criteria are valid for the entirety of Run 2.
\begin{table}[h]
\centering
\topcaption{Identification requirements for high-$\ET$ electrons in the barrel and in the endcaps.}
\cmsTable{
\begin{tabular}{ccc}
\hline
Variable                      & Barrel              &   Endcap  \\
\hline
$\ET$                         & $>$35\GeV           &   $>$35\GeV  \\
$\eta$ range                  & $\eta_{\text{SC}}<1.44$  &   $1.57 < \eta_{\text{SC}} < 2.50$ \\
$\abs{\Delta \eta_{\text{in}}^{\text{seed}}}$   & $<$0.004            &   $<$0.006 \\
$\abs{\Delta\phi_{\text{in}}}$           & $<$0.06 rad             &   $<$0.06 rad \\
$H/E$                           & $<$$1\GeV/E_{\text{SC}} + 0.05$       &   $<$$5\GeV/E_{\text{SC}} + 0.05$ \\
$\sigma_{i \eta i \eta}$  & \NA  &  $<$0.03  \\
$E_{2 {\times}5}/E_{5{\times}5}$  & $>$0.94 OR $E_{1{\times}5}/E_{5{\times}5} >  0.83$ & \NA \\
$I_{\text{ECAL}} + I_{\text{HCAL}}$   &  $<$$2.0\GeV + 0 .03 \ET + 0.28 \rho$   &   $<$$2.5\GeV + 0.28 \rho$ for $\ET<50\GeV$ \\
                           &                            & else $<$$2.5\GeV + 0.03 (\ET-50\GeV) + 0.28 \rho$ \\
$I_{\text{tracker}}$           &  $<$5\GeV              &  $<$5\GeV \\
Number of missing hits     &  ${\le}1$ & ${\le}1$  \\
$d_{\mathrm{xy}}$  & $<$0.02\cm & $<$0.05\cm \\
\hline
\end{tabular}
\label{tab:par1}
}
\end{table}

\clearpage
\subsection{Selection efficiency and scale factors}\label{sec:sf}
The electron and photon identification efficiencies, as well as the electron reconstruction efficiency,
 are measured in data using a tag-and-probe technique that utilizes $\PZ\to \Pe\Pe$ events~\cite{Khachatryan:2015hwa} described in detail in Sec.~\ref{sec:tnp}. The identification efficiency is measured for transverse energies above 10\GeV for electrons and above 20\GeV for photons. For photon identification efficiency no requirement is applied on the track and charge of the probe, pretending in this way to identify an electron from $\PZ$ boson decay as a photon.

\begin{figure}[hbtp]
  \centering
    \includegraphics[ width=\cmsFigWidth]{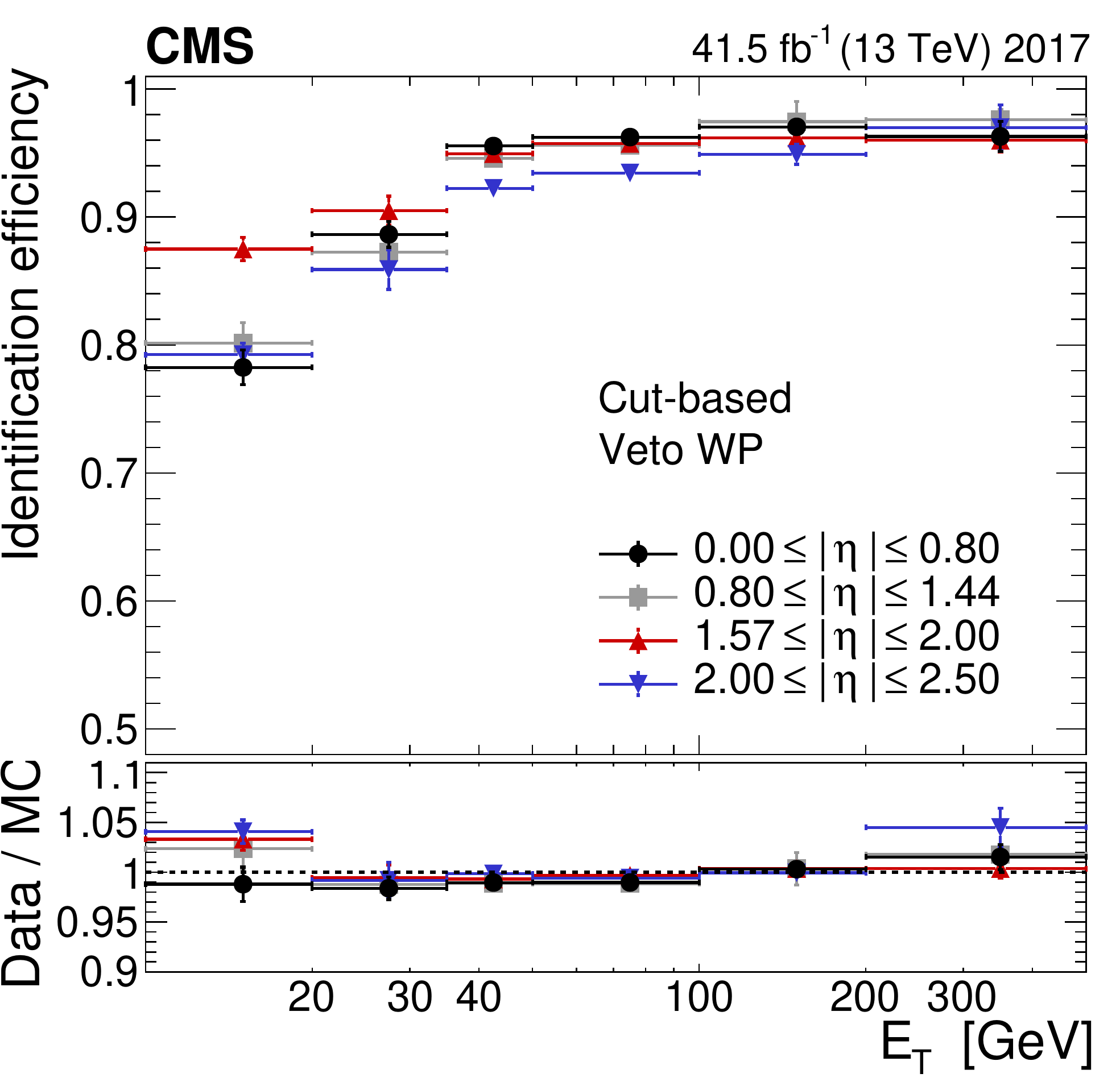}
    \includegraphics[ width=\cmsFigWidth]{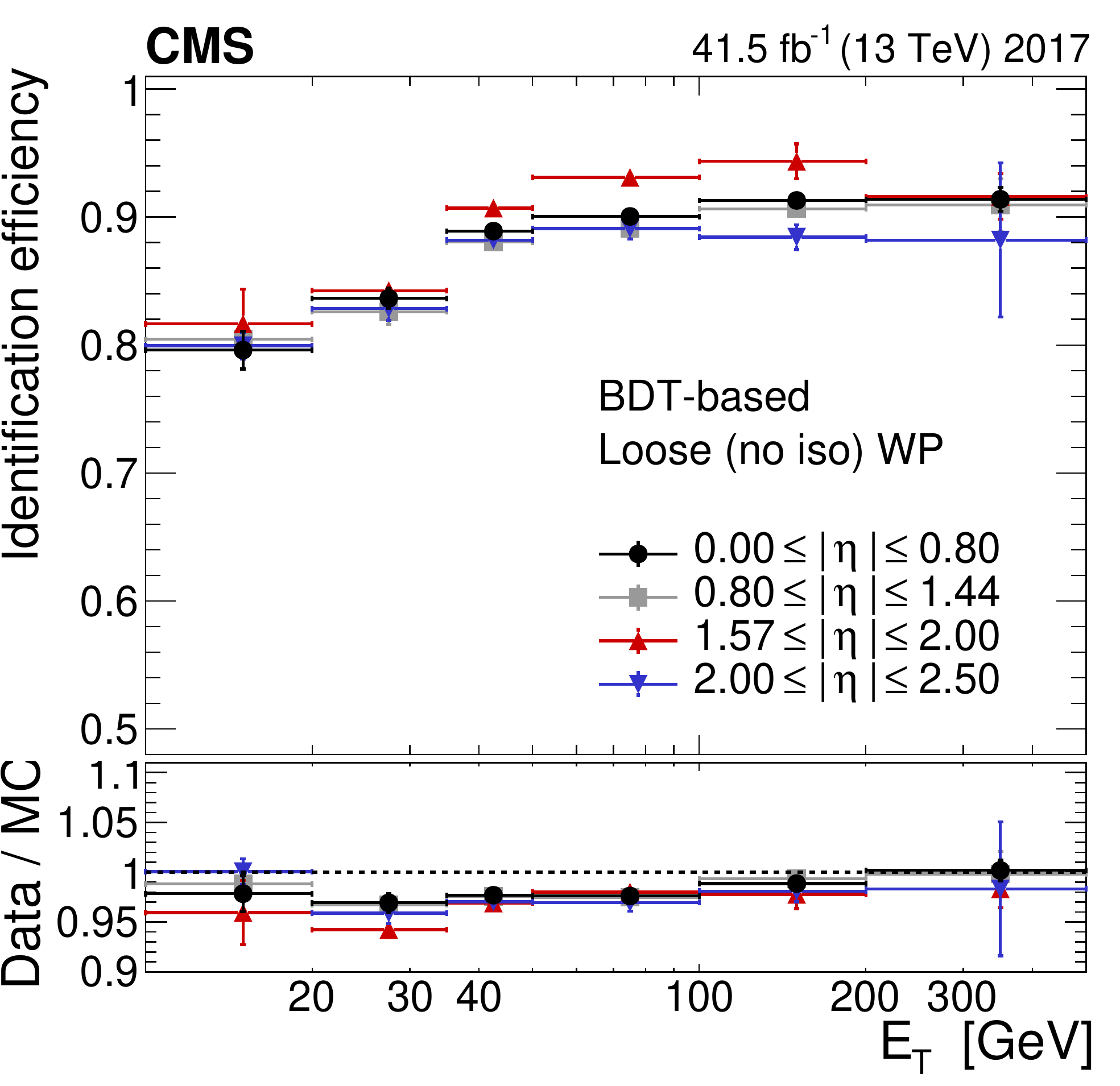}
     \caption{Electron identification efficiency measured in data (upper panels) and data-to-simulation efficiency ratios (lower panels), as a function of the electron $\ET$, for the cut-based identification veto working point (left) and the BDT-based (without isolation) loosest working point (right). The vertical bars on the markers represent combined statistical and systematic uncertainties.}
    \label{fig:electronIDeff_vsEta}
\end{figure}

The performance achieved for the two reference electron selections is shown in Fig.~\ref{fig:electronIDeff_vsEta}.
The electron identification efficiency in data (upper panels) and data-to-simulation efficiency ratios (lower panels) for the cut-based identification and the veto working point (left), and the BDT-based identification loosest working point (right) are shown as functions of the electron $\ET$.
The efficiency is shown in four $\eta$ ranges.
The vertical bars on the data-to-simulation ratios represent the combined statistical and systematic uncertainties.
For the three years of data-taking, efficiencies and scale factors are measured with total uncertainties of the same order of magnitude. For 2017 data, where the latest calibrations were used to reconstruct the data, the measured data-to-simulation efficiency ratios are closer to unity by 3--5\% over the entire energy range compared to 2016 and 2018 data.

\begin{figure}[hbtp]
  \centering
    \includegraphics[ width=\cmsFigWidth]{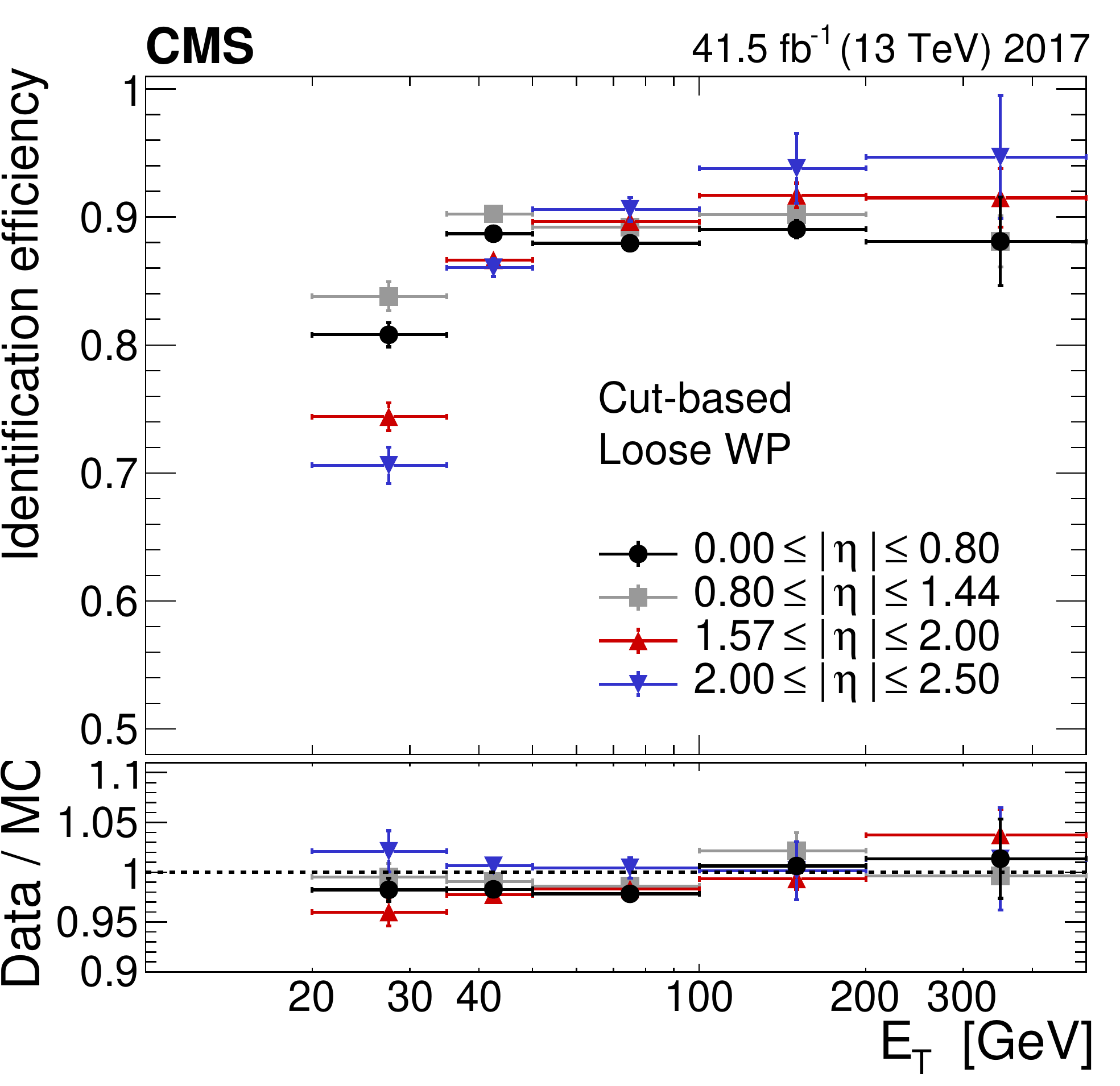}
    \includegraphics[ width=\cmsFigWidth]{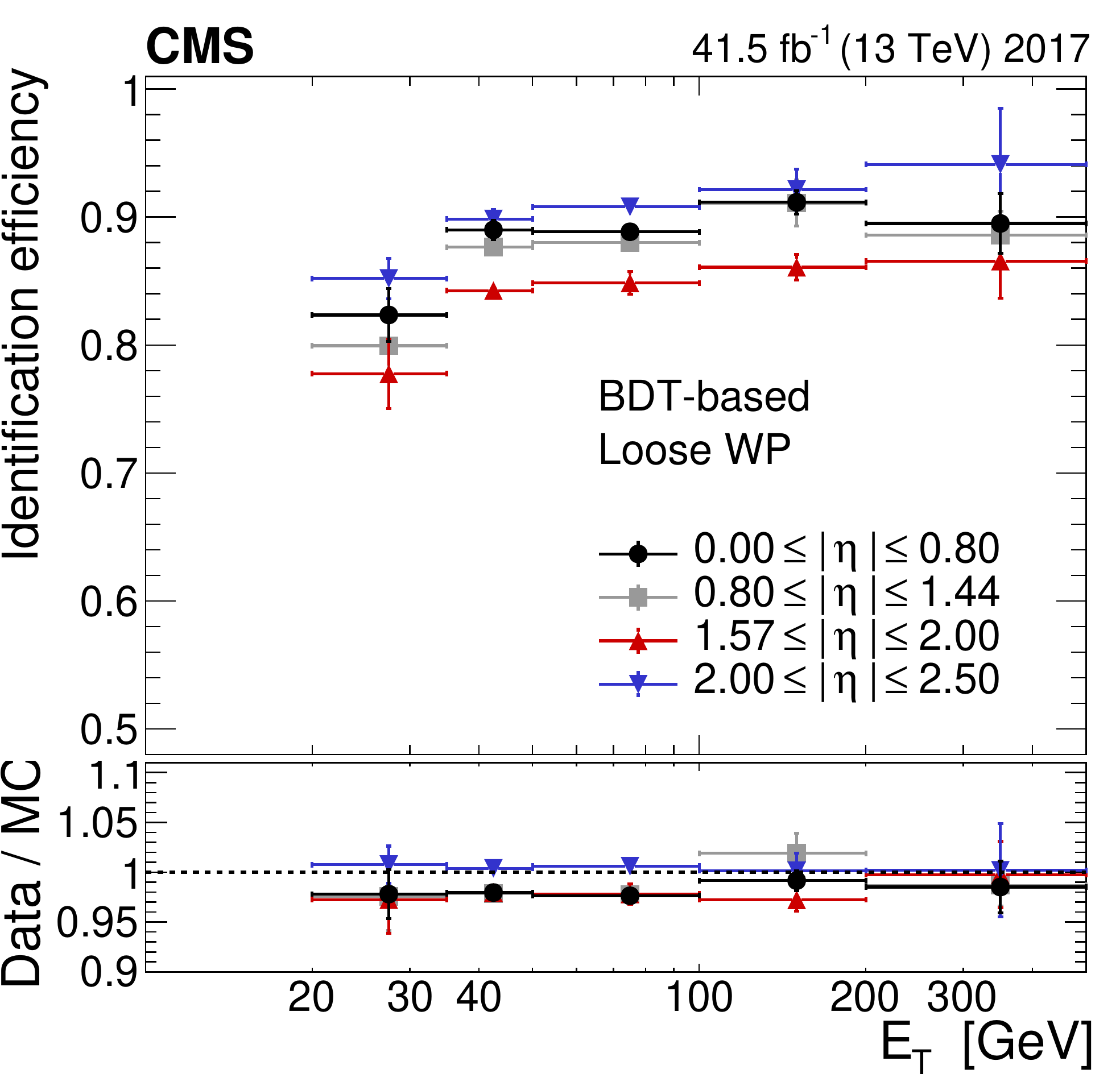}
     \caption{Photon identification efficiency in data (upper panels) and data-to-simulation efficiency ratios (lower panels) for the loose cut-based (left) and loosest BDT-based (right) identification working points, as functions of the photon $\ET$. The vertical bars on the markers represent combined statistical and systematic uncertainties.
     }
    \label{fig:photonIDeff_vsEta}
 \end{figure}

The photon identification efficiency in data and data-to-simulation efficiency ratios are shown in Fig.~\ref{fig:photonIDeff_vsEta}, for the loose cut-based (left) and loosest BDT-based (right) identification working points, as functions of the electron $\ET$.
The efficiency is shown in four $\eta$ ranges.
The vertical bars on the data-to-simulation ratio represent the combined statistical and systematic uncertainties.

The electron and photon identification efficiency in data is very well described by the MC simulations, and it is reflected in the fact that the ratios are within 5\% from unity for all the cut-based identification working points and the BDT-based ones. There are some effects that are very difficult to simulate. For example, the variation of ECAL noise with time, which affects the $\sigma_{i \eta i \eta}$ variable in low- and medium-$\ET$ electrons and photons in the endcap. Such effects are included in the correction factors.

\section{Performance of recalibrated data sets}
\label{sec:EOYvsUL}
\label{sec:perfUL2017}
\begin{figure}[hbtp]
   \centering
    \includegraphics[ width=\cmsFigWidth]{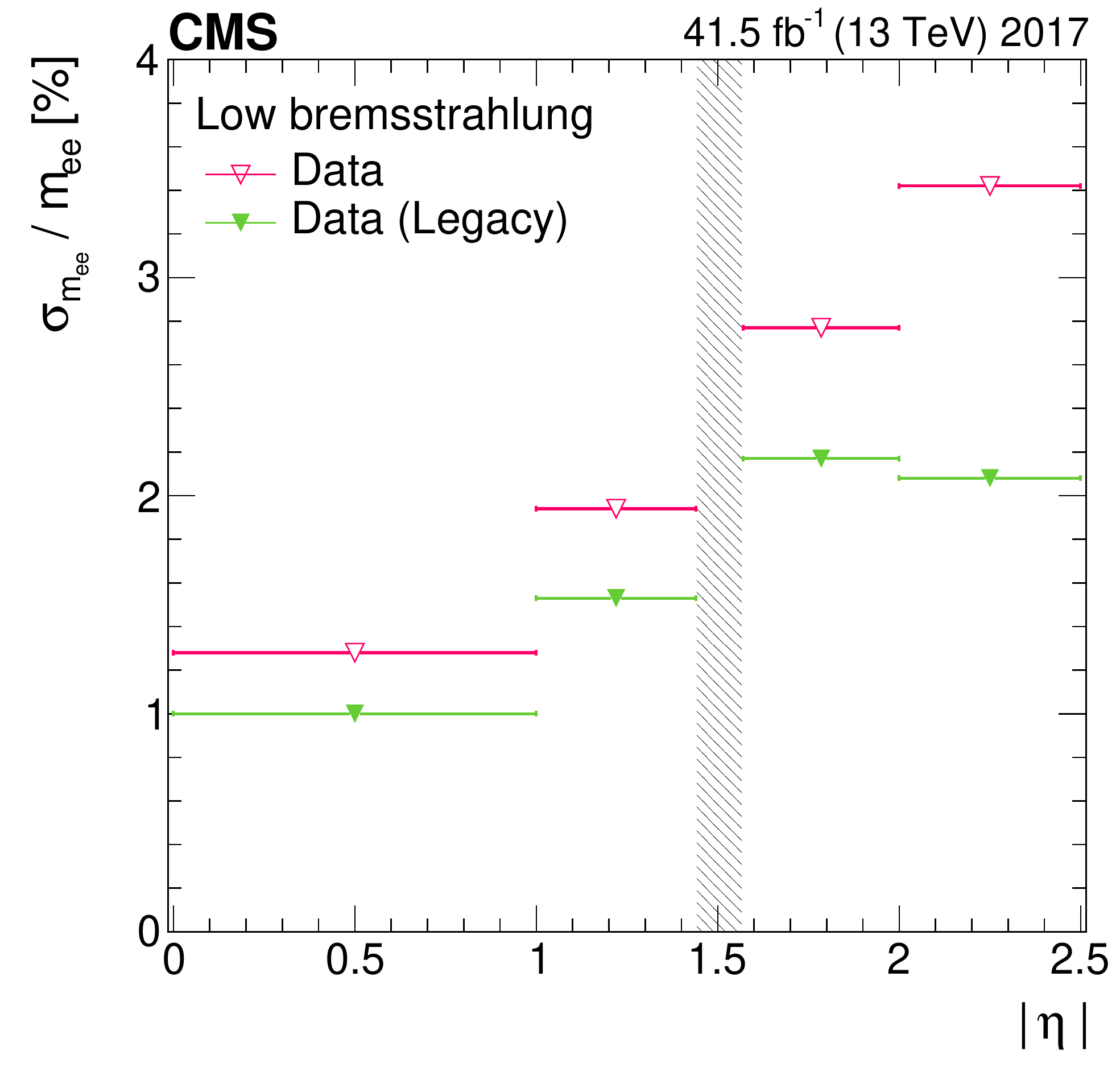}
    \caption{Dielectron mass resolution from $\PZ\to \Pe\Pe$ events in bins of $\eta$ of the electron for the barrel and the endcaps for the Legacy calibration (green filled markers) and the EOY calibration (pink empty markers) of the 2017 data set. The grey vertical band represents the region $1.44 < \abs{\eta} < 1.57$ and it is not included since it corresponds to the transition between barrel and endcap electromagnetic calorimeters.
             }
    \label{fig:resolution}
\end{figure}

During the LHC long shutdown 2 that began in 2019,
the CMS experiment initiated a program of recalibrating the full Run 2 data set.
It involved the realignment and
recalibration of all the subdetector components
and the related physics objects.
The simulation was also improved with a more accurate description of the
data in terms of dynamic inefficiencies, radiation damage,
and description of the detector noise.
Electron and photon reconstruction and identification performance strongly depend upon improvements on measurements with the ECAL subdetector.  An updated method to monitor and correct the ECAL crystal transparency loss due to radiation damage has been introduced. In parallel a more granular calibration of the crystals has been performed allowing to calibrate precisely also the highest pseudorapidity region of the calorimeter.
All these actions
have led to better resolution and better agreement
between data and simulation.
Figure~\ref{fig:resolution}
shows the improvement in resolution brought by the Legacy calibration of the ECAL for low showering electrons ($R_9 > 0.94$) as a function of pseudorapidity.
This resolution is estimated after all the corrections described in Section~\ref{sec:sec6} are applied, also including the scale and spreading corrections of Section~\ref{sec:SS}.
The relative resolution improves as a function of $\eta$ up to more than $50\%$ for $\abs{\eta} > 2.0$.

\begin{figure}[hbtp]
\centering
    \includegraphics[ width=\cmsFigWidth]{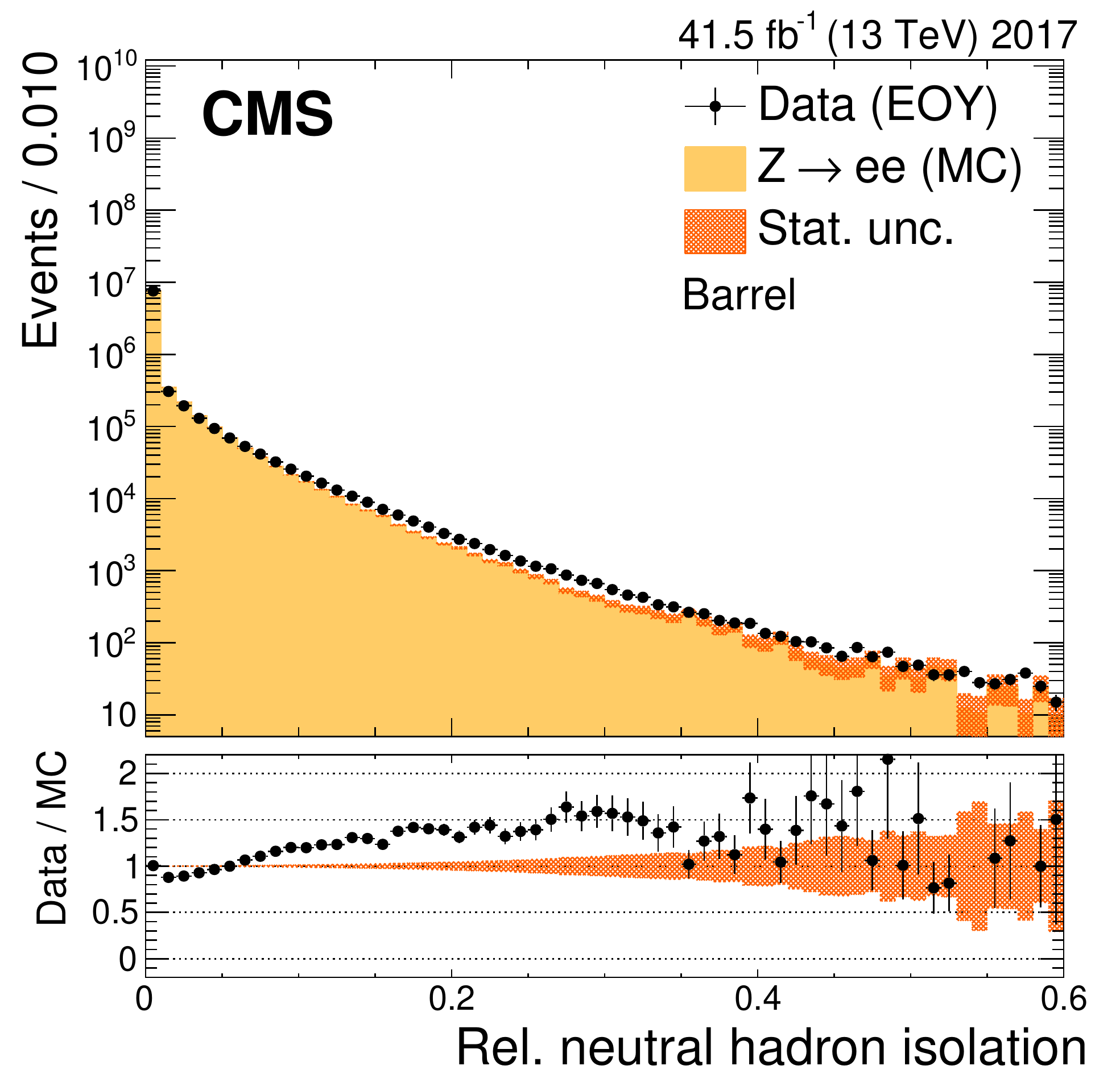}
    \includegraphics[ width=\cmsFigWidth]{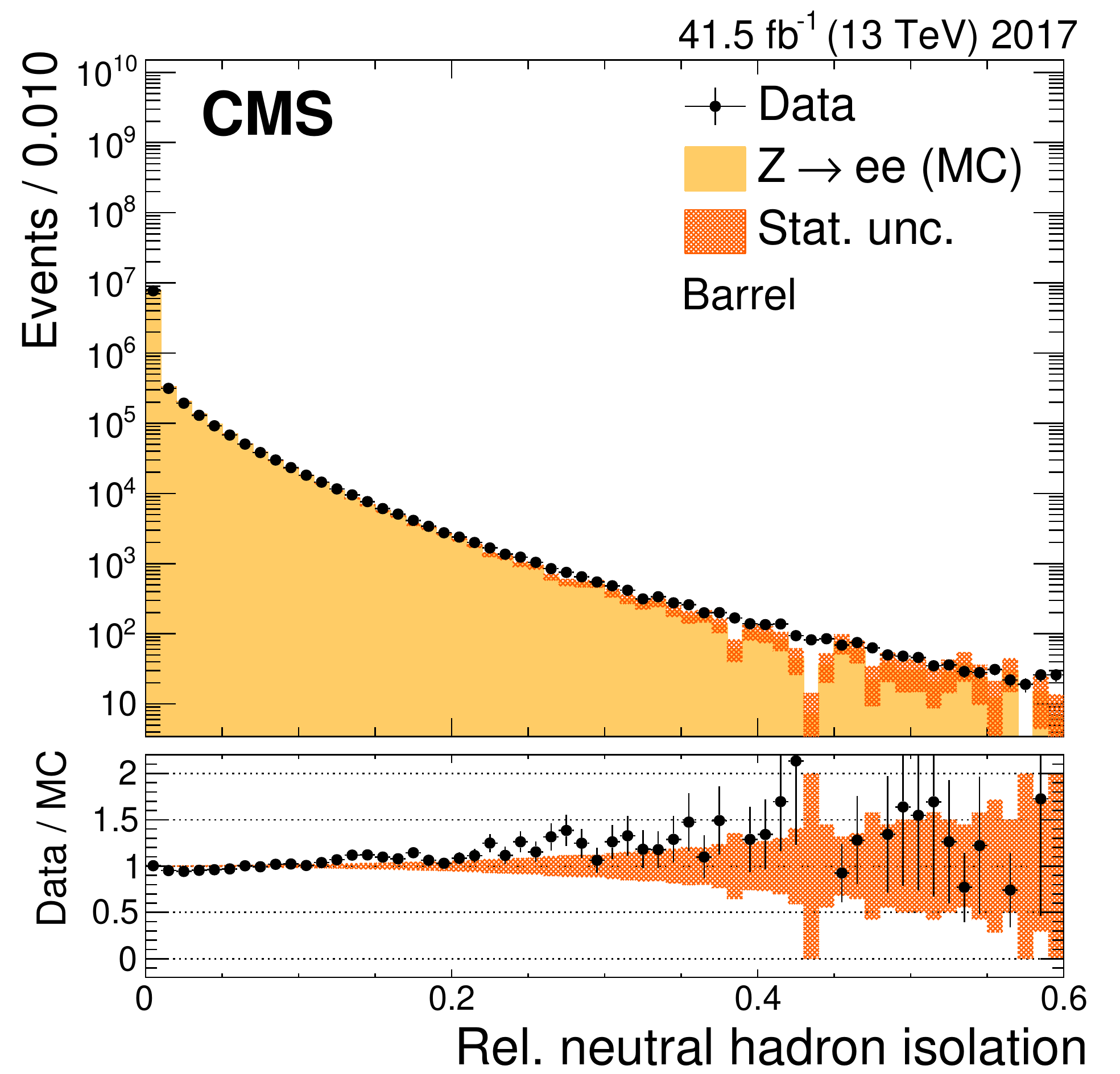} \\
    \caption{The PF-based neutral hadron isolation in the barrel. The left plot is from the EOY calibration and the right plot is from the Legacy calibration 2017 data set. The simulation is shown with the filled histograms and data are represented by the markers. The vertical bars on the markers represent combined statistical uncertainty in data. The hatched regions show the
statistical uncertainty in the simulation. The lower panels display the ratio of the data to the simulation, with the bands representing the uncertainties in the simulation, and show an improvement for the Legacy calibration compared to the EOY calibration.}
    \label{fig:ULdataMC}
\end{figure}

The improved detector calibration and simulation has
led to an improved agreement between data and simulation.
Figure~\ref{fig:ULdataMC} shows the improvement in the PF-based
relative neutral hadron isolation in the barrel.
These are obtained using $\PZ\to \Pe\Pe$ electrons, as described in Section~\ref{sec:AddVar}.

\begin{figure}[hbtp]
\centering
    \includegraphics[ width=\cmsFigWidth]{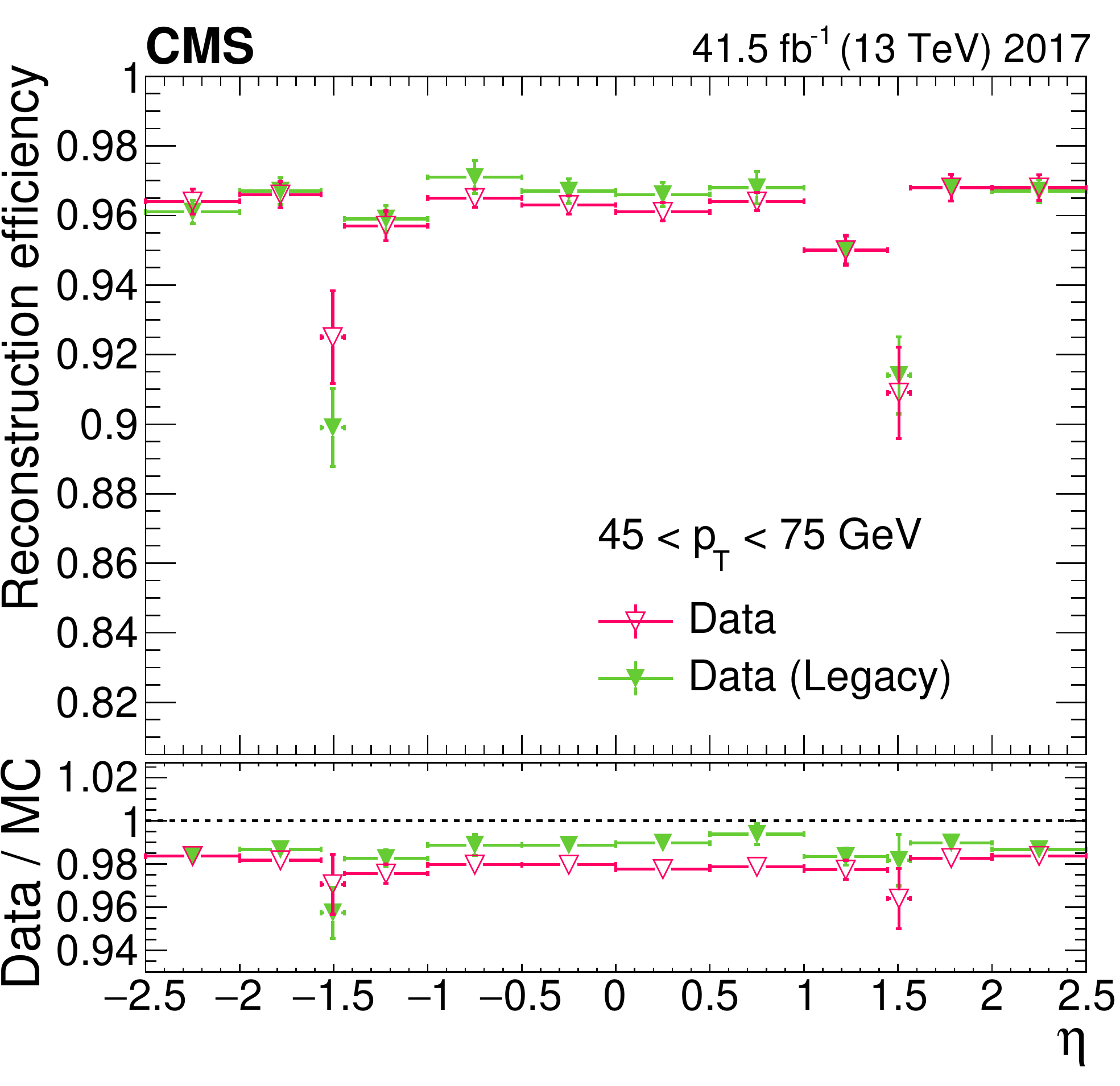}
    \caption{Electron reconstruction efficiency (upper panel) and data-to-simulation correction factors (lower panel) comparing the Legacy and the EOY calibrations of the 2017 data set for electrons from $\PZ\to \Pe\Pe$ decays with $\ET$ between 45 and 75\GeV.  The vertical bars on the markers represent combined statistical and systematic uncertainties.}
    \label{fig:eleRECO}
\end{figure}

\begin{figure}[h!]
\centering
    \includegraphics[ width=\cmsFigWidth]{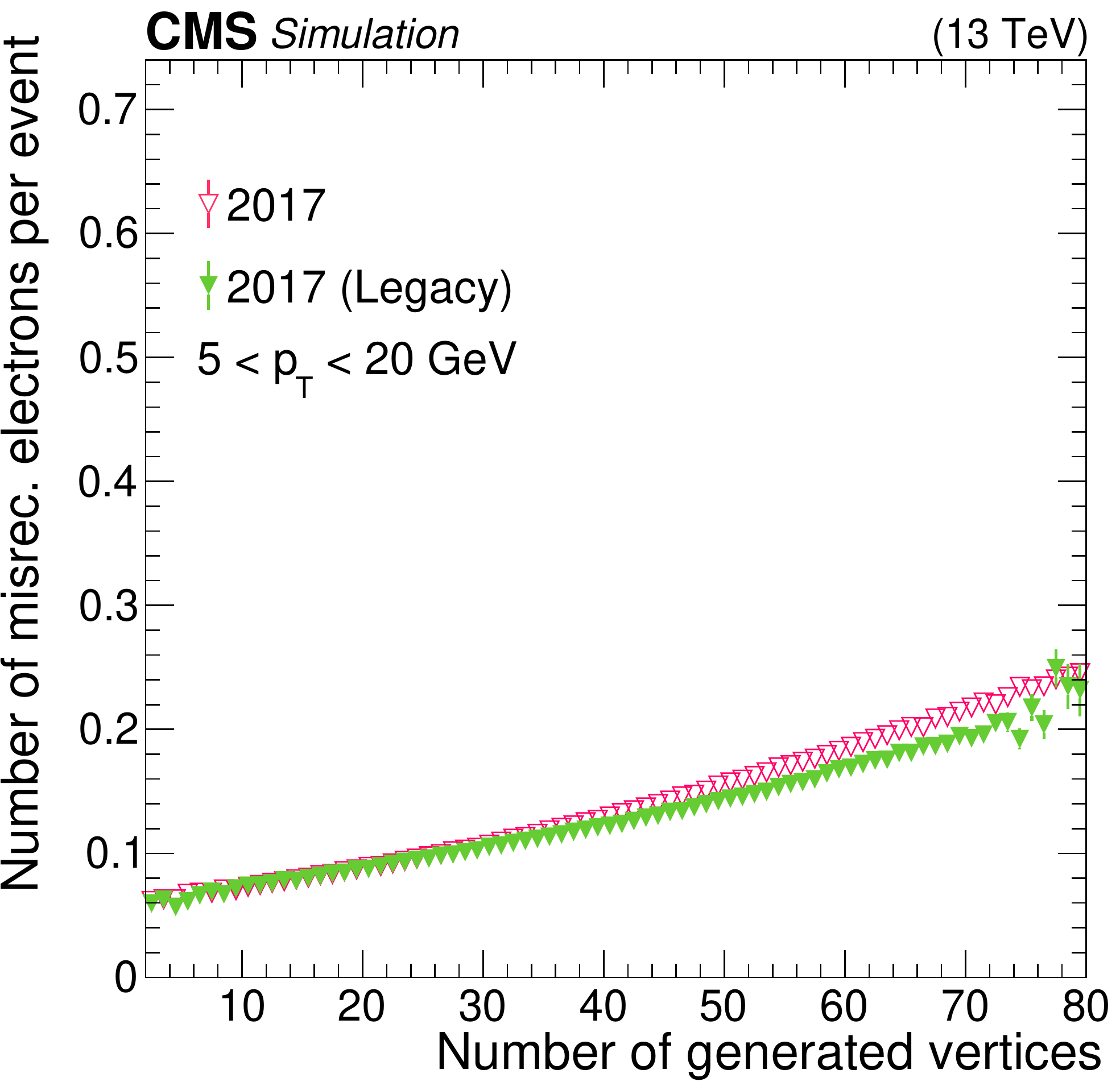}
    \includegraphics[ width=\cmsFigWidth]{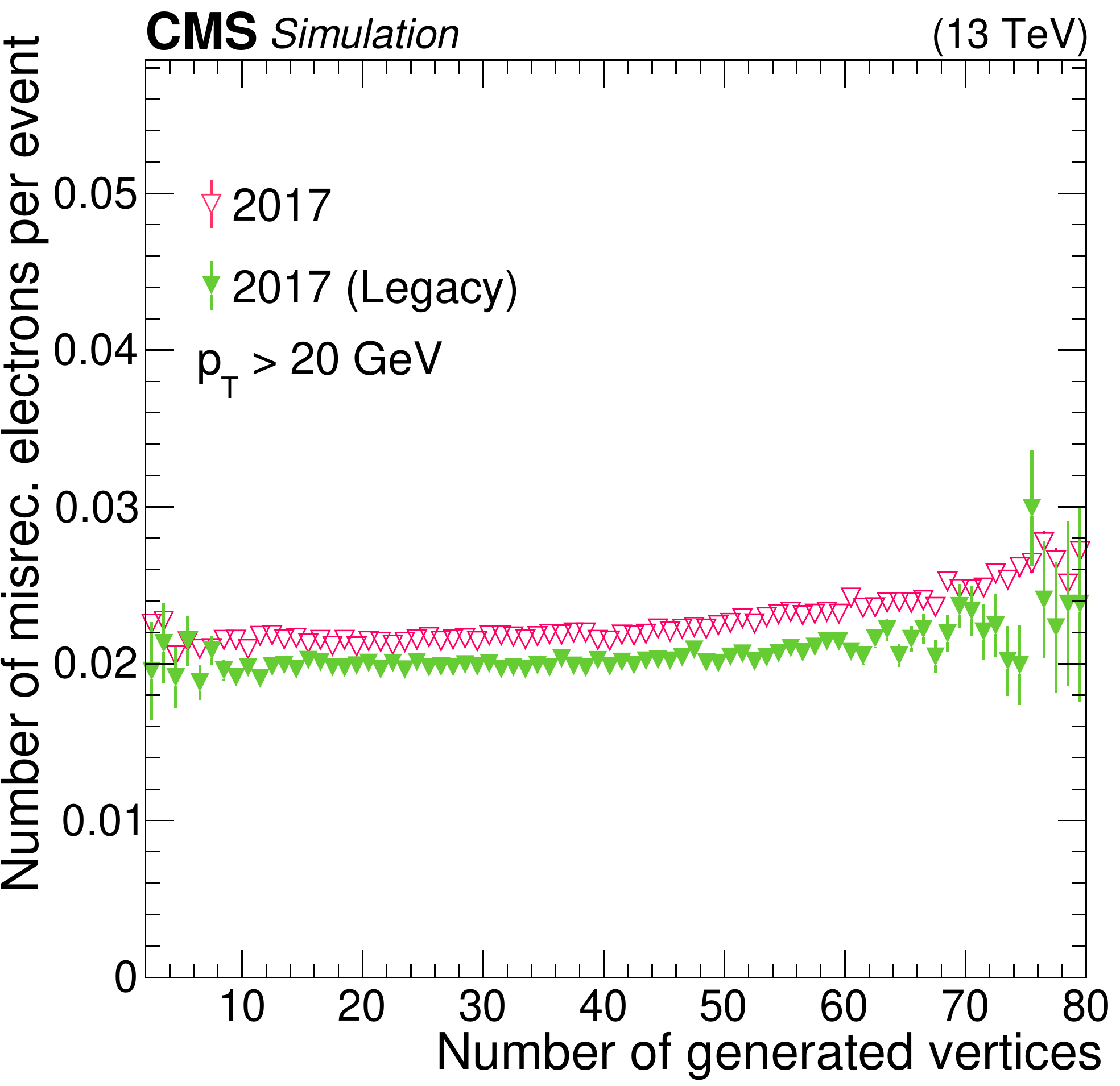}
    \caption{Number of misreconstructed electron candidates per event as a function of pileup in DY + jets MC events
      simulated with the Legacy and the EOY detector conditions of 2017. Results are shown for electrons
      with \pt within the 5--20\GeV range (left) and $\pt>20\GeV$ (right) before any selection criteria. Electrons in the region $1.44 <\abs{\eta}< 1.57$ are discarded. The vertical bars on the markers represent combined statistical uncertainties of the MC sample.}
    \label{fig:fr2017}
\end{figure}

Figure~\ref{fig:eleRECO} shows the improvement in the electron reconstruction data-to-simulation efficiency correction factors.  The magnitude of correction factors in the bottom panel is below 2\% for the Legacy calibration compared with 3\% for the EOY calibration. The number of misreconstructed electron candidates per event is reported in Fig.~\ref{fig:fr2017} and shows a slight decrease in the Legacy data set, due to the better conditions and calibrations.


\begin{figure}[hbtp]
\centering
    \includegraphics[ width=\cmsFigWidth]{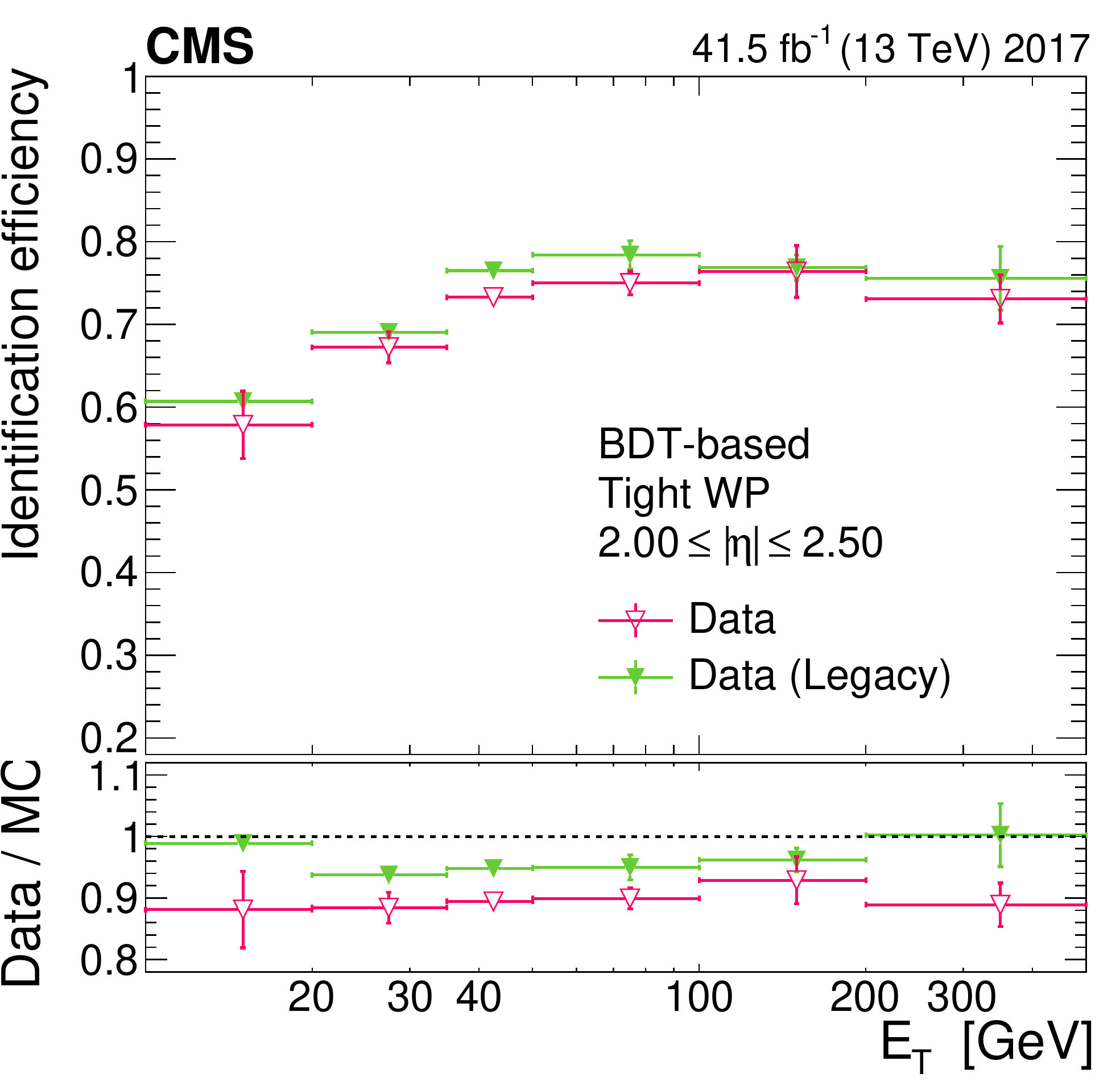}
    \caption{Tight BDT-based electron identification efficiency (upper panel) and data-to-simulation correction factors (lower panel) comparing the Legacy to the EOY calibration of the 2017 data set for electrons from $\PZ\to \Pe\Pe$ decays with $\eta$ between 2.00 and 2.50. The vertical bars on the markers represent combined statistical and systematic uncertainties.}
    \label{fig:eleID}
\end{figure}

Figure~\ref{fig:eleID} shows the improvement in the electron identification data-to-simulation efficiency corrections.  The correction factors in the bottom panel are above 0.95 for the Legacy calibration compared with 0.88--0.91 for the EOY calibration.

\subsection{Comparison among the three data-taking years of Run 2}
\label{sec:compare3years}

\begin{figure}[h]
\centering
    \includegraphics[ width=\cmsFigWidth]{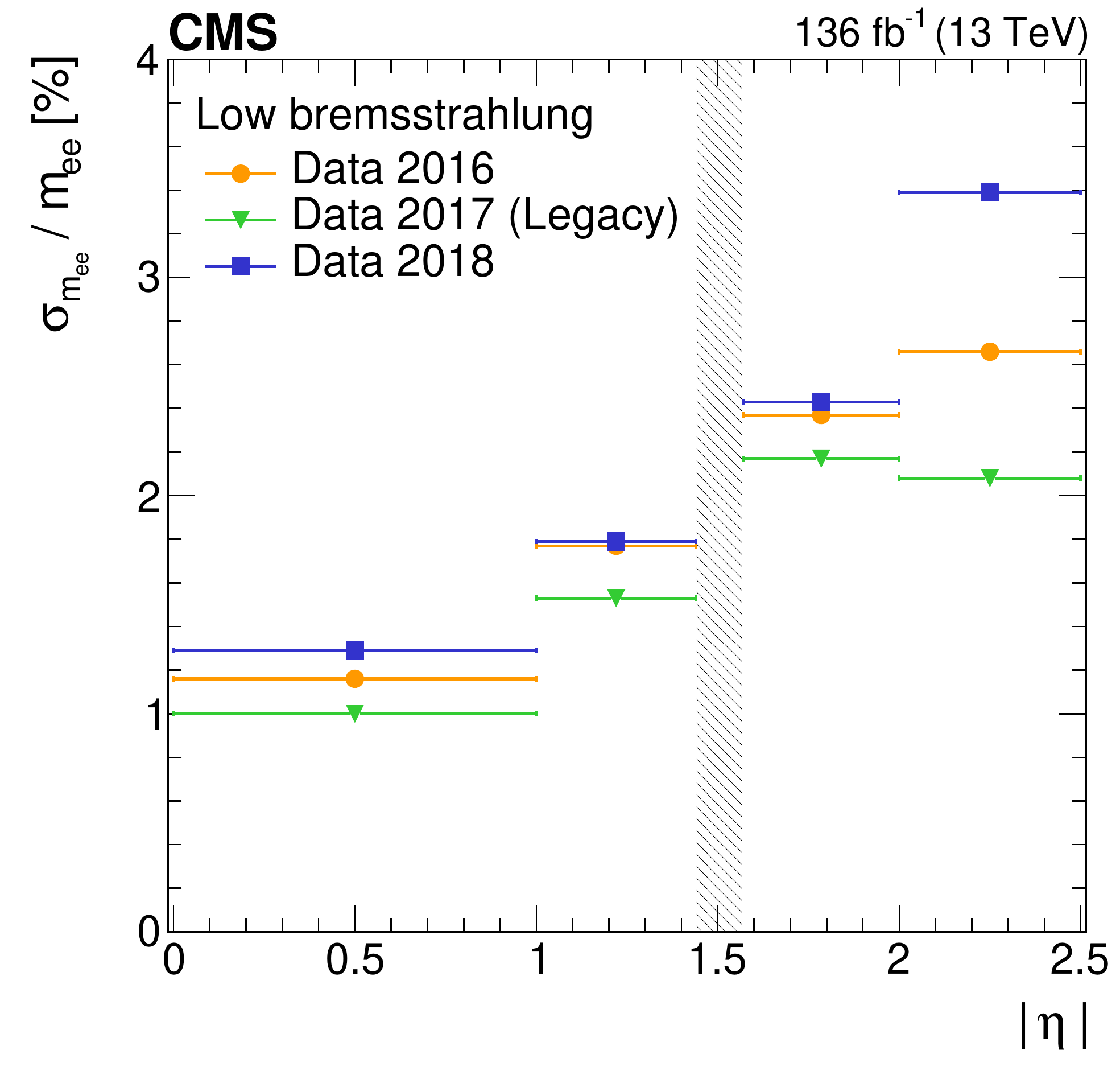}
     \caption{The relative $\PZ$ boson mass resolution from $\PZ\to \Pe\Pe$ decays in bins of the $\eta$ of the electron for the barrel and endcaps. Electrons from $\PZ\to \Pe\Pe$ decays are used and the resolution is shown for low-bremsstrahlung electrons. The plot compares the resolution achieved for the data collected at 13\TeV during Run 2 in 2016, 2017, and 2018. The 2016 and 2018 data are reconstructed with an initial calibration whereas the 2017 data are reconstructed with refined calibrations. The resolution is estimated after all the corrections are applied, as described in Section~\ref{sec:rescorr_Hgg}, including the scale and spreading corrections of Section~\ref{sec:SS}. The grey vertical band represents the region $1.44 < \abs{\eta} < 1.57$ and it is not included since it corresponds to the transition between barrel and endcap electromagnetic calorimeters.
     }
    \label{fig:resolution3years}
\end{figure}

Figure~\ref{fig:resolution3years} shows the $\PZ\to \Pe\Pe$ mass resolution as a function of $\eta$ of the low bremsstrahlung electrons for the three years of Run 2.
As expected, the energy resolution measured in 2017 is significantly better than for the other two years because of the Legacy calibration, and the impact of improved calibration is the dominant effect when comparing the 3 years in this manner. Overall, the energy resolution throughout the entire Run 2 period ranges from 1 to 3.4\%, depending on the $\eta$ region considered.

In general, the agreement between data and simulation depends on the
noise modeling in simulation and energy calibration for that object.
Better calibration of the
2017 data, together with a more appropriate simulation of the noise levels in MC,
have led to a better description of isolation variables.

\begin{figure}[h]
\centering
    \includegraphics[ width=\cmsFigWidth]{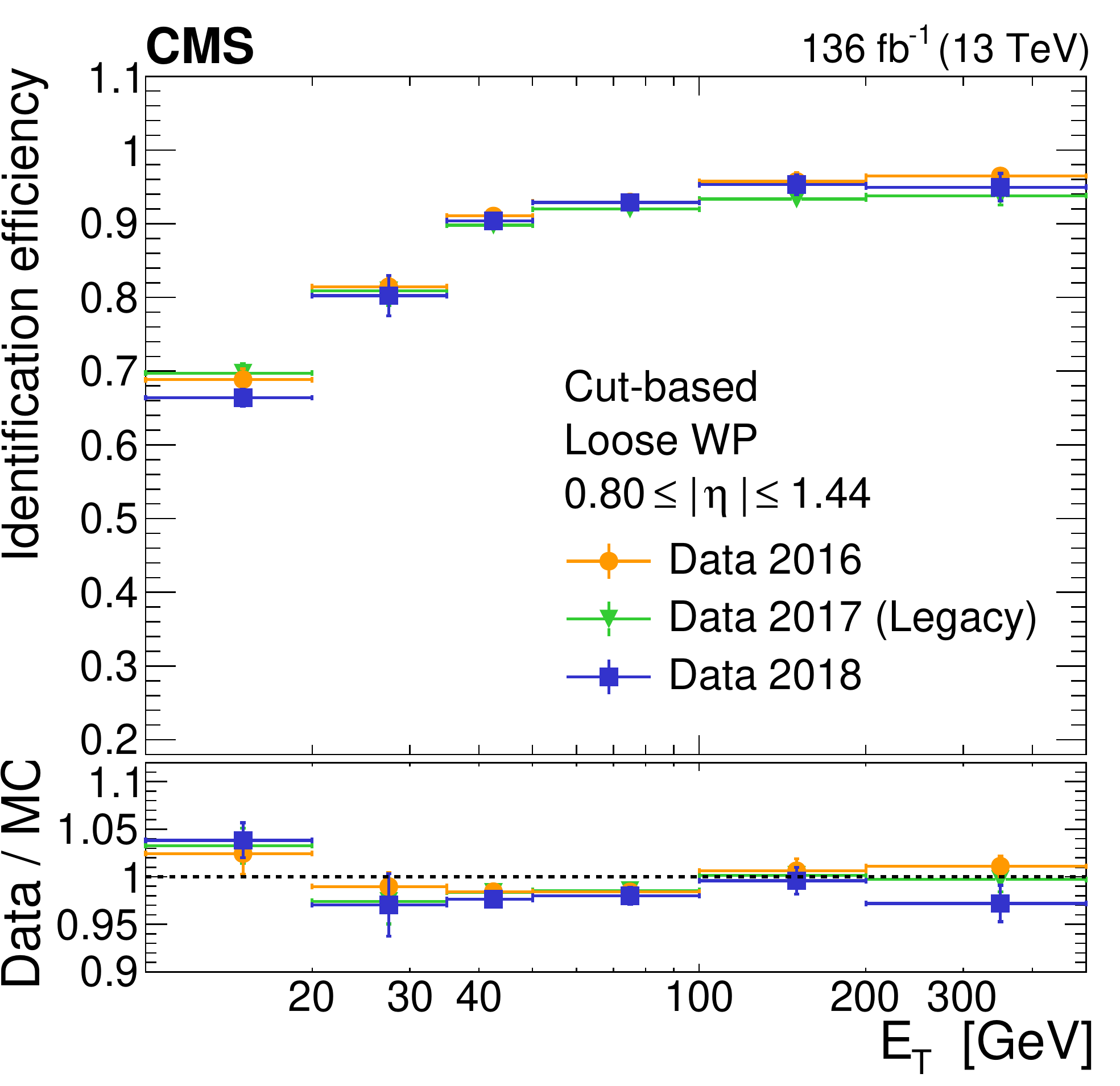}
     \caption{Cut-based loose electron identification efficiency in data (upper panel) and data-to-simulation efficiency ratios (lower panel) as a function of the electron \pt shown for 2016, 2017 ``Legacy'', and 2018 data-taking periods. The vertical bars on the markers represent combined statistical and systematic uncertainties.
     }
     \label{fig:compare3years_eleIDeff}
\end{figure}

The identification efficiencies for 2016, 2017, and 2018 are shown in Fig.~\ref{fig:compare3years_eleIDeff} for cut-based loose electron identification requirements. The efficiencies are stable within 5\% for the full range of $\ET$ of the electrons across the full ECAL. The correction factors are also stable within 3\% over the full three years.

\section{Timing performance} \label{sec:sec8timing}
In addition to the energy measurement, the ECAL provides a time of arrival for electromagnetic energy deposits that can separate prompt electrons and photons from backgrounds with a broader time of arrival distribution. The fast decay time of the PbWO$_{4}$ ECAL crystals, comparable to the LHC bunch crossing interval (80\% of the light is emitted in 25\unit{ns}), together with the use of electronic pulse shaping with an high sampling rate provides for an excellent timing resolution~\cite{Chatrchyan:2009aj}.
The better the precision and synchronization of the
timing measurement, the larger the rejection of the background.
Background sources with a broad time distribution include cosmic rays, beam halo muons, electronic noise, spikes (hadrons that directly ionize the EB photodetectors), and out-of-time $\Pp\Pp$ interactions.
A precise timing measurement also enables
the identification of particles predicted by different models beyond the standard model. Possible examples could be slow heavy charged R-hadrons~\cite{CMS-PAS-EXO-16-036}, which travel through the calorimeter and interact before decaying, and photons from the decay of long-lived new particles that reach the ECAL out-of-time with respect to particles traveling at the speed of light from the interaction point. For example, to identify neutralinos decaying into photons with decay lengths comparable to the ECAL radial size, a time measurement resolution better than 1\unit{ns} is necessary. The CMS Collaboration published results on searches for displaced photons from decays of long-lived neutralinos during Run 1~\cite{Chatrchyan:2012jwg} and Run 2~\cite{Sirunyan:2019wau}. These are  the most stringent limits to date on delayed photons at the LHC.

The ECAL timing performance has been measured prior to data-taking using electrons from test beams, cosmic muons, and beam splash events~\cite{Chatrchyan:2009aj}.
The resolution for large energy deposits ($E>50\GeV$) was estimated to be better than 30\unit{ps} and the linearity of the time response was also
verified. During collisions in the LHC, there are many additional effects that could worsen the performance, such as residual timing
jitter in the electronics or the clock distribution to the individual readout units, run-by-run variations, intercalibration effects, energy-dependent systematic uncertainties, and crystal damage due to radiation.
A detailed description of the method to measure the ECAL timing with the crystals is given in the next section.

\subsection{Time resolution measurement using \texorpdfstring{$\PZ \to \Pe\Pe$}{Zee} events}
The method used to extract the electron and photon time resolutions with the ECAL detector is based on comparisons of the time of arrival of the two electrons arising from $\PZ$ decays. The
time of arrival of each electron is the measured time of the most energetic hit of the energy deposit in the ECAL.
This time is corrected for the electron time-of-flight, which is determined
from the primary vertex position, obtained from the electron track. This correction is needed because the timestamp recorded by the ECAL crystal assumes a
time-of-flight from the origin of the detector, such that the distribution for the most energetic hit time is
centered around zero for all crystals.

\begin{figure}[hbtp]
\centering
    \includegraphics[ width=\cmsFigWidth]{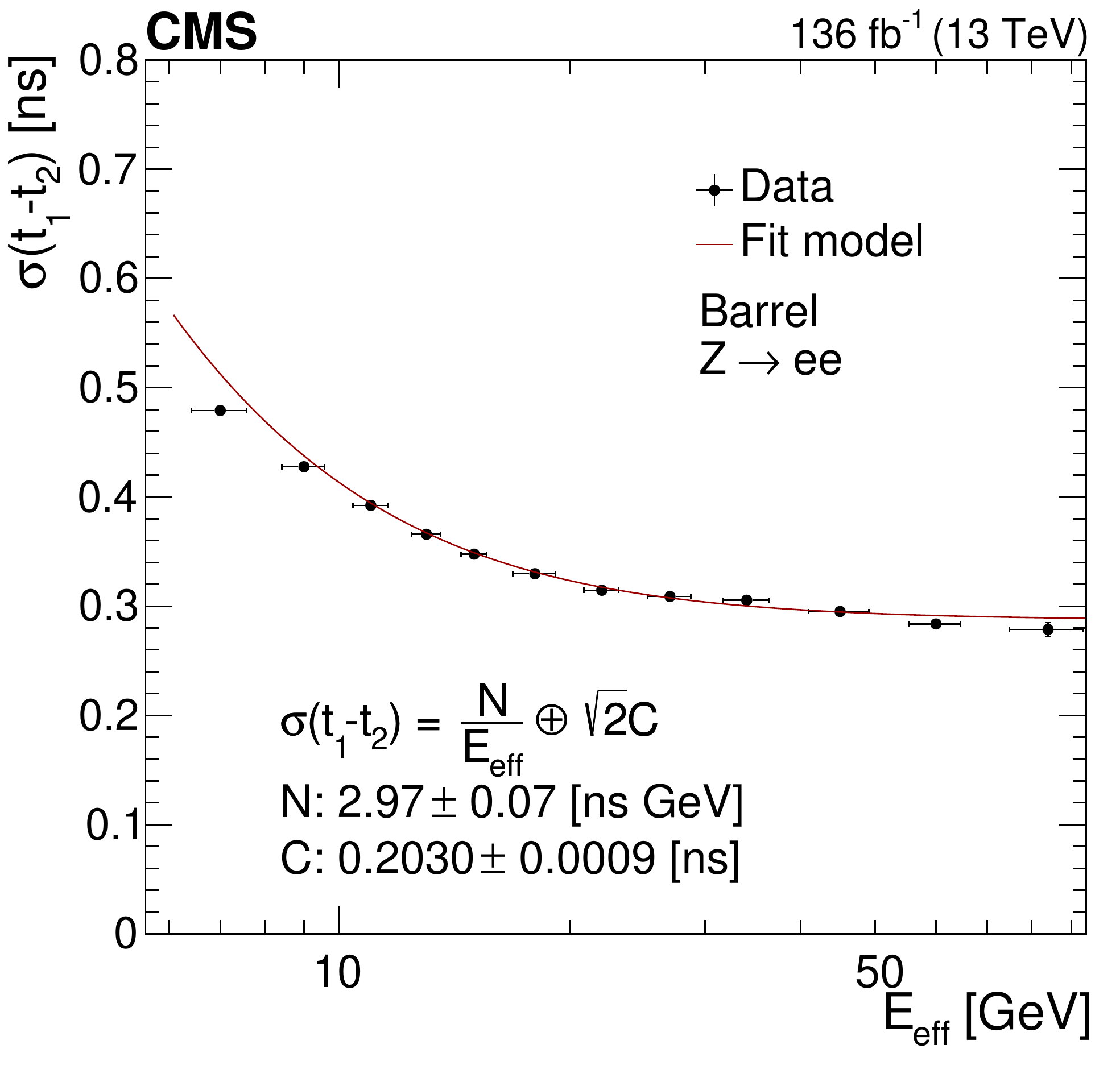}
    \caption{ECAL timing resolution as a function of the effective energy of the electron pairs as measured in 2016, 2017 (Legacy), and 2018 data. Vertical bars on the markers represent the uncertainties coming from the fit procedure used to determine the $\sigma(t_1-t_2)$ parameters. }
    \label{fig:TimeResZee}
\end{figure}

The two electron clusters are required to be in the EB and to pass loose identification criteria on the cluster shape. Their resulting invariant mass
has to be consistent with the $\PZ$ boson mass ($60 < m_{\Pe\Pe}< 120\GeV$). The energy of each
of the two hits must fall within the range $10 <E< 120\GeV$. The lower threshold is motivated by the minimal energy constraint applied to reconstruct good quality electrons, whereas the upper threshold is applied to include only ECAL signals below the lowest gain switch threshold described in Section~\ref{sec:SS}.
The resulting resolution for the full Run 2 inclusive data set is shown in Fig.~\ref{fig:TimeResZee} as a function of the effective energy of the dielectron system, which depends on the individual energies of the two electrons measured in the two seed crystals as $E_{\text{eff}} = E_1 E_2 /\sqrt{\smash[b]{E_1^2 +E_2^2}}$.
For 2017 data, the Legacy calibration is used.
The resolution is extracted as the $\sigma$ parameter of a Gaussian fit to the core of the distribution of the time difference between the two electrons.
The trend of the ECAL timing resolution as a function of $E_{\text{eff}}$ is modeled with a polynomial function as $\sigma(t_1-t_2) = N/E_{\text{eff}}\oplus \sqrt{2}C$ where $N$ represents the noise term and $C$ is the constant term and dominates at energies above 30-40\GeV.
The noise term $N$ is very similar to that obtained prior to collisions~\cite{Chatrchyan:2009aj} and the constant term $C$ is about 200\unit{ps}.

\section{Electron and photon reconstruction performance in PbPb collisions} \label{sec:sec10}
The quark-gluon plasma (QGP), a deconfined state of matter that is predicted~\cite{Bjorken} by quantum chromodynamics to exist at high temperatures and energy densities, is expected to be produced by colliding heavy nuclei at the LHC. Parton scattering reactions with large momentum transfer, which occur very early compared to QGP formation, provide tomographic probes of the plasma~\cite{dEnterria2010}. The outgoing partons interact strongly with the QGP and lose energy. This phenomenon has been observed at BNL RHIC~\cite{Adcox2005,Adams2005,Back2005,Arsene2005} and at CERN LHC~\cite{Chatrchyan2012, Aad2015,Aamodt2011,Aad2010,Chatrchyan2011} using measurements of hadrons with high $\pt$ and of jets, both created by the fragmentation of the high-momentum partons. Electroweak bosons such as photons and $\PZ$ bosons that decay into leptons do not interact strongly with the QGP. The electroweak boson $\pt$ reflects, on average, the initial energy of the associated parton that fragments into a jet, before any energy loss has occurred. Hence, the measurements of jets produced in the same hard scattering as a photon or a $\PZ$ boson have, in contrast to dijet measurements, a controlled configuration of the initial hard scattering.

The degree of overlap of the two colliding heavy nuclei (e.g., Pb) is defined using signals in the HF calorimeters, and is known as ``centrality''. Centrality is determined by the fraction of the total energy deposited in the HF, with 0\% centrality corresponding to the most central collisions.

The typical particle multiplicity in central PbPb collisions is $\mathcal{O}(10^4)$, giving rise to a dense underlying event (UE). For this reason, the reconstruction, identification, and energy correction algorithms must be revised and optimized to perform in the extreme conditions of central PbPb collisions. The PbPb collisions were recorded in 2018 at a nucleon-nucleon center-of-mass energy of $\sqrtsNN = 5.02$\TeV, corresponding to an integrated luminosity of 1.7\nbinv.

\subsection{Electron and photon reconstruction}\label{sec:HIN_pho_ele_reco}

Several changes have been made to the photon and electron reconstruction with respect to the algorithms used in $\Pp\Pp$ collisions. The out-of-time pileup in PbPb collisions is negligible, hence out-of-time hits and photons were excluded from the reconstruction.  

\begin{figure}[h!]
\centering
\includegraphics[ width=\cmsFigWidth]{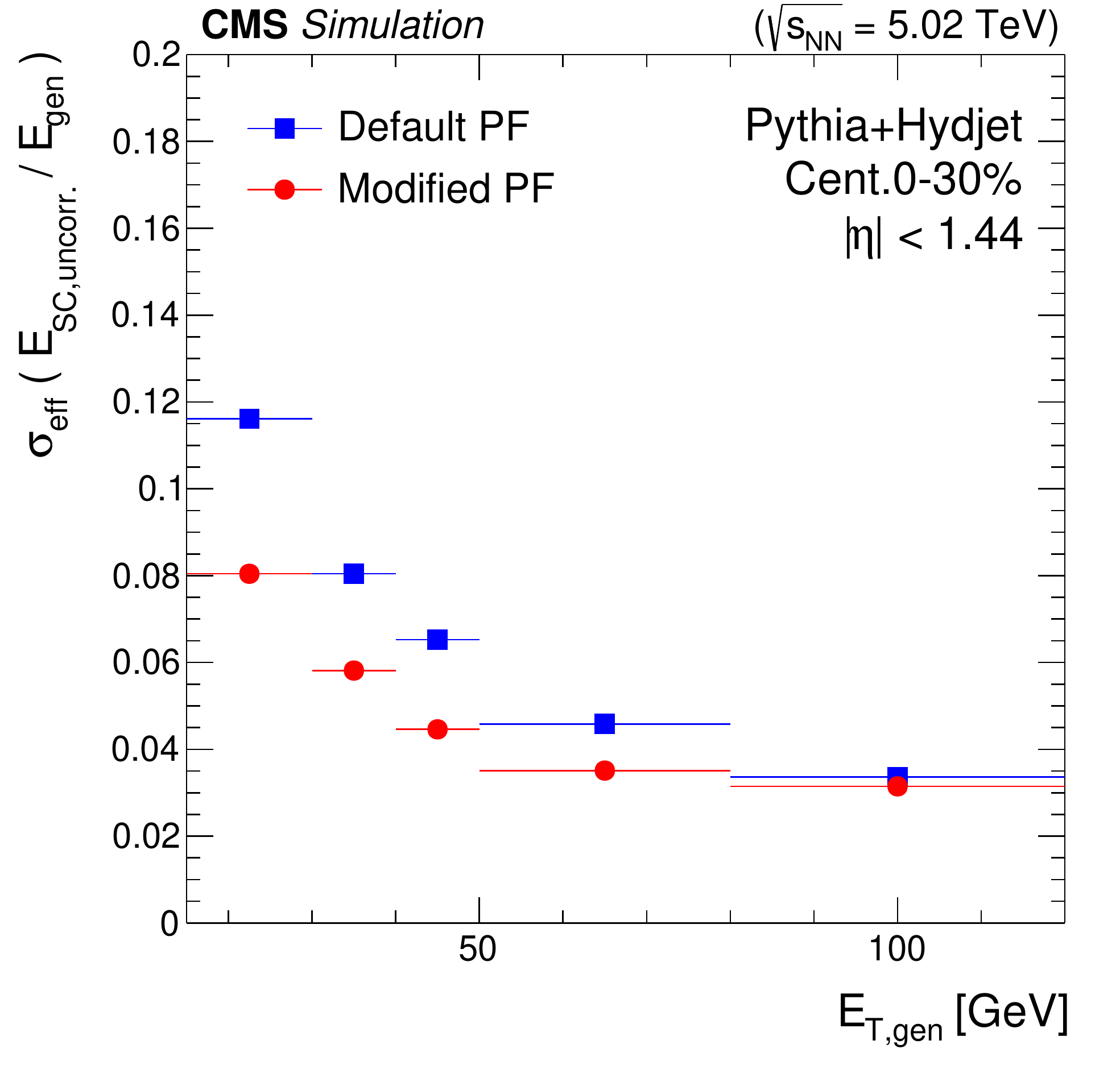}
\caption{Photon energy resolution comparison in simulation for different PF clustering algorithms in PbPb collisions as a function of the generated photon $\ET$. The default algorithm (in blue) has a dynamic $\eta$--$\phi$ window, and is compared to the modified PF clustering algorithm with an upper bound on the SC extent in the $\phi$ direction (in red). Barrel photons ($\abs{\eta} < 1.44$) are used in simulated PbPb events within the 0--30\% centrality range.}
\label{fig:HIN_E}
\end{figure}

The PF ECAL clustering algorithm described in Section~\ref{sec:ecal-clustering} uses a dynamic window in the $\phi$ direction that is dependent on the seed $\ET$ to recover the shower spreads in $\phi$, which are considerable at  low $\ET$. When applied to PbPb events, which have a denser environment of underlying events, this algorithm gives worse energy resolution and higher misidentification rates with respect to $\Pp\Pp$ collisions. To improve the performance, an upper bound of 0.2 was imposed on the extent of the SC in the $\phi$ direction. Following these changes, the misidentification rate decreased from 2.7\% to 0.5\% for photons with $40 < \ET< 60\GeV$, and the energy resolution, estimated from the effective width of the distribution of the ratio of the uncorrected SC energy ($E_{\text{SC,uncorr.}}$) to the true energy ($E_{\text{gen}}$), decreased: as shown in Fig.~\ref{fig:HIN_E}, the modified fixed-width PF algorithm resulted in an energy resolution between 8\% and 3\% at $\ET=20$ and $100\GeV$, respectively.
In the simulation, the effect of the PbPb UE is modeled by embedding the \PYTHIA output created with \textsc{CMS} \PYTHIA8 \textsc{CP5} set of parameters~\cite{Sirunyan2020} in events generated using 
\HYDJET~\cite{Lokhtin2006}, which is tuned to reproduce the charged-particle multiplicity and $\pt$ spectrum in PbPb data. This embedding is applied with an event-by-event weight factor, based on the
average number of nucleon-nucleon collisions calculated with a Glauber model~\cite{Loizides:2017ack} for each PbPb centrality interval.

In PbPb collisions, the large particle multiplicities involved often result in excessively long reconstruction times. As a result, the following modifications were made to the PbPb reconstruction algorithm to keep the reconstruction timing at a reasonable level: (i) the tracker-driven electron seeds were removed, (ii) the tracking regions were changed to be centered in a narrow region around the primary vertex, and (iii) the SC energy was required to be at least 15\GeV. These changes resulted in an improvement of the overall reconstruction timing by a factor of more than $5$. The reconstruction performance for electrons with $\ET > 20\GeV$, the kinematic region of interest to the majority of analyses, was not affected.

\subsection{Electron identification and selection efficiency}\label{sec:HIN_eleID_eff}\label{sec:HIN_eleID_enCorr}

As described in Section~\ref{sec:eleID}, several strategies are used in CMS to identify prompt isolated electrons and to separate them from background sources, such as photon conversions, jets misidentified as electrons, or electrons from semileptonic decays of \cPqc or \cPqb quarks. A cut-based technique was chosen for the electron identification in PbPb collisions, using shower-shape and track-related variables to separate the signal from the background. The selection requirements are optimized using the TMVA framework, with the working point target efficiencies remaining the same as in $\Pp\Pp$ collisions. The input variables are $\sigma_{i \eta i \eta}$, $H/E$ ratio computed from a single tower, $1/E - 1/p$, $\abs{\Delta \eta^{\text{seed}}_{\text{in}}}$ between the ECAL seed crystal and the associated  track, and $\abs{\Delta \phi_{\text{in}}}$ between the ECAL SC and the associated track. An optimization is performed in two centrality bins (0--30\% (central) and 30--100\% (peripheral)), since most of the included variables are centrality dependent. Variables that do not depend on centrality, i.e., the number of expected missing inner hits and three-dimensional impact parameter, were optimized in a second step.

The efficiency of electron reconstruction and identification selection requirements is estimated in data and simulation using the tag-and-probe method, as described in Section~\ref{sec:tnp}. Events are required to pass standard calorimeter noise filters, to trigger the single-electron HLT with an $\ET$ threshold of 20\GeV, and to have a primary vertex position $\abs{v_z} < 15\unit{cm}$. The event must contain at least two reconstructed electrons. Each electron has to be within the acceptance region ($20< \ET <200\GeV$, $\abs{\eta} < 2.1$), and should not be in the barrel-endcap transition region or in the problematic HCAL region for 2018 PbPb data, because in 2018, a 40$^{\circ}$ section of one end of the hadronic endcap calorimeter lost power during the data-taking period. All PbPb data is affected by this power loss.
Tag-and-probe electrons are defined as described in Section~\ref{sec:tnp}.
The tag-and-probe pairs are required to be oppositely charged and to have an invariant mass in the range 60--120\GeV. For the loose identification working point the data-to-simulation correction factor is smaller than 3\%, both in the barrel and the endcaps.

Several sources of systematic uncertainty are considered. The main uncertainty is related to the model used in the mass fit, and is estimated by comparing alternative distributions for signal and background. The second most important uncertainty is related to the tag requirement, varied from the tight to the medium working points. The total systematic uncertainty in the loose identification working point data-to-simulation correction factor is 2.0--4.5 (2.0--7.5)\% in the barrel (endcaps).

\subsection{Electron energy corrections}\label{sec:HIN_ele_enCorr}
In heavy ion collisions, the UE activity can vary greatly between the most central and peripheral collisions. The additional energy deposited by particles from the UE in the ECAL can be clustered together with the energy deposited from genuine electrons, and thus affect the energy scale of reconstructed electrons in a centrality-dependent manner. The electron energy scale is studied using control samples in data (as described in Section~\ref{sec:EnergyReg}) based on the invariant mass of the $\PZ$ boson, which is known precisely and within 5\% of the MC scale. The electron energy scale and resolution extracted from this study are used to correct the energy scale, and to smear the electron energy resolution in simulated samples, to match those observed in data.

The invariant mass of dielectron pairs from $\PZ\to \Pe\Pe$ decays is constructed from the ECAL energy component in three categories corresponding to the detector regions in which the two electrons are reconstructed, namely the EB and EE regions. The events are further subdivided into three centrality regions: 0--10, 10--30, and 30--100\%. The electrons are required to have a minimum $\ET$ of 20\GeV, pass the loose identification selection, and to be located outside the ECAL transition region or the problematic HCAL region. The invariant mass distributions are fitted with a DSCB distribution, from which the mean values are extracted. This is performed separately for data and simulation, and the ratio of the extracted mean values to the world-average $\PZ$ boson mass~\cite{Zyla2020} is used as a correction factor applied to the mean energy scale. The energy resolutions are extracted after first applying the scale factors derived to shift the invariant mass distributions back to the nominal $\PZ$ boson mass. The energy scale and resolution spreading correction factors are applied to the ECAL energy component of the reconstructed electrons, with the final electron momentum obtained by redoing the ECAL-tracker recombination.
The first two sources of systematic uncertainty are evaluated by constructing the invariant mass distributions after varying the electron selection criteria. The variations considered are to tighten the selection criteria from the loose to the medium working point, and increase the electron $\ET$ threshold from 20 to 25\GeV. The difference between the mean values of the nominal and the varied distributions is used as an estimate of the systematic uncertainty. The residual discrepancy between the corrected and the nominal $\PZ$ boson mass is also assumed as a systematic uncertainty, and is smaller than 1 (3)\% in the barrel (endcap) region.

\begin{figure}[h!]
\centering
\includegraphics[ width=\cmsFigWidth]{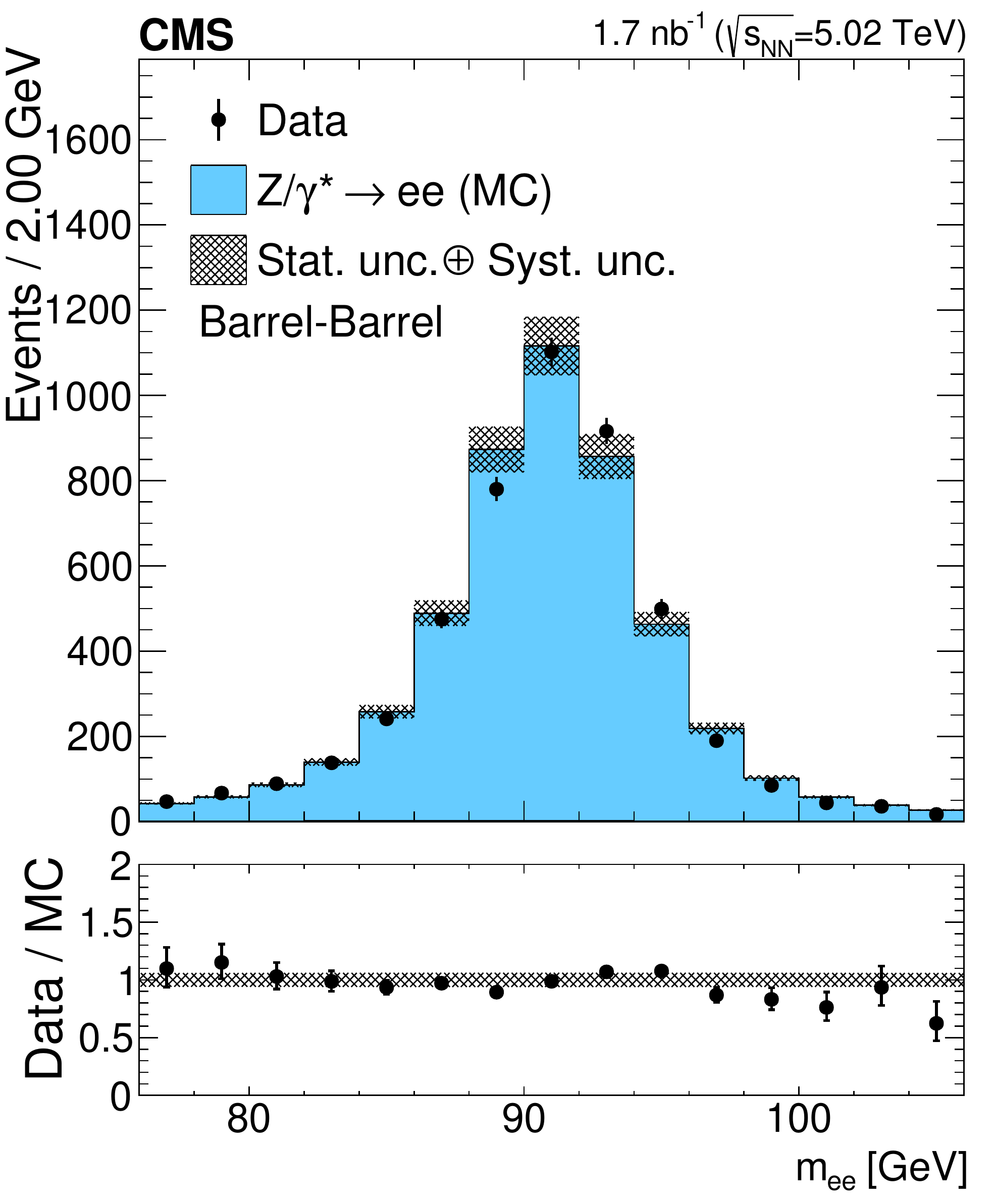}
\includegraphics[ width=\cmsFigWidth]{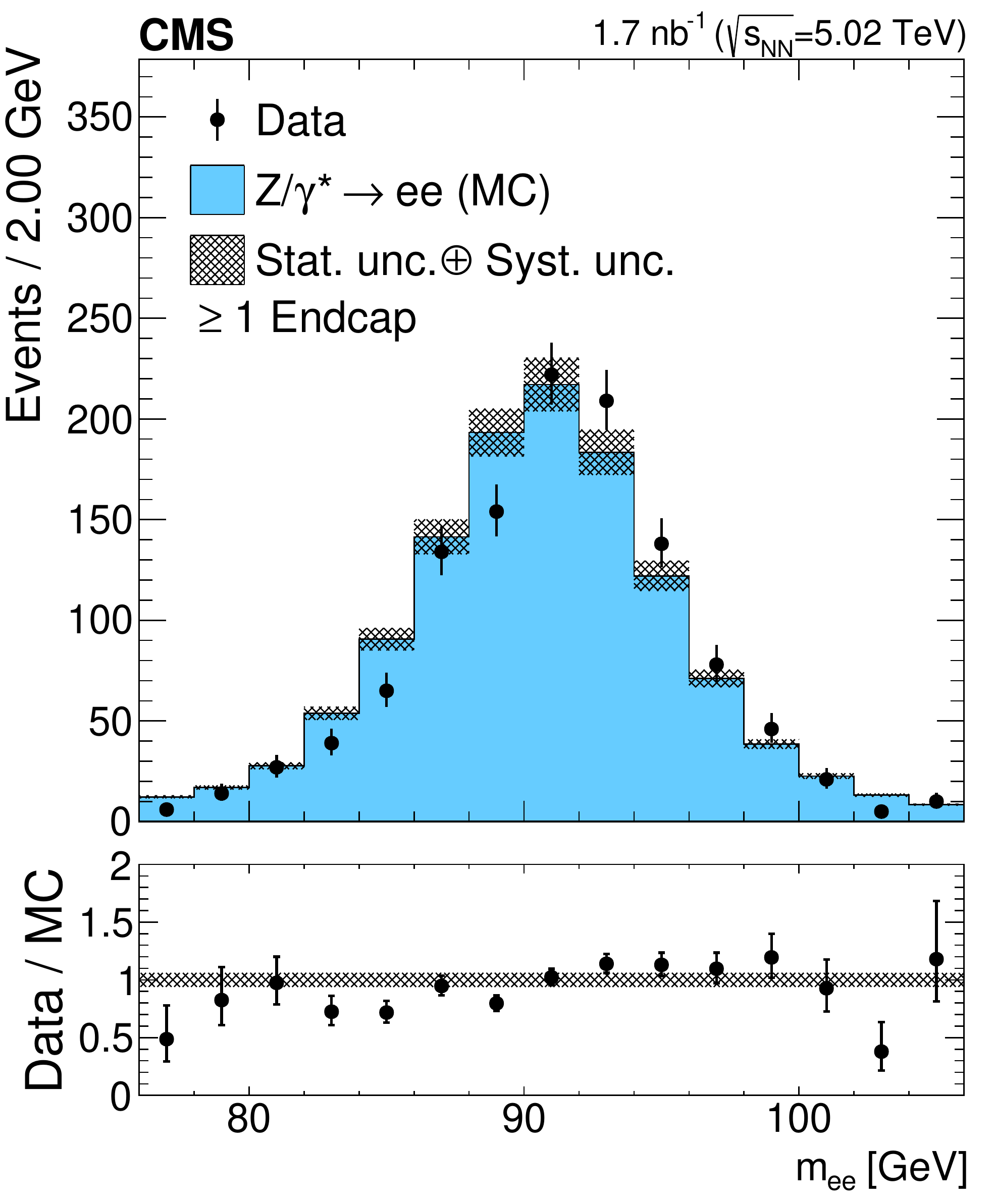}
\caption{Invariant mass distributions of the electron pairs from $\PZ \to \Pe\Pe$ decays selected in PbPb collisions, for the barrel-barrel electrons on the left, and with at least one electron in the endcaps on the right. The simulation is shown with the filled histograms and data are represented by the markers. The vertical bars on the markers represent the statistical uncertainties in data. 
The hatched regions show the combined statistical and systematic uncertainties in the simulation, with the uncertainty in the shapes of the predicted distributions  being the main contributor. The lower panels display the ratio of data to simulation.}
\label{fig:HINZee}
\end{figure}

A comparison of the $\PZ \to \Pe\Pe$ invariant mass peak between data and simulated Drell--Yan events generated with MadGraph5 at NLO~\cite{Alwall:2014hca} is shown in Fig.~\ref{fig:HINZee}. The electron energy in simulation has been corrected using scale corrections and resolution spreading, and electron reconstruction and identification efficiency corrections have also been applied.

\section{Summary}
The performance of electron and photon reconstruction and identification in CMS during LHC Run 2 was measured using data collected in proton-proton collisions at $\sqrt{s}=13$\TeV in 2016--2018 corresponding to a total integrated luminosity
of 136\fbinv.

A clustering algorithm developed to cope with the increasing pileup conditions is described, together with the use of the new pixel detector with one more layer and a reduced material budget. These are major changes in electron and photon performance with respect to Run 1.

Multivariate algorithms are used to correct the electron and photon energy measured in the electromagnetic calorimeter (ECAL), as well as to estimate the electron momentum by
combining independent measurements in the ECAL and in the tracker.
The overall energy scale and resolution are both calibrated using electrons from $\PZ \to \Pe\Pe$ decays.
The uncertainty in the electron and photon energy scale is within 0.1\% in the barrel, and 0.3\% in the endcaps in the transverse energy ($\ET$) range from 10 to 50\GeV. The stability of this calibration is estimated to be within 2--3\% for higher energies.
The measured energy resolution for electrons produced in $\PZ$ boson decays ranges from 2 to 5\%, depending on electron pseudorapidity and energy loss through bremsstrahlung in the detector material. The energy scale and resolution corrections have been checked for photons using $\PZ\to \mu\mu\gamma$ events and are adequate within the assigned systematic uncertainties.

The performance of electron and photon reconstruction and identification algorithms in data is studied with a tag-and-probe method using $\PZ\to \Pe\Pe$ events.
Good agreement is observed between data and simulation
for most of the variables relevant to both reconstruction and identification.
The reconstruction efficiency in data is better than 95\% in the $\ET$ range from 10
to 500\GeV. The data-to-simulation efficiency ratios, both for electron reconstruction and for
the various electron and photon selections, are compatible with unity within 2\% over the full $\ET$ range, down to an $\ET$ as low as 10 (20)\GeV for electrons (photons) when using the 2017 data reconstructed with the dedicated Legacy calibration.
Identification efficiencies target three working points with selection efficiencies of 70, 80, and 90\%, respectively.

The energy resolution and energy scale measurements, together with the relevant identification efficiencies, remain stable throughout the full Run 2 data-taking period (2016--2018). For the 2017 data-taking period, the dedicated Legacy calibration brings an improvement of up to 50\% in terms of relative energy resolution in the ECAL, as well as an improved agreement between data and simulation, leading to smaller reconstruction and identification efficiency correction over the entire $\ET$ and $\eta$ ranges. As a result of these calibrations the electron and photon reconstruction and identification performance at Run 2 are similar to that of Run 1, despite the increased pileup and radiation damage. The evident success of the dedicated Legacy calibration of 2017 data motivates a plan to pursue the same techniques for the 2016 and 2018 data.

The ECAL timing resolution is crucial at CMS to suppress noncollision backgrounds, as well as to perform dedicated searches for delayed photons or jets predicted in several models of physics beyond the standard model. A global timing resolution of 200\unit{ps} is measured  for electrons from $\PZ$ decays with the full Run 2 collision data.

Excellent performance in electron and photon reconstruction and identification has also been achieved in the case of lead-lead collisions at $\sqrtsNN = 5.02$\TeV in 2018 corresponding to a total integrated luminosity of 1.7\nbinv. Reconstruction, identification, and energy correction algorithms have been revised and optimized to perform in the extreme conditions of high underlying event activity in central lead-lead collisions. For electrons and photons reconstructed in lead-lead collisions, the uncertainty on the energy scale is estimated to be better than 1 (3)\% in the barrel (endcap) region.

\begin{acknowledgments}
\hyphenation{Bundes-ministerium Forschungs-gemeinschaft Forschungs-zentren Rachada-pisek} We congratulate our colleagues in the CERN accelerator departments for the excellent performance of the LHC and thank the technical and administrative staffs at CERN and at other CMS institutes for their contributions to the success of the CMS effort. In addition, we gratefully acknowledge the computing centers and personnel of the Worldwide LHC Computing Grid for delivering so effectively the computing infrastructure essential to our analyses. Finally, we acknowledge the enduring support for the construction and operation of the LHC and the CMS detector provided by the following funding agencies: the Austrian Federal Ministry of Education, Science and Research and the Austrian Science Fund; the Belgian Fonds de la Recherche Scientifique, and Fonds voor Wetenschappelijk Onderzoek; the Brazilian Funding Agencies (CNPq, CAPES, FAPERJ, FAPERGS, and FAPESP); the Bulgarian Ministry of Education and Science; CERN; the Chinese Academy of Sciences, Ministry of Science and Technology, and National Natural Science Foundation of China; the Colombian Funding Agency (COLCIENCIAS); the Croatian Ministry of Science, Education and Sport, and the Croatian Science Foundation; the Research and Innovation Foundation, Cyprus; the Secretariat for Higher Education, Science, Technology and Innovation, Ecuador; the Ministry of Education and Research, Estonian Research Council via PRG780, PRG803 and PRG445 and European Regional Development Fund, Estonia; the Academy of Finland, Finnish Ministry of Education and Culture, and Helsinki Institute of Physics; the Institut National de Physique Nucl\'eaire et de Physique des Particules~/~CNRS, and Commissariat \`a l'\'Energie Atomique et aux \'Energies Alternatives~/~CEA, France; the Bundesministerium f\"ur Bildung und Forschung, the Deutsche Forschungsgemeinschaft (DFG) under Germany's Excellence Strategy -- EXC 2121 ``Quantum Universe" -- 390833306, and Helmholtz-Gemeinschaft Deutscher Forschungszentren, Germany; the General Secretariat for Research and Technology, Greece; the National Research, Development and Innovation Fund, Hungary; the Department of Atomic Energy and the Department of Science and Technology, India; the Institute for Studies in Theoretical Physics and Mathematics, Iran; the Science Foundation, Ireland; the Istituto Nazionale di Fisica Nucleare, Italy; the Ministry of Science, ICT and Future Planning, and National Research Foundation (NRF), Republic of Korea; the Ministry of Education and Science of the Republic of Latvia; the Lithuanian Academy of Sciences; the Ministry of Education, and University of Malaya (Malaysia); the Ministry of Science of Montenegro; the Mexican Funding Agencies (BUAP, CINVESTAV, CONACYT, LNS, SEP, and UASLP-FAI); the Ministry of Business, Innovation and Employment, New Zealand; the Pakistan Atomic Energy Commission; the Ministry of Science and Higher Education and the National Science Centre, Poland; the Funda\c{c}\~ao para a Ci\^encia e a Tecnologia, Portugal; JINR, Dubna; the Ministry of Education and Science of the Russian Federation, the Federal Agency of Atomic Energy of the Russian Federation, Russian Academy of Sciences, the Russian Foundation for Basic Research, and the National Research Center ``Kurchatov Institute"; the Ministry of Education, Science and Technological Development of Serbia; the Secretar\'{\i}a de Estado de Investigaci\'on, Desarrollo e Innovaci\'on, Programa Consolider-Ingenio 2010, Plan Estatal de Investigaci\'on Cient\'{\i}fica y T\'ecnica y de Innovaci\'on 2017--2020, research project IDI-2018-000174 del Principado de Asturias, and Fondo Europeo de Desarrollo Regional, Spain; the Ministry of Science, Technology and Research, Sri Lanka; the Swiss Funding Agencies (ETH Board, ETH Zurich, PSI, SNF, UniZH, Canton Zurich, and SER); the Ministry of Science and Technology, Taipei; the Thailand Center of Excellence in Physics, the Institute for the Promotion of Teaching Science and Technology of Thailand, Special Task Force for Activating Research and the National Science and Technology Development Agency of Thailand; the Scientific and Technical Research Council of Turkey, and Turkish Atomic Energy Authority; the National Academy of Sciences of Ukraine; the Science and Technology Facilities Council, UK; the US Department of Energy, and the US National Science Foundation.

Individuals have received support from the Marie-Curie program and the European Research Council and Horizon 2020 Grant, contract Nos.\ 675440, 724704, 752730, and 765710 (European Union); the Leventis Foundation; the A.P.\ Sloan Foundation; the Alexander von Humboldt Foundation; the Belgian Federal Science Policy Office; the Fonds pour la Formation \`a la Recherche dans l'Industrie et dans l'Agriculture (FRIA-Belgium); the Agentschap voor Innovatie door Wetenschap en Technologie (IWT-Belgium); the F.R.S.-FNRS and FWO (Belgium) under the ``Excellence of Science -- EOS" -- be.h project n.\ 30820817; the Beijing Municipal Science \& Technology Commission, No. Z191100007219010; the Ministry of Education, Youth and Sports (MEYS) of the Czech Republic; the Lend\"ulet (``Momentum") Program and the J\'anos Bolyai Research Scholarship of the Hungarian Academy of Sciences, the New National Excellence Program \'UNKP, the NKFIA research grants 123842, 123959, 124845, 124850, 125105, 128713, 128786, and 129058 (Hungary); the Council of Scientific and Industrial Research, India; the HOMING PLUS program of the Foundation for Polish Science, cofinanced from European Union, Regional Development Fund, the Mobility Plus program of the Ministry of Science and Higher Education, the National Science Center (Poland), contracts Harmonia 2014/14/M/ST2/00428, Opus 2014/13/B/ST2/02543, 2014/15/B/ST2/03998, and 2015/19/B/ST2/02861, Sonata-bis 2012/07/E/ST2/01406; the National Priorities Research Program by Qatar National Research Fund; the Ministry of Science and Higher Education, project no. 0723-2020-0041 (Russia); the Tomsk Polytechnic University Competitiveness Enhancement Program; the Programa de Excelencia Mar\'{i}a de Maeztu, and the Programa Severo Ochoa del Principado de Asturias; the Thalis and Aristeia programs cofinanced by EU-ESF, and the Greek NSRF; the Rachadapisek Sompot Fund for Postdoctoral Fellowship, Chulalongkorn University, and the Chulalongkorn Academic into Its 2nd Century Project Advancement Project (Thailand); the Kavli Foundation; the Nvidia Corporation; the SuperMicro Corporation; the Welch Foundation, contract C-1845; and the Weston Havens Foundation (USA).  
\end{acknowledgments}
\bibliography{auto_generated} 

\providecommand{\href}[2]{#2}\begingroup\raggedright\begin{thebibliography}{10}%
\makeatletter
\providecommand{\hrefCMSnoop }[0]{\@secondoftwo}%
\makeatother
\providecommand{\doi}{\texttt{doi:}\begingroup \urlstyle{tt}\Url}

\bibitem{Evans:2008zzb}
\hrefCMSnoop {}{``{LHC Machine}'',} \textit{ JINST} \textbf{ 3} (2008) S08001,
  \href{http://dx.doi.org/10.1088/1748-0221/3/08/S08001}{\doi{10.1088/1748-0221/3/08/S08001}}.

\bibitem{Chatrchyan:2013lba}
\hrefCMSnoop {}{{CMS Collaboration}, ``{Observation of a new boson with mass
  near 125 GeV in pp collisions at $\sqrt{s}$ = 7 and 8 TeV}'',} \textit{ JHEP}
  \textbf{ 06} (2013) 081,
  \href{http://dx.doi.org/10.1007/JHEP06(2013)081}{\doi{10.1007/JHEP06(2013)081}},
  \href{http://www.arXiv.org/abs/1303.4571}{\texttt{arXiv:1303.4571}}.

\bibitem{Aad:2012tfa}
\hrefCMSnoop {}{{ATLAS Collaboration}, ``{Observation of a new particle in the
  search for the standard model Higgs boson with the ATLAS detector at the
  LHC}'',} \textit{ Phys. Lett. B} \textbf{ 716} (2012) 1,
  \href{http://dx.doi.org/10.1016/j.physletb.2012.08.020}{\doi{10.1016/j.physletb.2012.08.020}},
  \href{http://www.arXiv.org/abs/1207.7214}{\texttt{arXiv:1207.7214}}.

\bibitem{Khachatryan:2015hwa}
\hrefCMSnoop {}{{CMS Collaboration}, ``{Performance of electron reconstruction
  and selection with the CMS detector in proton-proton collisions at $\sqrt{s}$
  = 8 TeV}'',} \textit{ JINST} \textbf{ 10} (2015) P06005,
  \href{http://dx.doi.org/10.1088/1748-0221/10/06/P06005}{\doi{10.1088/1748-0221/10/06/P06005}},
  \href{http://www.arXiv.org/abs/1502.02701}{\texttt{arXiv:1502.02701}}.

\bibitem{Khachatryan:2015iwa}
\hrefCMSnoop {}{{CMS Collaboration}, ``{Performance of photon reconstruction
  and identification with the CMS detector in proton-proton collisions at
  $\sqrt{s}$ = 8 TeV}'',} \textit{ JINST} \textbf{ 10} (2015) P08010,
  \href{http://dx.doi.org/10.1088/1748-0221/10/08/P08010}{\doi{10.1088/1748-0221/10/08/P08010}},
  \href{http://www.arXiv.org/abs/1502.02702}{\texttt{arXiv:1502.02702}}.

\bibitem{CMS-PAS-LUM-17-001}
\href {https://cds.cern.ch/record/2257069}{{CMS Collaboration}, ``{CMS}
  luminosity measurements for the 2016 data-taking period'',} CMS Physics
  Analysis Summary CMS-PAS-LUM-17-001, 2017.

\bibitem{CMS-PAS-LUM-17-004}
\href {https://cds.cern.ch/record/2621960}{{{CMS}} Collaboration, ``{CMS}
  luminosity measurement for the 2017 data-taking period at $\sqrt{s} = 13$
  {TeV}'',} CMS Physics Analysis Summary CMS-PAS-LUM-17-004, 2018.

\bibitem{CMS-PAS-LUM-18-002}
\href {https://cds.cern.ch/record/2676164}{{{CMS}} Collaboration, ``{CMS}
  luminosity measurement for the 2018 data-taking period at $\sqrt{s} = 13$
  {TeV}'',} CMS Physics Analysis Summary CMS-PAS-LUM-18-002, 2019.

\bibitem{Chatrchyan:2008zzk}
\hrefCMSnoop {}{{CMS Collaboration}, ``{The {CMS} experiment at the {CERN}
  {LHC}}'',} \textit{ JINST} \textbf{ 3} (2008) S08004,
  \href{http://dx.doi.org/10.1088/1748-0221/3/08/S08004}{\doi{10.1088/1748-0221/3/08/S08004}}.

\bibitem{Dominguez:1481838}
\href {https://cds.cern.ch/record/1481838}{{CMS Collaboration}, ``{CMS
  technical design report for the pixel detector upgrade}'',} Technical Report
  CERN-LHCC-2012-016, 2012.

\bibitem{Chatrchyan:2014fea}
\hrefCMSnoop {}{{CMS Collaboration}, ``{Description and performance of track
  and primary-vertex reconstruction with the CMS tracker}'',} \textit{ JINST}
  \textbf{ 9} (2014) P10009,
  \href{http://dx.doi.org/10.1088/1748-0221/9/10/P10009}{\doi{10.1088/1748-0221/9/10/P10009}},
  \href{http://www.arXiv.org/abs/1405.6569}{\texttt{arXiv:1405.6569}}.

\bibitem{CERN-LHCC-97-033}
\href {https://cds.cern.ch/record/349375}{{CMS Collaboration}, ``{The CMS
  electromagnetic calorimeter project: technical design report}'',} Technical
  Report CERN-LHCC-97-033, 1997.

\bibitem{Sirunyan:2017ulk}
\hrefCMSnoop {}{{CMS Collaboration}, ``{Particle-flow reconstruction and global
  event description with the CMS detector}'',} \textit{ JINST} \textbf{ 12}
  (2017) P10003,
  \href{http://dx.doi.org/10.1088/1748-0221/12/10/P10003}{\doi{10.1088/1748-0221/12/10/P10003}},
  \href{http://www.arXiv.org/abs/1706.04965}{\texttt{arXiv:1706.04965}}.

\bibitem{Cacciari:2008gp}
\hrefCMSnoop {}{M.~Cacciari, G.~P. Salam, and G.~Soyez, ``{The anti-$\kt$ jet
  clustering algorithm}'',} \textit{ JHEP} \textbf{ 04} (2008) 063,
  \href{http://dx.doi.org/10.1088/1126-6708/2008/04/063}{\doi{10.1088/1126-6708/2008/04/063}},
  \href{http://www.arXiv.org/abs/0802.1189}{\texttt{arXiv:0802.1189}}.

\bibitem{Cacciari:2011ma}
\hrefCMSnoop {}{M.~Cacciari, G.~P. Salam, and G.~Soyez, ``{FastJet user
  manual}'',} \textit{ Eur. Phys. J. C} \textbf{ 72} (2012) 1896,
  \href{http://dx.doi.org/10.1140/epjc/s10052-012-1896-2}{\doi{10.1140/epjc/s10052-012-1896-2}},
  \href{http://www.arXiv.org/abs/1111.6097}{\texttt{arXiv:1111.6097}}.

\bibitem{Khachatryan:2016bia}
\hrefCMSnoop {}{{CMS Collaboration}, ``{The CMS trigger system}'',} \textit{
  JINST} \textbf{ 12} (2017) P01020,
  \href{http://dx.doi.org/10.1088/1748-0221/12/01/P01020}{\doi{10.1088/1748-0221/12/01/P01020}},
  \href{http://www.arXiv.org/abs/1609.02366}{\texttt{arXiv:1609.02366}}.

\bibitem{Perrotta:2015jyu}
\hrefCMSnoop {}{A.~Perrotta, ``{Performance of the CMS high level trigger}'',}
  \textit{ J. Phys. Conf. Ser.} \textbf{ 664} (2015) 082044,
  \href{http://dx.doi.org/10.1088/1742-6596/664/8/082044}{\doi{10.1088/1742-6596/664/8/082044}}.

\bibitem{Sirunyan:2020foa}
\hrefCMSnoop {}{{CMS Collaboration}, ``{Pileup mitigation at CMS in 13 TeV
  data}'',} \textit{ JINST} \textbf{ 15} (2020) P09018,
  \href{http://dx.doi.org/10.1088/1748-0221/15/09/P09018}{\doi{10.1088/1748-0221/15/09/P09018}},
  \href{http://www.arXiv.org/abs/2003.00503}{\texttt{arXiv:2003.00503}}.

\bibitem{Chatrchyan:2013dga}
\hrefCMSnoop {}{{CMS Collaboration}, ``{Energy calibration and resolution of
  the CMS electromagnetic calorimeter in pp collisions at $\sqrt{s} = 7$
  TeV}'',} \textit{ JINST} \textbf{ 8} (2013) P09009,
  \href{http://dx.doi.org/10.1088/1748-0221/8/09/P09009}{\doi{10.1088/1748-0221/8/09/P09009}},
  \href{http://www.arXiv.org/abs/1306.2016}{\texttt{arXiv:1306.2016}}.

\bibitem{Chatrchyan:2014wfa}
\hrefCMSnoop {}{{CMS Collaboration}, ``{Alignment of the CMS tracker with LHC
  and cosmic ray data}'',} \textit{ JINST} \textbf{ 9} (2014) P06009,
  \href{http://dx.doi.org/10.1088/1748-0221/9/06/P06009}{\doi{10.1088/1748-0221/9/06/P06009}},
  \href{http://www.arXiv.org/abs/1403.2286}{\texttt{arXiv:1403.2286}}.

\bibitem{Alwall:2014hca}
J.~Alwall\hrefCMSnoop {}{ {et~al.}, ``{The automated computation of tree-level
  and next-to-leading order differential cross sections, and their matching to
  parton shower simulations}'',} \textit{ JHEP} \textbf{ 07} (2014) 079,
  \href{http://dx.doi.org/10.1007/JHEP07(2014)079}{\doi{10.1007/JHEP07(2014)079}},
  \href{http://www.arXiv.org/abs/1405.0301}{\texttt{arXiv:1405.0301}}.

\bibitem{Sjostrand:2014zea}
T.~Sj{\"o}strand\hrefCMSnoop {}{ {et~al.}, ``{An introduction to PYTHIA
  8.2}'',} \textit{ Comput. Phys. Commun.} \textbf{ 191} (2015) 159,
  \href{http://dx.doi.org/10.1016/j.cpc.2015.01.024}{\doi{10.1016/j.cpc.2015.01.024}},
  \href{http://www.arXiv.org/abs/1410.3012}{\texttt{arXiv:1410.3012}}.

\bibitem{Khachatryan:2015pea}
\hrefCMSnoop {}{{CMS Collaboration}, ``{Event generator tunes obtained from
  underlying event and multiparton scattering measurements}'',} \textit{ Eur.
  Phys. J. C} \textbf{ 76} (2016) 155,
  \href{http://dx.doi.org/10.1140/epjc/s10052-016-3988-x}{\doi{10.1140/epjc/s10052-016-3988-x}},
  \href{http://www.arXiv.org/abs/1512.00815}{\texttt{arXiv:1512.00815}}.

\bibitem{Sirunyan2020}
\hrefCMSnoop {}{{CMS Collaboration}, ``{Extraction and validation of a new set
  of CMS PYTHIA8 tunes from underlying-event measurements}'',} \textit{ Eur.
  Phys. J. C} \textbf{ 80} (2020), no.~1, 4,
  \href{http://dx.doi.org/10.1140/epjc/s10052-019-7499-4}{\doi{10.1140/epjc/s10052-019-7499-4}},
  \href{http://www.arXiv.org/abs/1903.12179}{\texttt{arXiv:1903.12179}}.

\bibitem{Frederix:2012ps}
\hrefCMSnoop {}{R.~Frederix and S.~Frixione, ``{Merging meets matching in
  MC@NLO}'',} \textit{ JHEP} \textbf{ 12} (2012) 061,
  \href{http://dx.doi.org/10.1007/JHEP12(2012)061}{\doi{10.1007/JHEP12(2012)061}},
  \href{http://www.arXiv.org/abs/1209.6215}{\texttt{arXiv:1209.6215}}.

\bibitem{Ball:2014uwa}
\hrefCMSnoop {}{{NNPDF} Collaboration, ``{Parton distributions for the LHC Run
  II}'',} \textit{ JHEP} \textbf{ 04} (2015) 040,
  \href{http://dx.doi.org/10.1007/JHEP04(2015)040}{\doi{10.1007/JHEP04(2015)040}},
  \href{http://www.arXiv.org/abs/1410.8849}{\texttt{arXiv:1410.8849}}.

\bibitem{Ball:2012cx}
\hrefCMSnoop {}{{NNPDF} Collaboration, ``{Parton distributions with LHC
  data}'',} \textit{ Nucl. Phys. B} \textbf{ 867} (2013) 244,
  \href{http://dx.doi.org/10.1016/j.nuclphysb.2012.10.003}{\doi{10.1016/j.nuclphysb.2012.10.003}},
\href{http://www.arXiv.org/abs/1207.1303}{\texttt{arXiv:1207.1303}}.

\bibitem{Agostinelli:2002hh}
\hrefCMSnoop {}{{GEANT4} Collaboration, ``{\GEANTfour} --- a simulation
  toolkit'',} \textit{ Nucl. Instrum. Meth. A} \textbf{ 506} (2003) 250,
  \href{http://dx.doi.org/10.1016/S0168-9002(03)01368-8}{\doi{10.1016/S0168-9002(03)01368-8}}.

\bibitem{Sirunyan:2020pmc}
\hrefCMSnoop {}{{CMS Collaboration}, ``{Reconstruction of signal amplitudes in
  the CMS electromagnetic calorimeter in the presence of overlapping
  proton-proton interactions}'',} \textit{ JINST} \textbf{ 15} (2020) P10002,
  \href{http://dx.doi.org/10.1088/1748-0221/15/10/P10002}{\doi{10.1088/1748-0221/15/10/P10002}},
  \href{http://www.arXiv.org/abs/2006.14359}{\texttt{arXiv:2006.14359}}.

\bibitem{Adam_2005}
\hrefCMSnoop {}{W.~Adam, R.~Fr{\"u}hwirth, A.~Strandlie, and T.~Todorov,
  ``{Reconstruction of electrons with the Gaussian-sum filter in the CMS
  tracker at the LHC}'',} \textit{ J. Phys. G} \textbf{ 31} (2005) 9,
  \href{http://dx.doi.org/10.1088/0954-3899/31/9/n01}{\doi{10.1088/0954-3899/31/9/n01}},
  \href{http://www.arXiv.org/abs/physics/0306087}{\texttt{arXiv:physics/0306087}}.

\bibitem{Hocker:2007ht}
\hrefCMSnoop {}{H.~Voss, A.~H{\"o}cker, J.~Stelzer, and F.~Tegenfeldt,
  ``{TMVA}, the toolkit for multivariate data analysis with {ROOT}'',} in
  \textit{ XIth International Workshop on Advanced Computing and Analysis
  Techniques in Physics Research (ACAT)}, p.~40.
\newblock 2007.
\newblock
  \href{http://www.arXiv.org/abs/physics/0703039}{\texttt{arXiv:physics/0703039}}.
\newblock {[PoS(ACAT)040]}.
\href{http://dx.doi.org/10.22323/1.050.0040}{\doi{10.22323/1.050.0040}}.

\bibitem{QUINLAN1987221}
\hrefCMSnoop {}{J.~R. Quinlan, ``Simplifying decision trees'',} \textit{ Int.
  J. Man-Mach. Stud.} \textbf{ 27} (1987) 221,
  \href{http://dx.doi.org/10.1016/S0020-7373(87)80053-6}{\doi{10.1016/S0020-7373(87)80053-6}}.

\bibitem{PhysRev.93.768}
\hrefCMSnoop {}{H.~A. Bethe and L.~C. Maximon, ``{Theory of bremsstrahlung and
  pair production. I. differential cross section}'',} \textit{ Phys. Rev.}
  \textbf{ 93} (1954) 768,
  \href{http://dx.doi.org/10.1103/PhysRev.93.768}{\doi{10.1103/PhysRev.93.768}}.

\bibitem{Khachatryan:2010xn}
\hrefCMSnoop {}{{CMS Collaboration}, ``{Measurements of inclusive $W$ and $Z$
  cross sections in pp collisions at $\sqrt{s}=7$ TeV}'',} \textit{ JHEP}
  \textbf{ 01} (2011) 080,
  \href{http://dx.doi.org/10.1007/JHEP01(2011)080}{\doi{10.1007/JHEP01(2011)080}},
  \href{http://www.arXiv.org/abs/1012.2466}{\texttt{arXiv:1012.2466}}.

\bibitem{Bohm:2004zi}
\hrefCMSnoop {}{A.~R. Bohm and Y.~Sato, ``{Relativistic resonances: their
  masses, widths, lifetimes, superposition, and causal evolution}'',} \textit{
  Phys. Rev. D} \textbf{ 71} (2005) 085018,
  \href{http://dx.doi.org/10.1103/PhysRevD.71.085018}{\doi{10.1103/PhysRevD.71.085018}},
  \href{http://www.arXiv.org/abs/hep-ph/0412106}{\texttt{arXiv:hep-ph/0412106}}.

\bibitem{Zyla2020}
\hrefCMSnoop {}{{Particle Data Group}, P.~A. Zyla {et~al.}, ``Review of
  particle physics'',} \textit{ Prog. Theor. Exp. Phys.} \textbf{ 2020} (2020)
  083C01,
  \href{http://dx.doi.org/10.1093/ptep/ptaa104}{\doi{10.1093/ptep/ptaa104}}.

\bibitem{Oreglia:1980cs}
\href {http://www.slac.stanford.edu/cgi-wrap/getdoc/slac-r-236.pdf}{M.~J.
  Oreglia, ``A study of the reactions $\psi^\prime \to \gamma \gamma \psi$''}.
\newblock PhD thesis, Stanford University, 1980.
\newblock {SLAC} Report {SLAC-R-236}.

\bibitem{CMS-PAS-TRG-17-001}
\hrefCMSnoop {}{{CMS Collaboration}, ``{Performance of the CMS Level-1 trigger
  in proton-proton collisions at $\sqrt{s} =$ 13 TeV}'',} \textit{ JINST}
  \textbf{ 15} (2020) P10017,
  \href{http://dx.doi.org/10.1088/1748-0221/15/10/P10017}{\doi{10.1088/1748-0221/15/10/P10017}},
  \href{http://www.arXiv.org/abs/2006.10165}{\texttt{arXiv:2006.10165}}.

\bibitem{Khachatryan:2014ira}
\hrefCMSnoop {}{{CMS Collaboration}, ``{Observation of the diphoton decay of
  the Higgs boson and measurement of its properties}'',} \textit{ Eur. Phys. J.
  C} \textbf{ 74} (2014) 3076,
  \href{http://dx.doi.org/10.1140/epjc/s10052-014-3076-z}{\doi{10.1140/epjc/s10052-014-3076-z}},
  \href{http://www.arXiv.org/abs/1407.0558}{\texttt{arXiv:1407.0558}}.

\bibitem{Sirunyan:2018fpa}
\hrefCMSnoop {}{{CMS Collaboration}, ``{Performance of the CMS muon detector
  and muon reconstruction with proton-proton collisions at $\sqrt{s}=$ 13
  TeV}'',} \textit{ JINST} \textbf{ 13} (2018) P06015,
  \href{http://dx.doi.org/10.1088/1748-0221/13/06/P06015}{\doi{10.1088/1748-0221/13/06/P06015}},
  \href{http://www.arXiv.org/abs/1804.04528}{\texttt{arXiv:1804.04528}}.

\bibitem{Sirunyan:2020xwk}
\hrefCMSnoop {}{{CMS Collaboration}, ``{A measurement of the Higgs boson mass
  in the diphoton decay channel}'',} \textit{ Phys. Lett. B} \textbf{ 805}
  (2020) 135423,
  \href{http://dx.doi.org/10.1016.j.physletb.2020.135425}{\doi{10.1016.j.physletb.2020.135425}}.

\bibitem{Chen:2016btl}
\hrefCMSnoop {}{T.~Chen and C.~Guestrin, ``{XGBoost: a scalable tree boosting
  system}'',} 2016.
  \href{http://www.arXiv.org/abs/1603.02754}{\texttt{arXiv:1603.02754}}.

\bibitem{CMS-PAS-EXO-19-019}
\hrefCMSnoop {}{{CMS Collaboration}, ``{Search for contact interactions and
  large extra dimensions in the dilepton mass spectra from proton-proton
  collisions at $\sqrt{s} =$ 13 TeV}'',} \textit{ JHEP} \textbf{ 04} (2019)
  114,
  \href{http://dx.doi.org/10.1007/JHEP04(2019)114}{\doi{10.1007/JHEP04(2019)114}},
  \href{http://www.arXiv.org/abs/1812.10443}{\texttt{arXiv:1812.10443}}.

\bibitem{Chatrchyan:2009aj}
\hrefCMSnoop {}{{CMS Collaboration}, ``{Time reconstruction and performance of
  the CMS electromagnetic calorimeter}'',} \textit{ JINST} \textbf{ 5} (2010)
  T03011,
  \href{http://dx.doi.org/10.1088/1748-0221/5/03/T03011}{\doi{10.1088/1748-0221/5/03/T03011}},
  \href{http://www.arXiv.org/abs/0911.4044}{\texttt{arXiv:0911.4044}}.

\bibitem{CMS-PAS-EXO-16-036}
\hrefCMSnoop {}{{CMS Collaboration}, ``{Search for long-lived charged particles
  in proton-proton collisions at $\sqrt{s}=$ 13 TeV}'',} \textit{ Phys. Rev. D}
  \textbf{ 94} (2016) 112004,
  \href{http://dx.doi.org/10.1103/PhysRevD.94.112004}{\doi{10.1103/PhysRevD.94.112004}},
  \href{http://www.arXiv.org/abs/1609.08382}{\texttt{arXiv:1609.08382}}.

\bibitem{Chatrchyan:2012jwg}
\hrefCMSnoop {}{{CMS Collaboration}, ``{Search for long-lived particles
  decaying to photons and missing energy in proton-proton collisions at
  $\sqrt{s}=7$ TeV}'',} \textit{ Phys. Lett. B} \textbf{ 722} (2013) 273,
  \href{http://dx.doi.org/10.1016/j.physletb.2013.04.027}{\doi{10.1016/j.physletb.2013.04.027}},
  \href{http://www.arXiv.org/abs/1212.1838}{\texttt{arXiv:1212.1838}}.

\bibitem{Sirunyan:2019wau}
\hrefCMSnoop {}{{CMS Collaboration}, ``{Search for long-lived particles using
  delayed photons in proton-proton collisions at $\sqrt{s}=$ 13 TeV}'',}
  \textit{ Phys. Rev. D} \textbf{ 100} (2019) 112003,
  \href{http://dx.doi.org/10.1103/PhysRevD.100.112003}{\doi{10.1103/PhysRevD.100.112003}},
  \href{http://www.arXiv.org/abs/1909.06166}{\texttt{arXiv:1909.06166}}.

\bibitem{Bjorken}
\href {http://cds.cern.ch/record/141477}{J.~D. Bjorken, ``{Energy loss of
  energetic partons in quark-gluon plasma: possible extinction of high p$_{T}$
  jets in hadron-hadron collisions}'',} Technical Report
  FERMILAB-PUB-82-59-THY, FERMILAB, Batavia, IL, Aug, 1982.

\bibitem{dEnterria2010}
\hrefCMSnoop {}{D.~d'Enterria, ``Jet quenching'',} in \textit{ Relativistic
  Heavy Ion Physics}, p.~471.
\newblock Springer Berlin Heidelberg, 2010.
\newblock
  \href{http://dx.doi.org/10.1007/978-3-642-01539-7_16}{\doi{10.1007/978-3-642-01539-7_16}}.

\bibitem{Adcox2005}
\hrefCMSnoop {}{{PHENIX} Collaboration, ``Formation of dense partonic matter in
  relativistic nucleus{\textendash}nucleus collisions at {RHIC}: Experimental
  evaluation by the {PHENIX} collaboration'',} \textit{ Nucl. Phys. A} \textbf{
  757} (2005) 184,
  \href{http://dx.doi.org/10.1016/j.nuclphysa.2005.03.086}{\doi{10.1016/j.nuclphysa.2005.03.086}},
  \href{http://www.arXiv.org/abs/nucl-ex/0410003}{\texttt{arXiv:nucl-ex/0410003}}.

\bibitem{Adams2005}
\hrefCMSnoop {}{{STAR} Collaboration, ``Experimental and theoretical challenges
  in the search for the quark gluon plasma: The {STAR} collaboration critical
  assessment of the evidence from {RHIC} collisions'',} \textit{ Nucl. Phys. A}
  \textbf{ 757} (2005) 102,
  \href{http://dx.doi.org/10.1016/j.nuclphysa.2005.03.085}{\doi{10.1016/j.nuclphysa.2005.03.085}},
  \href{http://www.arXiv.org/abs/nucl-ex/0501009}{\texttt{arXiv:nucl-ex/0501009}}.

\bibitem{Back2005}
\hrefCMSnoop {}{{PHOBOS} Collaboration, ``The {PHOBOS} perspective on
  discoveries at {RHIC}'',} \textit{ Nucl. Phys. A} \textbf{ 757} (2005) 28,
  \href{http://dx.doi.org/10.1016/j.nuclphysa.2005.03.084}{\doi{10.1016/j.nuclphysa.2005.03.084}},
  \href{http://www.arXiv.org/abs/nucl-ex/0410022}{\texttt{arXiv:nucl-ex/0410022}}.

\bibitem{Arsene2005}
\hrefCMSnoop {}{{BRAHMS} Collaboration, ``Quark gluon plasma and color glass
  condensate at {RHIC}? {T}he perspective from the {BRAHMS} experiment'',}
  \textit{ Nucl. Phys. A} \textbf{ 757} (2005) 1,
  \href{http://dx.doi.org/10.1016/j.nuclphysa.2005.02.130}{\doi{10.1016/j.nuclphysa.2005.02.130}},
  \href{http://www.arXiv.org/abs/nucl-ex/0410020}{\texttt{arXiv:nucl-ex/0410020}}.

\bibitem{Chatrchyan2012}
\hrefCMSnoop {}{{CMS Collaboration}, ``Study of high-$\pt$ charged particle
  suppression in {PbPb} compared to pp collisions at {$\sqrt{s_{\text{NN}}} =
  5.02$\TeV}'',} \textit{ Euro. Phys. J. C} \textbf{ 72} (2012) 1945,
  \href{http://dx.doi.org/10.1140/epjc/s10052-012-1945-x}{\doi{10.1140/epjc/s10052-012-1945-x}},
  \href{http://www.arXiv.org/abs/1202.2554}{\texttt{arXiv:1202.2554}}.

\bibitem{Aad2015}
\hrefCMSnoop {}{{ATLAS Collaboration}, ``Measurement of charged particle
  spectra in {PbPb} collisions at {$\sqrt{s_{\text{NN}}} = 5.02$\TeV} with the
  {ATLAS} detector at the {LHC}'',} \textit{ JHEP} \textbf{ 09} (2015) 050,
  \href{http://dx.doi.org/10.1007/jhep09(2015)050}{\doi{10.1007/jhep09(2015)050}},
  \href{http://www.arXiv.org/abs/1504.04337}{\texttt{arXiv:1504.04337}}.

\bibitem{Aamodt2011}
\hrefCMSnoop {}{{ALICE Collaboration}, ``Suppression of charged particle
  production at large transverse momentum in central {PbPb} collisions at
  {$\sqrt{s_{\text{NN}}} = 5.02$\TeV}'',} \textit{ Phys. Lett. B} \textbf{ 696}
  (2011) 30,
  \href{http://dx.doi.org/10.1016/j.physletb.2010.12.020}{\doi{10.1016/j.physletb.2010.12.020}},
  \href{http://www.arXiv.org/abs/1012.1004}{\texttt{arXiv:1012.1004}}.

\bibitem{Aad2010}
\hrefCMSnoop {}{{ATLAS Collaboration}, ``Observation of a centrality-dependent
  dijet asymmetry in lead-lead collisions {$\sqrt{s_{\text{NN}}} = 5.02$\TeV}
  with the {ATLAS} detector at the {LHC}'',} \textit{ Phys. Rev. Lett.}
  \textbf{ 105} (2010) 252303,
  \href{http://dx.doi.org/10.1103/physrevlett.105.252303}{\doi{10.1103/physrevlett.105.252303}},
  \href{http://www.arXiv.org/abs/1011.6182}{\texttt{arXiv:1011.6182}}.

\bibitem{Chatrchyan2011}
\hrefCMSnoop {}{{CMS Collaboration}, ``{Observation and studies of jet
  quenching in PbPb collisions at $\sqrt{s_{\text{NN}}} = 5.02$ TeV}'',}
  \textit{ Phys. Rev. C} \textbf{ 84} (2011) 024906,
  \href{http://dx.doi.org/10.1103/PhysRevC.84.024906}{\doi{10.1103/PhysRevC.84.024906}},
  \href{http://www.arXiv.org/abs/1102.1957}{\texttt{arXiv:1102.1957}}.

\bibitem{Lokhtin2006}
\hrefCMSnoop {}{I.~P. Lokhtin and A.~M. Snigirev, ``A model of jet quenching in
  ultrarelativistic heavy ion collisions and high-$\pt$ hadron spectra at
  {RHIC}'',} \textit{ Eur. Phys. J. C} \textbf{ 45} (2006) 211,
  \href{http://dx.doi.org/10.1140/epjc/s2005-02426-3}{\doi{10.1140/epjc/s2005-02426-3}},
  \href{http://www.arXiv.org/abs/hep-ph/0506189}{\texttt{arXiv:hep-ph/0506189}}.

\bibitem{Loizides:2017ack}
\hrefCMSnoop {}{C.~Loizides, J.~Kamin, and D.~d'Enterria, ``{Improved Monte
  Carlo Glauber predictions at present and future nuclear colliders}'',}
  \textit{ Phys. Rev. C} \textbf{ 97} (2018) 054910,
  \href{http://dx.doi.org/10.1103/PhysRevC.97.054910}{\doi{10.1103/PhysRevC.97.054910}},
  \href{http://www.arXiv.org/abs/1710.07098}{\texttt{arXiv:1710.07098}}.
  [Erratum: \DOI{10.1103/physrevc.99.019901}].

\end{thebibliography}\endgroup
\cleardoublepage \appendix\section{The CMS Collaboration \label{app:collab}}\begin{sloppypar}\hyphenpenalty=5000\widowpenalty=500\clubpenalty=5000\vskip\cmsinstskip
\textbf{Yerevan Physics Institute, Yerevan, Armenia}\\*[0pt]
A.M.~Sirunyan$^{\textrm{\dag}}$, A.~Tumasyan
\vskip\cmsinstskip
\textbf{Institut f\"{u}r Hochenergiephysik, Wien, Austria}\\*[0pt]
W.~Adam, T.~Bergauer, M.~Dragicevic, A.~Escalante~Del~Valle, R.~Fr\"{u}hwirth\cmsAuthorMark{1}, M.~Jeitler\cmsAuthorMark{1}, N.~Krammer, L.~Lechner, D.~Liko, I.~Mikulec, F.M.~Pitters, N.~Rad, J.~Schieck\cmsAuthorMark{1}, R.~Sch\"{o}fbeck, M.~Spanring, S.~Templ, W.~Waltenberger, C.-E.~Wulz\cmsAuthorMark{1}, M.~Zarucki
% ...
\vskip\cmsinstskip
\dag: Deceased\\
1:  Also at Vienna University of Technology, Vienna, Austria\\
% ...
\end{sloppypar}
\end{document}